\documentclass[sigconf]{acmart}

\AtBeginDocument{%
  \providecommand\BibTeX{{%
    \normalfont B\kern-0.5em{\scshape i\kern-0.25em b}\kern-0.8em\TeX}}}

\copyrightyear{2025}
\acmYear{2025}
\setcopyright{cc}
\setcctype{by}
\acmConference[CHI '25]{CHI Conference on Human Factors in Computing Systems}{April 26-May 1, 2025}{Yokohama, Japan}
\acmBooktitle{CHI Conference on Human Factors in Computing Systems (CHI '25), April 26-May 1, 2025, Yokohama, Japan}\acmDOI{10.1145/3706598.3714319}
\acmISBN{979-8-4007-1394-1/25/04}

\ccsdesc[500]{Human-centered computing~Empirical studies in HCI}
\ccsdesc[500]{Human-centered computing~User studies}
\ccsdesc[500]{Human-centered computing~User interface management systems}
\ccsdesc[500]{Applied computing~Annotation}
\ccsdesc[500]{Computing methodologies~Natural language processing}
\ccsdesc[500]{Human-centered computing~User interface programming}

\keywords{Data Annotation; Data Labeling; Large Language Model; Iterative labeling; End-User Programming}
\title[Prompting in the Dark]{Prompting in the Dark: Assessing Human Performance in Prompt Engineering for Data Labeling When Gold Labels Are Absent}
\usepackage{xspace}
\usepackage{multirow}
\usepackage{subcaption}

\newcommand{\system}{\mbox{\textsc{PromptingSheet}}\xspace}

\newcommand{\kenneth}[1]{{\small\textcolor{blue}{\bf [#1 --Ken]}}}
\newcommand{\steven}[1]{{\small\textcolor{red}{\bf [#1 --Steven]}}}
\newcommand{\saniya}[1]{{\small\textcolor{green}{\bf [#1 --Saniya]}}}

\newcommand{\eg}{{\it e.g.}}
\newcommand{\ie}{{\it i.e.}}

\begin{document}



\author{Zeyu He}
\affiliation{%
  \institution{The Pennsylvania State University}
  %\streetaddress{P.O. Box 1212}
  \city{University Park}
  \state{PA}
  \country{USA}
}
\email{zmh5268@psu.edu}

\author{Saniya Naphade}
\affiliation{%
  \institution{GumGum Inc.}
  %\streetaddress{P.O. Box 1212}
  \city{Tempe}
  \state{AZ}
  \country{USA}
}
\email{saniya.naphade@gumgum.com}

\author{Ting-Hao `Kenneth' Huang}
\affiliation{%
  \institution{The Pennsylvania State University}
  %\streetaddress{P.O. Box 1212}
  \city{University Park}
  \state{PA}
  \country{USA}
}
\email{txh710@psu.edu}

\renewcommand{\shortauthors}{He et al.}

\begin{abstract}
Millions of users prompt large language models (LLMs) for various tasks, but how good are people at prompt engineering?
Do users actually get closer to their desired outcome over multiple iterations of their prompts? 
These questions are crucial when no gold-standard labels are available to measure progress. 
This paper investigates a scenario in LLM-powered data labeling, ``prompting in the dark,'' where users iteratively prompt LLMs to label data without using manually-labeled benchmarks.
We developed \system, a Google Sheets add-on that enables users to compose, revise, and iteratively label data through spreadsheets.
Through a study with 20 participants, we found that prompting in the dark was highly unreliable---only 9 participants improved labeling accuracy after four or more iterations.
% \kenneth{Update numbers}\steven{Added! Will update after user study finished}
Automated prompt optimization tools like DSPy also struggled when few gold labels were available.
%Our findings highlight the importance of gold labels and automated support in human prompt engineering, informing future tool design.
Our findings highlight the importance of gold labels and the needs, as well as the risks, of automated support in human prompt engineering, providing insights for future tool design.

\end{abstract}

\maketitle

\section{Introduction}\label{sec:intro}
\begin{figure*}[t]
    \centering
    \includegraphics[width=0.99\linewidth]{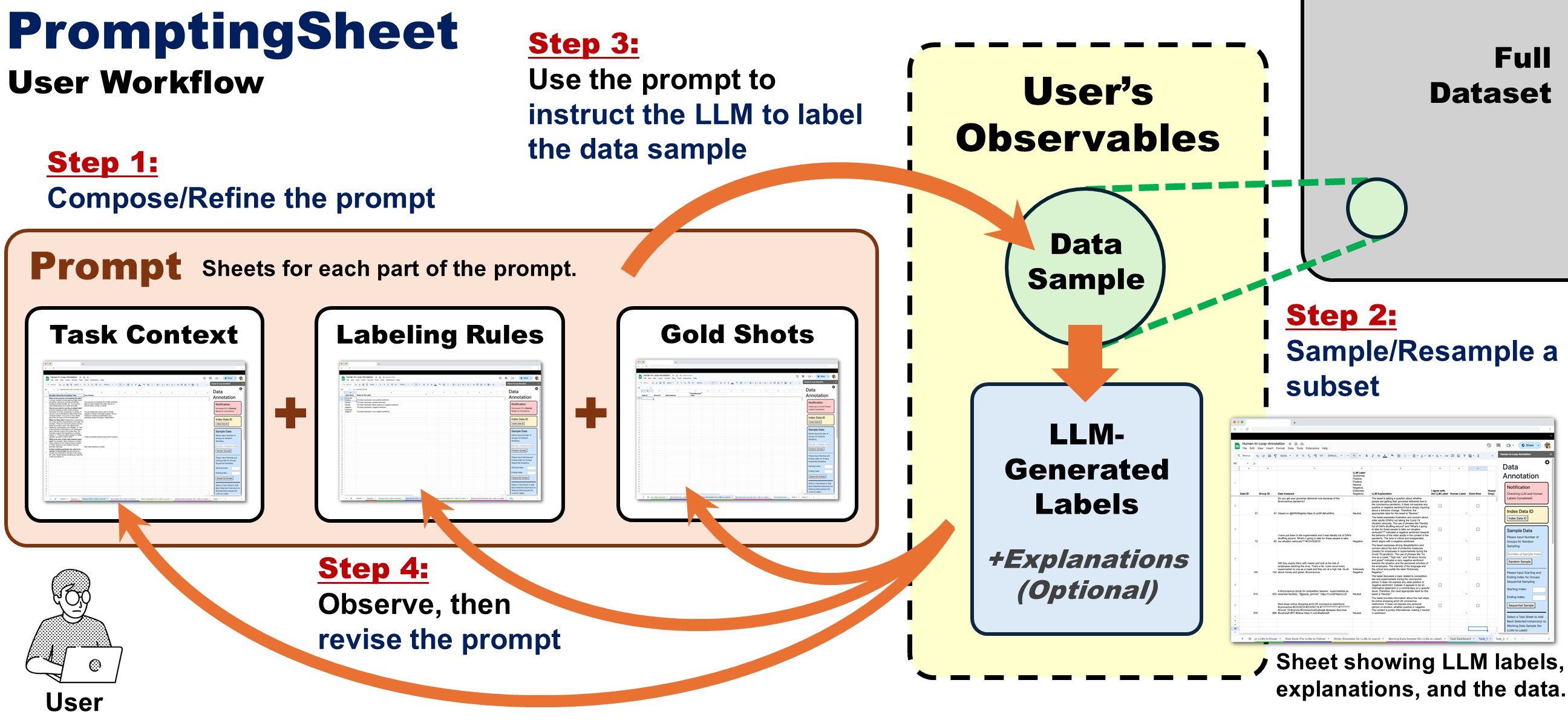}\Description{This is our user workflow which includes four iterative steps. Before the steps, users need to have a full dataset imported. In Step 1, users need to compose/refine the prompt by working on Context, Rule Book, and Shots sheets. In Step 2, users need to sample/resample a subset of data from the dataset. In Step 3, after users click Start Annotation, the system will use the prompt from step 1 to instruct the LLM to label the data sample. Then annotated results will be saved in a new tab. In Step 4, users will observe the annotation results. If they want to keep improving the LLM performance, they will need to go back to Step 1 to refine the prompt. If they are satisfied with the results, they can end the annotation workflow.}
    \caption{\system is a Google Sheets add-on that allows users to compose prompts~(Step 1), use those prompts to instruct LLMs to label data~(Steps 2 and 3), review the resulting labels and optional explanations, and iteratively revise and relabel data~(Step 4)---all within the same Google Sheets document. The process does not begin with users manually labeling data; instead, users' understanding of the data and labeling scheme evolves as they interact with the LLM and review its outputs.}
    \label{fig:system-workflow-fig-v2}
\end{figure*}

Large language models (LLMs) have empowered millions, if not billions, to perform a wide range of programming and data science tasks, even without formal technical backgrounds.
%\kenneth{Maybe do not say without formal technical backgrounds. Maybe people work in tech but without formal software engineering background....? hmm}\steven{sounds good}
People can ask LLMs to teach them step-by-step how to build a web app from scratch~\cite{voronin2024development}, %\kenneth{TODO: Add one more example}\steven{added}
have LLMs analyze data and generate insights~\cite{ma-etal-2023-insightpilot,laradji2023capture}, %\kenneth{TODO: Add one more example here} \steven{added}
or instruct LLMs to label thousands of data items~\cite{10.1145/3613904.3642834,horych2024promises,he-etal-2024-annollm}. %\kenneth{TODO: Add two more examples} \steven{added}
All these were made possible by LLMs' ability to converse in natural language and to follow users' instructions, also known as ``prompts.'' 
The practice of ``prompt engineering'' has emerged to describe the process of developing and optimizing prompts to use LLMs efficiently~\cite{chen2023unleashing,wang2023review,giray2023prompt}.
Tools like LangChain\footnote{LangChain: https://www.langchain.com/} and ChainForge~\cite{arawjo2024chainforge} were developed to facilitate complex chaining of prompts; technologies such as DSPy~\cite{khattab2023dspy} were also created to automate prompt optimization.
\textbf{However, how effective are people at prompt engineering in practice? 
Do users really get closer to their desired outcome over multiple iterations of their prompts?}
These questions are especially important when gold-standard labels are unavailable.
Decades of research indicates that gold-standard labels---data inputs paired with known correct predictions or ideal outputs (also referred to as ``gold labels'' or simply ``gold''~\cite{abdalla2023hurdles,wang2024impact, snow2008cheap,kilgarriff1998gold,silveira2014gold, wiebe1999development,sorokin2008utility}---are critical for evaluating annotation quality~\cite{han2020crowd,gadiraju2015training,nahum2024llms} and system performance~\cite{daka2014survey,ellims2004unit,hamill2004unit}. 
%\kenneth{TODO: We kinda sai this is ``decades of research'', so please find some older reference with a high citation count}\steven{Sure. Done}
However, in real-world scenarios, gold labels can be absent, expensive to obtain at scale, or difficult to use.
Millions of users iterate on prompts daily using ChatGPT's text-based chat interface~\cite{mortensen2024chatgpt, cnbc2024openai}, which does not provide any way to upload or evaluate against gold labels. 
%\kenneth{TODO: Cite 1 or 2 credible sources on how many users does ChatGPT have in 2024}\steven{this news said 300 million active users per week, should we rephrase saying weekly to be more accurate? } -- Kenneth: this should be fine
Researchers conducting qualitative coding with LLMs often do not begin with stabilized gold codes~\cite{dai2023llm,de2024further,gao2024collabcoder,zhang-etal-2024-glape, pang2023language}.
%\kenneth{TODO Kenneth: Maybe add LLM for explotory task} \steven{added! one is to evaluate prompt without gold label; another is to generate answers to unlabeled question and evaluate the quality of these answer, without pre-set standard.}
%\kenneth{TODO: Need 1 or 2 more ref on qualitative coding with LLMs WITH NO gold labels}\steven{done}
%\steven{zhang-etal-2024-glape work on prompt evaluation without gold labels; pang2023language presents an unsupervised method where LLMs act as both student and teacher. The model generates answers to unlabeled questions and evaluates the quality of these answers, assigning scores accordingly.}
%Many automated data annotation efforts face insufficient pre-labeled data due to 
%cost, \kenneth{TODO: Cite 1 or 2 papers that do not have gold labels because it's too expensive}
%privacy, or \kenneth{TODO: Cite 1 or 2 papers that do not have gold labels due to user privacy}
%expertise constraints.
%\steven{Several studies have noted that pre-labeled data was absent in real-world scenarios~\cite{liu2019deep, yang2019evaluating, oikarinen2021detecting, slote2024unlocking}. }
%\steven{It often presumes that practitioners have access to static, labeled datasets for evaluation and training. However, this assumption frequently diverges from real-world scenarios, where data is private, encrypted, or unlabeled~\cite{slote2024unlocking}.}
Many automated data annotation efforts face insufficient gold labels due to 
the high cost of gathering labels~\cite{wang2020learnability,liu2024zero}, %\kenneth{TODO: Cite 1 or 2 automatic labeling papers that do not have gold labels because it's too expensive} \steven{done}
difficulty recruiting experts for large-scale annotation tasks~\cite{caiafa2020decomposition, pais2024overcoming,wolf2023dealing}, %\kenneth{TODO: Cite 1 or 2 automatic labeling papers in which the task are very specialized and hard to have large scale dataset}\steven{done}
privacy restrictions~\cite{hathurusinghe2021privacy, yao2023labeling}, or %\kenneth{TODO: Cite 1 or 2 automatic labeling papers in which the gold labels are users' personal information and thus protected}\steven{done}
the lack of universally agreed standards~\cite{wang2024end, organisciak2014crowd}. %\kenneth{TODO: MAYBE cite Amy's paper + maybe 1 more paper in which the task is very subjective or personalized}\steven{done}
%\kenneth{I kinda feel that I put too many examples for automatic annotation cases..... Not sure.}
In such cases, without reliable benchmarks to track prompting progress, users can only rely on their own prompting ability to drive toward desired outcomes.
Yet, this ability is difficult to measure and thus remains understudied.
Without understanding how well users can prompt LLMs through iterations, it is hard to determine how much support users need---and when they need it---to improve their prompt engineering efforts effectively.

This paper investigates a particular scenario in LLM-powered data labeling, which we refer to as ``\textbf{prompting in the dark}.'' 
In these scenarios, \textbf{users' understanding of the data and their desired labeling scheme evolves through their interactions with the LLM and its outputs rather than through manual labeling.}
The process typically unfolds as follows: 
Users begin by writing a prompt to instruct the LLM to label unannotated text data. 
They then observe the outcomes, sometimes reviewing the LLM's explanations for its labels, and iteratively refine their prompt until satisfied with the results.
This iterative refinement process is a common practice in research contexts, where ``the prompt provided to the LLM is iteratively revised'' to analyze or label data~\cite{shah2024prompt}.
Although this prompt refinement process is often undocumented~\cite{shah2024prompt}, we have observed a growing trend of bypassing manual labeling entirely in favor of starting directly with LLM prompting. 
This approach, admittedly, has some usability advantages: 
it eliminates the upfront effort of manual labeling, as the LLM takes on the task, leaving users to focus on reviewing and validating the results. 
Furthermore, at any point in the process, users have an up-to-date prompt that reflects their evolving understanding of the task and the data, which can be immediately used for additional data labeling.
However, despite these seemingly appealing benefits, the core assumption underlying this prompting-in-the-dark approach---with each iteration, the LLM's performance improves, gradually converging on an outcome that aligns with the user's expectations---has yet to be rigorously validated.

To study users' prompt engineering practices for data labeling, we developed \textbf{\system}, 
a \textbf{spreadsheet-based end-user programming tool for prompt engineering in text data labeling tasks}.
%a \textbf{spreadsheet-based end-user programming tool for prompt engineering}.
\system was built as a Google Sheets add-on that allows users to compose prompts, use those prompts to instruct LLMs to label data, review the resulting labels and optional explanations, and iteratively revise and relabel data---all within the same Google Sheets document.
Figure~\ref{fig:system-workflow-fig-v2} shows \system's workflow.
Equipped with the system, we conducted an in-lab user study with 20 participants, who used the system to perform a 5-point sentiment labeling task on a tweet dataset.
%, rating sentiments on a 5-point scale (from Very Positive to Very Negative). 
We found that ``prompting in the dark'' was often unreliable and, at times, counterproductive. 
Of the 20 participants, only 9 improved their labeling accuracy when compared to the labels they manually annotated at the end of the session.
%\kenneth{TODO: This this still true?}\steven{Yeah, this is still true. We only move 2 participants from explanation group to non-explanation group. Participant's overall results do not change.}
% \kenneth{TODO Steven: Update the numbers.}
Even though providing %LLM-generated explanations and 
a larger data sample for users to review sometimes helped, many participants ended the session with worse labeling outcomes.
%many participants had worse labeling results by the session's end.
%in terms of both accuracy and mean squared error (MSE) 
We also investigated the effectiveness of an automated prompt optimization tool, DSPy~\cite{khattab2023dspy}, and found that although it helped in some cases, it struggled to reliably optimize prompts with the small number of user-validated labels generated during the process.

This paper contributes to the rapidly evolving discourse on human-LLM interaction, specifically focusing on measuring human performance in prompt engineering. 
Our findings highlight the necessity of starting with at least a small set of manually labeled data, which can act as a critical beacon when prompting in the dark, as well as the need for more automated support, such as typo checks and rule suggestions,
while ensuring designs that mitigate potential overreliance on LLMs' predictions. 
These insights can inform the development of future tools, enabling a wider range of users to prompt LLMs more effectively.
%to enhance the reliability of human prompt engineering. 
%These insights can guide the design of future tools, helping a broader range of users prompt LLMs more effectively.

\subsection{Research Questions\label{sec:rq}}

The central research question guiding our study is:

\begin{itemize}
    \item \textbf{RQ 1-1: How effective are people at prompt engineering in the ``prompting in the dark'' scenario?}
\end{itemize}

In addition to examining prompt engineering effectiveness, we aim to understand how key factors influence human performance when prompting in the dark. 
Among the many variables that could impact outcomes~\cite{kulesza2012tell}, two stand out:
(1) the size of the sample data in each iteration, and 
(2) whether the LLM's explanations for its labels are shown to users.
Accordingly, we address the following additional research questions:

\begin{itemize}
    \item \textbf{RQ 1-2: How does sample size affect human performance in prompt engineering?}
    \item \textbf{RQ 1-3: How does displaying LLM explanations impact human performance in prompt engineering?}
\end{itemize}

We acknowledge that many variables could influence the quality of labels in LLM-powered data annotation. 
These include variables studied in crowdsourced data labeling research, such as 
task type~\cite{hettiachchi2022survey, zhen2021crowdsourcing}, %\kenneth{TODO: Add references for all of these}\steven{added}
task difficulty~\cite{zheng2022virtual, zhang2021personalized}, 
instruction quality~\cite{zamfirescu2023herding, zhang2023llmaaa}, 
label diversity~\cite{kazai2012face}, and 
label aggregation methods~\cite{drutsa2020crowdsourcing, li2024comparative}.\footnote{Some variables in crowdsourced data annotation, such as workers' financial or intrinsic incentives, do not apply to LLM-powered data annotation.}
%We focus on the two key variables that we believe are widely applicable when iteratively prompting LLMs to label text data. 
%Prior research highlights the significance of sample size and explanations, with sample size influencing human-AI workflows and explanations serving as a primary method for LLMs to communicate reasoning. 
%These variables have been extensively studied for their impact on human-AI collaboration. 
We focus on these two key variables because they are widely applicable to any iterative prompting process with LLMs. Prior research shows that sample size significantly impacts human-AI workflows, while explanations are a primary means for LLMs to convey their reasoning to users.
Below, we provide additional context for these variables:

\begin{itemize}
    \item

%We focus on these two variables based on the following rationale:

\textbf{Data Sample Size in Each Iteration.}
In iterative labeling research~\cite{lee2024clarify}, particularly within active learning~\cite{bernard2018vial}, a recurring challenge is the trade-off between the limited time, attention, and effort of human annotators and the need for AI models to learn effectively from human-labeled instances~\cite{pandey2022modeling,wang2016human}. 
%A common strategy is to identify and label the most challenging or ambiguous data points, as these provide the greatest learning benefit for the model.
This trade-off is a fundamental issue in many human-AI collaboration scenarios. 
In iterative prompt refinement, larger sample sizes give users more data to inform prompt revisions, potentially improving performance. 
However, reviewing large datasets in every iteration is labor-intensive and time-consuming. 
Striking a balance between providing sufficient data for effective prompt improvement and minimizing user effort is critical. 
Given the universal relevance of this variable in human-AI collaboration, we examine it as a key factor in this study.

\item
\textbf{LLM Explanations for Predictions.}
%Another key variable is whether users are provided with the LLM's explanations for its predictions. 
%Prior research suggests that 
Bi-directional communication between users and LLMs---where users receive feedback from the model rather than only giving instructions---is essential for effective collaboration~\cite{shen2024towards}.
%\kenneth{TODO: Cite Hua's bi-directionality AI alignment paper.}\steven{added}
When prompting in the dark, gold labels and metrics like accuracy are unavailable, LLM-generated explanations become one of the few ways the model can provide feedback to users~\cite{teso2023leveraging,kulesza2015principles}.
However, prior studies show mixed results regarding the effectiveness of AI-generated explanations in human-AI collaboration. 
Some studies suggest that explanations may hinder human understanding~\cite{turpin2024language,shen2020useful}, while others indicate they can enhance decision-making~\cite{singh2024rethinking,10.1145/3579605}.
A cost-benefit framework has been proposed to make sense of these mixed results, suggesting that explanations fail to help in some cases because they do not reduce the cognitive cost of verifying AI predictions~\cite{10.1145/3579605}.
In most text data labeling tasks, except for cases with highly complex text, we hypothesize that reading the entire text and then reviewing explanations to validate the AI label generally requires more effort than carefully reading the text and deciding the label directly. 
Based on this framework, LLM explanations may offer limited utility. 
To evaluate whether this applies to iterative prompt refinement for text labeling and to contribute to the broader discourse on human-AI collaboration, we studied explanations as a key variable.

%\kenneth{TODO: Cite the Explanation Can Reduce.... paper}\steven{added}
%In text data labeling, unfortunately, reading the entire text item and reviewing the text explanations often require similar cognitive effort. 
%To determine whether this is the case for iterative prompt refinement aiming to label text data, and also to contribute to the ongoing discourse on human-AI collaboration, we decided to study explanations as one of our key variables.

\end{itemize}

Lastly, we are interested in how much automatic prompt optimization tools, such as DSPy~\cite{khattab2023dspy}, can enhance human performance in prompting-in-the-dark scenarios. 
This investigation is important because, even when users struggle with effective prompting, tools like DSPy might still be able to refine human-crafted prompts and achieve satisfactory results.
%However, the main challenge is that prompting in the dark often involves too few labeled data points for optimization tools to learn effectively. 
%Even when users label a few data points for sanity checks, the dataset may be insufficient for optimization tools to learn effectively.
However, 
the main challenge is that prompting in the dark typically provides too few labeled data points for optimization tools to learn effectively, even with a few sanity-check labels from users.
To address this, we pose the following research question:

%A key challenge in these scenarios is that the labeling process often fails to generate enough labeled data for optimization tools to effectively learn from. 
%Therefore, we aim to address the following research question:

\begin{itemize}
    \item \textbf{RQ 2: Can automatic prompt optimization tools like DSPy improve human performance in ``prompting in the dark'' scenarios?}
\end{itemize}

\section{Related Work}
%\subsection{End-User Programming}

%\kenneth{The way I like to think about Related Work is that this section should (sometimes subtly, not explicitly, but effectively!) answer some underlying questions that reviewers might want to ask. So, here we go:}\steven{sounds good!}

\subsection{Ways of Optimizing Prompts for LLMs}
%\subsection{Prompt Engineering and How Good Humans Are at It}
Prompts are the primary means by which users interact with, utilize, and instruct LLMs. 
Since the emergence of these models, researchers and developers have invested significant effort into understanding how to craft better prompts for more effective use. 

\paragraph{Automatic Prompt Optimization.}
Much of the prior work has focused on automatically optimizing prompts. 
A common theme across these studies is the use of gold-standard labels to guide the optimization process.
For example, \citet{pryzant2023automatic} introduced an automatic prompt optimization method inspired by gradient descent; 
\citet{manas2024improving} presented an approach that begins with a user prompt and iteratively generates revised prompts to maximize consistency between the generated image and prompt, without human intervention; 
\citet{wan2024teach} explored two types of prompt optimization, instruction and exemplar, and suggested that combining both can yield optimal results; 
\citet{sun2023autohint} combined zero-shot and few-shot learning to optimize prompts automatically; %eliminating the need for manual prompt engineering; 
and \citet{levi2024intent} improved prompt optimization through synthetic data generation and iterative refinement, focusing on aligning prompts with user intent by creating challenging boundary cases for iterative prompt refinement.
While these studies were interesting and relevant, they generally assumed the availability of gold-standard labels and did not address situations where labels are absent or where standards are constantly evolving.

\paragraph{User-Driven Prompt Optimization.}
In addition to automatic prompt optimization, some research has focused on human capabilities in optimizing prompts. 
\citet{zhou2023revisiting} found that manual prompting often outperforms automated methods in various scenarios; 
\citet{10.1145/3544548.3581388} discovered that people tend to design prompts opportunistically rather than systematically, which often leads to lower success rates. 
To the best of our knowledge, the most relevant prior work is by \citet{wang2024end}, who developed an iterative refinement system that enables users to prompt LLMs to build a personalized classifier for social media content. 
Their study explored three user strategies for improving prompts and measured their effectiveness. 
While conceptually related to our work, their focus was not on how users evolve their understanding and expectations when interacting with LLMs. 
Instead, participants labeled ground truth at the beginning of the study, prior to using the system.

\subsection{Tools for Prompt Engineering}
With the advances in LLMs, numerous tools have been developed to assist with prompt engineering. 
Most of these tools follow a software-engineering paradigm, where testing (such as unit tests or integration tests) is a central concept, and thus often assume the existence of gold-standard labels.
For example, PromptIDE is an interactive tool that helps experts iteratively refine prompts by providing various inputs, visualizing their performance on small validation datasets, and optimizing them based on quantitative feedback~\cite{strobelt2022interactive}; 
PromptAid is a visual analytics system for interactively creating, refining, testing, and iterating prompts while tracking accuracy changes~\cite{mishra2023promptaid};
%It allows users to adjust prompts through keyword modifications, paraphrasing, and adding few-shot examples; 
ChainForge is an open-source visual toolkit for prompt engineering and on-demand hypothesis testing of text-generation LLMs~\cite{arawjo2024chainforge};
and, promptfoo applies a test-driven approach to LLM development, producing matrix views that enable quick evaluation of outputs across multiple prompts~\cite{webster2023promptfoo}.
While these tools are inspiring and valuable, the scenarios we focus on do not rely on the constant availability of gold labels.

\subsection{Human-LLM Collaborative Data Annotation}
%Another relevant area of research involves using LLMs for data annotation. 
Beyond simply treating LLMs as automatic labelers---common in countless NLP projects~\cite{tan2024large}---a growing body of work explores how to combine human and LLM efforts to achieve better annotation outcomes, such as improved accuracy or speed.
Even as LLMs outperform humans in many labeling tasks, human-AI collaboration often produces better results than either alone~\cite{vaccaro2024combinations}.
For example, \citet{kim2024meganno+} introduced a human-LLM collaborative annotation system where LLMs handle bulk annotation tasks, while humans selectively verify and refine the annotations. 
%\steven{However, this system was limited to deployment within Jupyter Notebook, lacking an end-to-end solution. This design imposed significant barriers, as it required users to possess technical expertise for system setup before using the tool, limiting accessibility and scalability in non-technical domains.}
\citet{goel2023llms} proposed an approach that combines LLMs with human expertise to efficiently generate ground truth labels for medical text annotation.
Additionally, \citet{10.1145/3613904.3642834} demonstrated how aggregating crowd workers' labels with GPT-4's output can achieve higher labeling accuracy than either source alone.
These studies generally aim to split the workflow of data labeling between humans and LLMs in a smart way, making the task more effective or efficient. 

In contrast, our work does not focus on dividing or combining the workload, but on how humans can teach LLMs---through prompt refinement---to better label the specific type of data.
Few prior studies have emphasized iterative prompt refinement in human-LLM collaborative data annotation.
For example, \citet{liu2024harnessing} developed a workflow for video content analysis, refining prompts to improve LLM-generated annotations and align them with human judgment.
Additionally, \citet{zhang2023llmaaa} proposed LLMAAA, which uses LLMs as annotators in a feedback loop to label data efficiently.
Their study shows that poorly designed prompts result in subpar performance, especially in complex tasks. %while incorporating demonstrations and aligning label descriptions with natural language significantly enhances accuracy and reliability.
Our work advances this relatively understudied area of human-LLM collaborative annotation research.

\subsection{Gold-Standard Labels in Annotation Tasks}\label{sec:related-work-gold-label}
Decades of research have shown that gold-standard labels play a critical role in quality control for data annotation pipelines~\cite{han2020crowd,gadiraju2015training,le2010ensuring,doroudi2016toward,hettiachchi2021challenge}.
Embedding items with known labels into the data annotation process allows requesters to reliably capture quality signals, 
such as workers' level of expertise~\cite{abraham2016many, abassi2019worker, yang2018improving} %\kenneth{TODO: Add refs about using gold labels to decide workers' expertise level}\steven{added}
or attentiveness to tasks~\cite{hettiachchi2021challenge, oleson2011programmatic}. %\kenneth{TODO: Add refs about using gold labels to do attention checks for workers}\steven{added}
These insights enable requesters to take appropriate actions, such as 
retraining annotators~\cite{le2010ensuring, doroudi2016toward,hettiachchi2021challenge}, %\kenneth{TODO: Add refs about retraining workers}\steven{added}
removing low-performing workers~\cite{10.1145/3613904.3642834, snow2008cheap,downs2010your,le2010ensuring}, %\kenneth{TODO: Add refs about removing or blocking low-performing workers}\steven{added}
or identifying potential issues in the annotation interfaces~\cite{toomim2011utility,10.1145/3613904.3642834, rahmanian2014user, komarov2013crowdsourcing}. %\kenneth{TODO: Add refs for crowd worker interfaces. At least cite: Toomim, M., Kriplean, T., Pörtner, C., \& Landay, J. (2011, May). Utility of human-computer interactions: Toward a science of preference measurement. In Proceedings of the SIGCHI Conference on Human Factors in Computing Systems (pp. 2275-2284).}\steven{added}
Gold labels are also beneficial for requesters during post-annotation data processing. 
They can be used to weight labels from different workers in label aggregation~\cite{abassi2017gold,abassi2019worker}, %\kenneth{TODO: Add label aggregation methods that use gold labels particularly to weight different workers}\steven{added}
improve label aggregation strategies~\cite{khattak2011quality, snow2008cheap},  %\kenneth{TODO: Add label aggregation methods that learn whatever from gold labels}\steven{added}
or 
exclude unreliable workers' outputs entirely~\cite{abassi2019worker}. %\kenneth{TODO: Cite ref using gold labels to remove workers from label aggregation}\steven{added}
Beyond requesters, gold labels are also beneficial for data labelers like crowd workers. 
Gold labels can be used to train workers~\cite{doroudi2016toward, le2010ensuring, gadiraju2015training,han2020crowd}, %\kenneth{TODO: Cite ref that uses gold labels for worker training}\steven{added}
provide real-time feedback to help them recalibrate their understanding of the task~\cite{le2010ensuring,hettiachchi2021challenge}, %\kenneth{TODO: Cite the visible gold paper from Amazon}\steven{added}
or remind them to pay more attention~\cite{ hettiachchi2021challenge,oleson2011programmatic}. %\kenneth{TODO: Cite attention check papers}\steven{amazon paper also warn workers in real time}

While gold labels are useful for quality control, as stated in the Introduction (Section~\ref{sec:intro}), %\kenneth{TODO: Update references}\steven{done}
they are not always available in real-world scenarios due to constraints such as data privacy or the cost of gathering gold labels~\cite{liu2019deep, yang2019evaluating, oikarinen2021detecting, slote2024unlocking}.
To address these challenges, researchers have developed methods to generate (approximations of) quality signals without gold labels. 
In the realm of LLM-powered data annotation, for instance, CoPrompter evaluates how well an LLM's output aligns with user-specified requirements as a feedback mechanism~\cite{joshi2024coprompter}. %\kenneth{TODO: Cite: Joshi, I., Shahid, S., Venneti, S., Vasu, M., Zheng, Y., Li, Y., ... \& Chan, G. Y. Y. (2024). CoPrompter: User-Centric Evaluation of LLM Instruction Alignment for Improved Prompt Engineering. arXiv preprint arXiv:2411.06099.}\steven{added}
Other studies also leverage the stability~\cite{chiang2023can} %\kenneth{TODO: Add ref}\steven{added}
%chiang2023can found LLM evaluation are stable over different formatting
or confidence~\cite{gligoric2024can} %\kenneth{TODO: Add ref}\steven{added}
%gligoric2024can introduce CONFIDENCEDRIVEN INFERENCE: a method that combines LLM annotations and LLM confidence indicators to strategically select which human annotations should be collected
of LLM outputs to infer quality signals.
%Our research investigates how effectively humans can iteratively refine prompts to guide LLMs in labeling data when gold-standard labels are unavailable, providing alternative quality signals.
Our research examines how effectively humans can refine prompts to guide LLMs in labeling data without gold-standard labels, providing insights into human prompting capabilities in the absence of reliable guidance signals.

\section{\system: A Spreadsheet-Based End-User Prompt Engineering Tool}

%\kenneth{Note that Figure 1 is a bit simplified, e.g., label verification, keep items for next iteration, is dismissed to make sure clear communication.}
%The \system is a Google Spreadsheet add-on with a user interface for guiding LLMs by providing rules or exemplary data to improve LLM performance. The system could be adapted to various single-class data annotation tasks. 
%Users can easily operate the system with minimal learning time and without the need for complex environment setup, unlike other applications.~\steven{find some other annotation systems}
In this paper, we present \system, a Google Sheets add-on that allows users to load a dataset into a Dataset spreadsheet, manually compose each part of a prompt within Task Context, Labeling Rules, and Shots sheets, and use the composed prompt to instruct an LLM to annotate data---all within the same Google Sheets document.
Motivated by the need to enable general users to prompt LLMs without installing and configuring professional tools like integrated development environments (IDEs) or Jupyter Notebooks, we decided to build a tool based on spreadsheets, which most computer users are already familiar with.
This section overviews \system's design and workflows.
%\kenneth{TODO: Maybe cite other tools and name what setup or configs they require}

\subsection{Design Goals}\label{sec:3-system-design}
%\kenneth{(1) Support PITD so we don't have gold label and don't calculate accuracy etc becasyue we think they're not reliable and evovling, and (2) using spreadsheet because it's just very easy to use and practically provide many flexibilities to users.}

%Prompting in the Dark: 
\paragraph{Adapting to Evolving Labeling Goals}
%\paragraph{Prompting in the Dark and Users' Needs for Evolving, Open-Ended Labeling Schemes.}
The goal of \system is to enable general users to create and refine prompts iteratively for LLM-powered data labeling, particularly in situations where they start without any labeled gold data or manual labeling, \ie, the ``prompting in the dark'' scenario. 
In these cases, users' understanding of the data and desired labeling scheme evolves through their interactions with the LLM, based on its predicted labels and explanations, rather than through their own manual efforts. 
The lack of gold labels (or sufficient labeled data) introduces the core challenge of the prompt-in-the-dark process: aside from users' observations and judgments about the labeling results, there is no concrete way to provide quick and comprehensive feedback on the progress of prompting.
We view this as a trade-off between two user needs in prompt engineering: 
{\em (i)} allowing users' understanding of the data and labeling goals to evolve, and 
{\em (ii)} providing clear guidance and reliable feedback to assess progress toward a defined annotation goal. 
Previously, supervised learning-based classifiers required labeled data, so users focused heavily on the second need, as manual labeling was always needed and assumed to finalize the coding scheme. 
The rise of LLMs has reduced the need for pre-labeled data, allowing users to put more focus on their first needs. 
The growing popularity of the prompting-in-the-dark approach reflects users' need for evolving and dynamic labeling practices~\cite{austin2024grad,zhang-etal-2024-glape,wang2024human}.
%zhang2023labelvizier to facilitate the validation and relabeling of large-scale technical text annotations. Its interactive, visual analytic interface allows users to detect and correct three main types of labeling errors: duplicate, wrong, and missing labels. 
%\kenneth{TODO: Do we have any reference to support it's a popular practice now?}\steven{those are three study that have no gold label, they iteratively use user-defined criteria to evaluate and refine.}\kenneth{Oh and did they manage to improve the accuracy over time??}\steven{Yeah, they got a higher rating and accuracy}\kenneth{Hmmmmmm so what's the deal? How are their systems and approaches different?} %, rather than simply saving time on manual labeling. 
% dudley2018review describe the interative machine leanring paradigm that user iterative build and refine model. The model refinement is driven by user input. It more focused on the human input to refine. kim2024evallm is a prompt refining tool by evaluating outputs on user-defined criteria. It is not a labeling task
Our design goal for \system is to offer users greater flexibility and freedom in defining how their data should be labeled.

%\kenneth{I revised this following paragraph to tailor it more close to our target users. Please take a look.}
\paragraph{Supporting a Wide Range of ``Newcomers'' Brought in By LLMs.}
%\paragraph{End-User Prompting Tools.}
%Our second design goal for \system is to create a tool for general users, including people with limited or no programming skills.
Our second design goal for \system is to develop a tool for users with \textbf{little to no experience in large-scale text data labeling}, including but not limited to those with limited or no programming skills.
The rationale behind this goal is two-fold.
On a practical level, 
people new to large-scale data annotation---empowered by LLMs to undertake such tasks with greater ease---are more likely to adopt approaches that diverge from conventional practices. 
In the crowdsourcing literature, many papers emphasize best practices for data annotation~\cite{hsueh2009data, sabou2014corpus, vondrick2013efficiently, drutsa2019practice, wang2013perspectives}, %\kenneth{TODO: Add ref on crowdsourcing best practices}\steven{added}
such as ethical pay rates~\cite{fort2011amazon,shmueli2021beyond}, %\kenneth{TODO: Add refs on crowd workers' ethical pay}\steven{added}
usable worker interfaces~\cite{toomim2011utility,10.1145/3613904.3642834, rahmanian2014user, komarov2013crowdsourcing}, %\kenneth{TODO: Add refs to crowd interface's impact on crowdsourced data quality--- Maybe cite our own CHI paper too}\steven{added}
and 
gold labels for quality control~\cite{han2020crowd,gadiraju2015training,le2010ensuring,doroudi2016toward,hettiachchi2021challenge}. %\kenneth{TODO: Again, add ref IN CROWDSOURCING for gold labls}\steven{added}
However, these practices are often neglected in real-world scenarios. 
For instance, many tasks on MTurk still offer very low pay~\cite{AI_workers_low_wages} %\kenneth{Add ref: A data-driven analysis of workers' earnings on Amazon Mechanical Turk}\steven{added}
or rely on poorly designed interfaces~\cite{fowler2023frustration}. %\kenneth{Add ref: Frustration and ennui among Amazon MTurk workers}\steven{added}
Newcomers to large-scale data annotation are even less likely to be familiar with these best practices, including carefully establishing gold labels before prompting LLMs.
%users without programming experience are more likely to prompt in the dark, struggling to interact effectively with LLMs. 
%In contrast, those with software engineering backgrounds are familiar with using established frameworks and tools, where automatic testing---such as unit and integration testing---is standard. 
The bigger picture is that LLMs are adding many ``new members'' to the world of programming and data science.
%---people with little or no coding experience. 
This group brings new practices, user needs, challenges, and research questions to HCI, requiring more focused attention.

%\bigskip
\sloppy
Based on these two design goals, we decided to build \system based on spreadsheets, a format that most computer users are already familiar with.
We distinguish our goals from existing efforts in two significant ways. 
First, while projects like LangChain or ChainForge focus on developers or those with programming backgrounds, requiring software installations or configurations, we aim to focus on general users who do not necessarily have such expertise. 
Second, some projects explore new interactions enabled by LLMs~\cite{10.1145/3586183.3606833}, but our project is concerned with understanding how effectively users can use familiar interfaces, such as spreadsheets, to interact with LLMs.

%--------------- dead kitten ----------

\begin{comment}

At the practical level, users with little or no programming background might be more likely to prompt in the dark. 
In contrast, individuals familiar with software engineering practices are accustomed to using existing frameworks or tools, where automatic testing---such as unit and integration testing---is standard.
These users have more experience as well as technological support for creating gold labels for testing.
At a deeper level, what is happening is that LLMs bring many people with limited or no programming skills into the realm of data science or programming tasks.
This group of people brings new and interesting challenges and research questions to HCI and thus deserves more attention.

%we believe that the greatest value of LLMs lies in the new possibilities they offer to general users. 
%For people without coding skills, LLMs enable tasks such as building websites from scratch, creating classifiers for automating email filtering, and labeling data to extract insights---activities that were previously within the realm of programmers. 

\end{comment}

\begin{figure*}[t]
    \centering
    \includegraphics[width=0.9\linewidth]{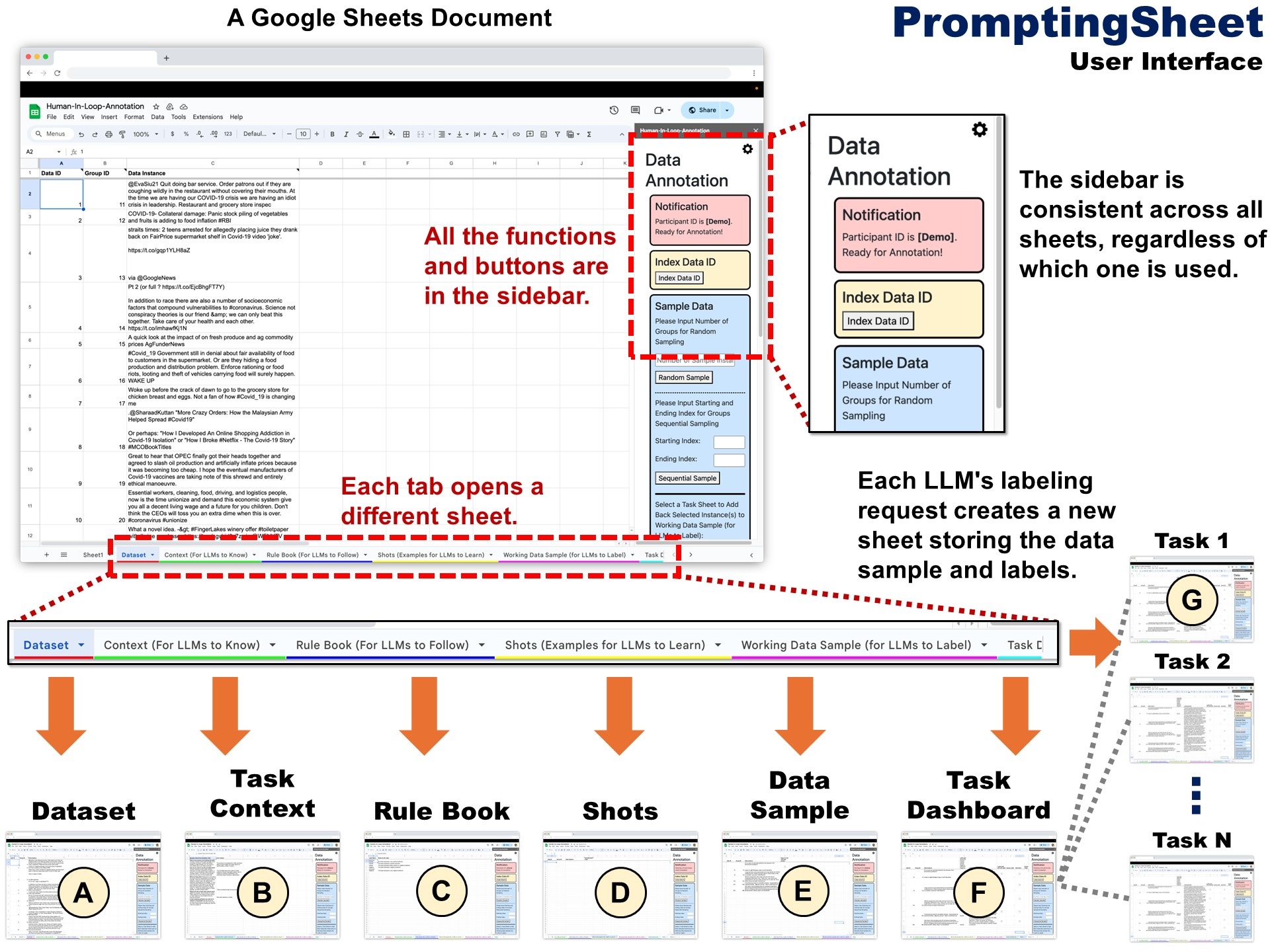}\Description{This is the user interface layout and pre-defined spreadsheet tab explanation. Each predefined sheet has a set of predefined columns. PromptSheet allows users to load a dataset into the Dataset tab, which is the starting tab of the system. By clicking each tab at the bottom of the Google Sheet, users can navigate to Task Context tab, Rule Book tab, Shots tab, Working Data Sample tab, task dashboard tab and task results tabs. The task result tabs will be generated after each new annotation round and store all new annotated results.}
    \caption{The user interface and all the predefined sheets of \system, where each sheet has a set of pre-defined columns. 
    \system allows users to load a dataset into a Dataset (A) sheet, manually compose each part of a prompt within Task Context (B), Labeling Rules (C), and Shots (D) sheets, and use the composed prompt to instruct an LLM to annotate data and store the labeling results in a separate task sheet (G). All functions are presented as manuals and buttons within the sidebar on the right.}
    \label{fig:system-interface-ui-kenneth}
\end{figure*}

\subsection{User Interface and Pre-Defined Sheets}

\system is a Google Sheets add-on that enables users to load data, sample a subset for labeling, compose and edit prompts, use these prompts to request LLMs for data labeling, and iteratively revise the prompts. 
Figure~\ref{fig:system-interface-ui-kenneth} shows the interface of \system.

\paragraph{Sidebar.}
Following Google Sheets' design constraints, all functions are presented as manuals and buttons within the sidebar on the right. 
The sidebar remains consistent across all sheets, regardless of which sheet is in use. 
At the top of the sidebar, \system provides a real-time notification that keeps users informed about its ongoing processes, such as ``Data Indexing,'' ``Data Sampling,'' ``Generating the Instructional Prompt,'' or ``Annotating.''

\paragraph{Pre-Defined Sheets.}
\system includes a set of predefined spreadsheets, each with a set of pre-defined columns. 
At the bottom of the interface, a series of tabs allows users to switch between sheets, with each sheet dedicated to a different part of the data labeling process. 
The following describes each sheet in detail. 
(To help readers easily identify which sheet we are referring to, we indexed each sheet as A, B, C, ..., and G in all the figures. 
These indexes were not present in the actual system to users.)

%\subsubsection{All the Sheets and What They Do}

%\hyperref[fig:system-interface-1]{Figure 1} overviews the \system's user interface. 
%The system consists of seven main components:

\begin{itemize}

\item \textbf{Dataset (Sheet A)}:
%The ``Dataset'' includes three columns: Data ID, Group ID, and Data Instance. 
%Each data instance is uniquely identified by a corresponding Data ID.
%A single Group ID may encompass one or multiple Data Instances.
This spreadsheet stores the full dataset.
Users can copy and paste the dataset into this sheet or use any supported Google Sheets import method (in Step 0).
The sheet includes three key predefined columns: (1) Data ID, (2) Group ID, and (3) Data Instance. 
Each data instance is uniquely indexed by its corresponding Data ID, which users can generate by clicking the ``Index Data ID'' function in the sidebar. 
The Group ID is used for annotating sequential data, such as when each sentence in an article is treated as a separate data instance, but all sentences from the same article share the same Group ID.
In our design, this sheet is intended to serve as a static data source, and we anticipate that users will not modify it after loading the data.

%\kenneth{TODO: Maybe add words to mention we don't expect people touch it after Step 0 and basicallyy serve as a database.}

\item \textbf{Task Context (Sheet B)}:
%The ``Context'' tab provides information to help describe the annotation task users are working on. It addresses questions related to the purpose and application of the data annotation task and the origin and size of each data instance. The LLM will use the information provided in this tab to generate an instructional prompt for a later step.
This spreadsheet stores the meta-information and context for the labeling task, which will later be incorporated into the prompt. 
The sheet includes predefined questions that characterize the task, such as the purpose of the data labeling, how the labels will be used, the source of the data, and the size of each data instance.
Table~\ref{tab:task-sheet-questions} in Appendix~\ref{sec:context-question-appendix} shows all the questions.
%\kenneth{TODO: Maybe add all the questions to Appendix.}
Users provide answers to these questions (in Step 1 or 4), and \system automatically incorporates both the questions and their answers into the prompt used for LLMs to label the data.

\item \textbf{Rule Book (Sheet C)}: 
%The ``Rule book'' tab is where users define the criteria and definitions for each label used during the annotation process.
This spreadsheet contains the labeling rules that the LLM will follow.
It includes two key predefined columns: (1) Label Name and (2) Rules for the Label. 
Users manually define the criteria and descriptions for each label in free text (in Step 1 or 4), detailing the guidelines for the annotation process. 
Multiple rules can be added for a single label, providing flexibility in defining the labeling criteria.

\item \textbf{Shots (Sheet D)}: 
%In the ``Shots'' tab, users can enter gold standard labels from their iterations or manually provide them as reference points.
This spreadsheet stores all the high-quality examples, including data instances and their corresponding labels, which will be included in the prompt to guide the LLM in labeling the data. 
These examples, commonly referred to as ``shots'' in prompts, follow the same predefined column structure, with an additional ``Gold-Standard Label'' column. 
Users can add these examples manually (in Step 1) or use \system's function to do so (in Step 4).

\item \textbf{Working Data Sample (Sheet E)}:
%Users can sample the data from the ``Dataset'' tab to the ``Working Data Sample'' tab. In the annotation process, \textit{only} data instance in the ``Working Data Sample'' tab will be annotated.
% by the LLMs using the instructional prompt, provided rules, and gold shots.
This spreadsheet stores the current subset of data selected from the full dataset, ready for the LLM to label.
Users can sample data from the Dataset sheet by clicking the corresponding buttons in the sidebar; users can choose between random sampling or selecting a specific range (Step 2). 
During the annotation process, only the data instances in the Working Data Sample sheet will be labeled. 
\system will copy the entire data sample from the Working Data Sample sheet to create a new sheet to label (Step 3).

\item \textbf{Task Dashboard (Sheet F)}:
%The ``Task Dashboard'' tab records all iteration task details such as task number, timestamp, used prompt, and total costs. 
This spreadsheet tracks all labeling tasks performed so far.
When the user clicks the ``Start Annotation'' button in the sidebar (in Step 3), \system creates a new sheet for the task (e.g., Task 1 sheet) and adds a new row in the Task Dashboard to record the labeling activities.
Task Dashboard sheet (Figure~\ref{fig:task-dashboard-new})
logs task details such as task number, timestamp, the prompt used, and total costs.

\item \textbf{Task 1 (Sheet G), Task 2, ..., Task N}:
%After each annotation, the annotation results will be saved in a new tab (e.g., Task\_1, Task\_2, etc) corresponding to that specific iteration. 
Each of these sheets stores the annotation results for each labeling request, including data samples, LLM-generated labels, and LLM explanations (optional).
These sheets also include columns that allow users to validate or correct the LLM labels and optionally add them to the Shots sheet (in Step 4).
When the user clicks the ``Start Annotation'' button in the sidebar (in Step 3), \system generates a new task sheet to handle the specific labeling task.

\end{itemize}

%\kenneth{Users are allowed to add new columns.}

Notably, while users must follow our guidelines for using the predefined columns in each sheet and inputting data correctly, they are free to add more columns or even additional sheets, just as they would in a regular Google Sheets document. 
For instance, when pasting a dataset into the Dataset sheet, it is common for the dataset to include its own IDs or additional information for each data entry. 
Users can easily store this extra information by creating new columns within the Dataset sheet.

%---------- dead kitten ----------

\begin{comment}

\subsubsection{Other Features} \steven{todo: add figures in the appendix. screenshots for different notification messages. interface screenshots for removal and clearing.}
\begin{itemize}
    \item \textbf{Real-Time System Notification: }\system provides a notification feature that informs users of its current processes, such as ``Data Indexing'', ``Data Sampling'', ``Generating the Instructional Prompt'', ``Annotating'', etc.
    \item \textbf{Remove Unselected Data Instance (Figure \ref{fig:remove-clear}): }This function will remove data instances that do not have the ``Keep it in the next data sample'' checked in the ``Working Data Sample'' tab.
    \item \textbf{Clear Data Instance (Figure \ref{fig:remove-clear}): }This function will clear all data instances in the ``Working Data Sample'' tab.
\end{itemize}

\end{comment}

\subsection{User Workflow}
\begin{figure*}[t]
    \centering
    \includegraphics[width=0.99\linewidth]{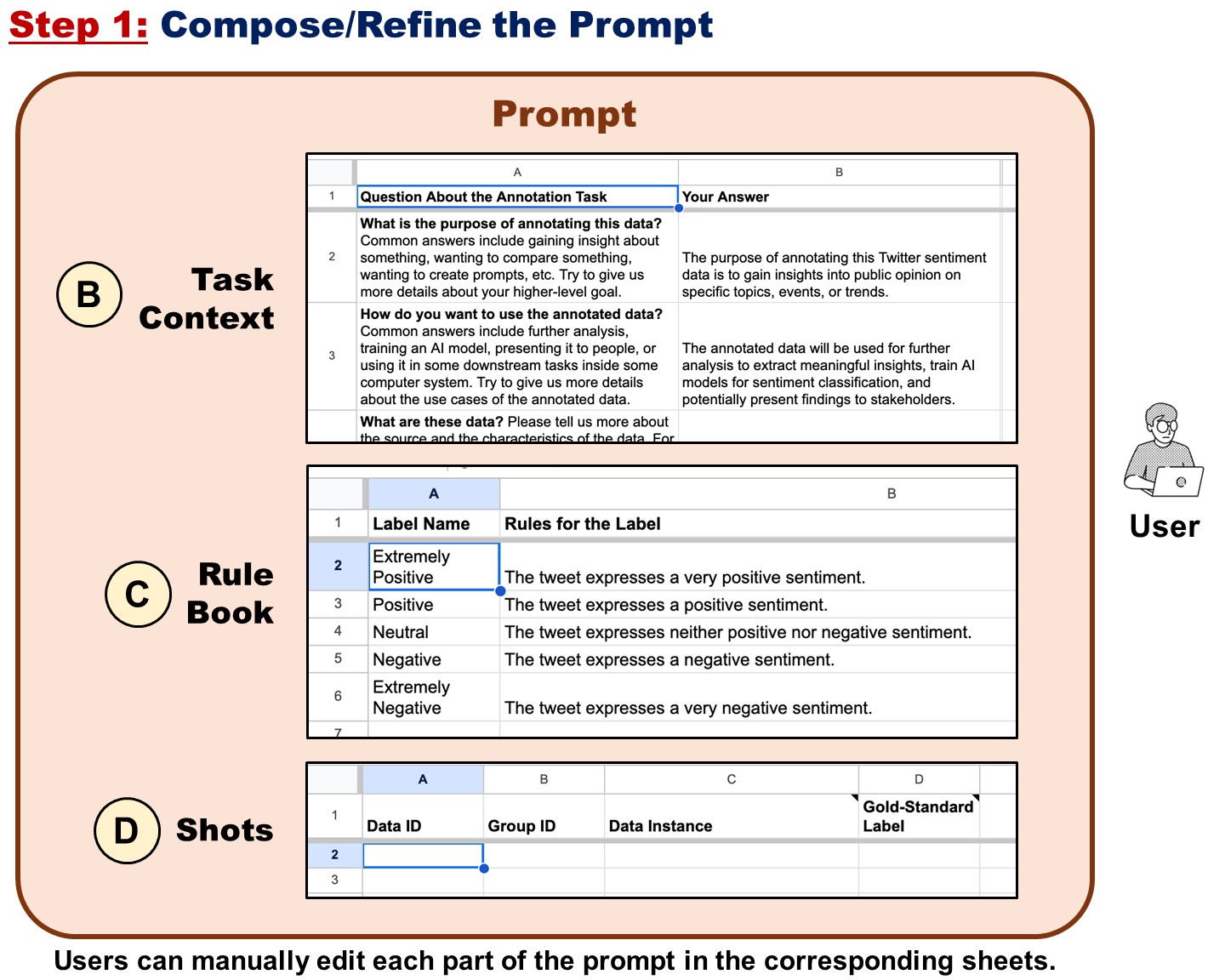}\Description{This is Step 1 described in Figure 1. Users can provide data annotation context in the Context Tab, provide their rule and definition in the Rule Book tab, and add gold standard labels in the Shots tab. These tabs will compose prompts for later GPT to use.}
    \caption{The overview of step 1 of the data labeling process, compose or refine the prompt.
This is the most critical step, where the user composes and refines prompts for the LLM to label the data. In \system, the prompt consists of three parts, each corresponding to a separate sheet: Context (B), Rule Book (C), and Shots (D). 
At the beginning of this labeling process, the user has only a vague idea of what they want to label and will continuously refine that idea. 
Each time the prompt is revised, it reflects an evolution of their understanding and approach to the labeling task.}
    \label{fig:step-1}
\end{figure*}

\begin{figure*}[t]
    \centering
    \includegraphics[width=0.99\linewidth]{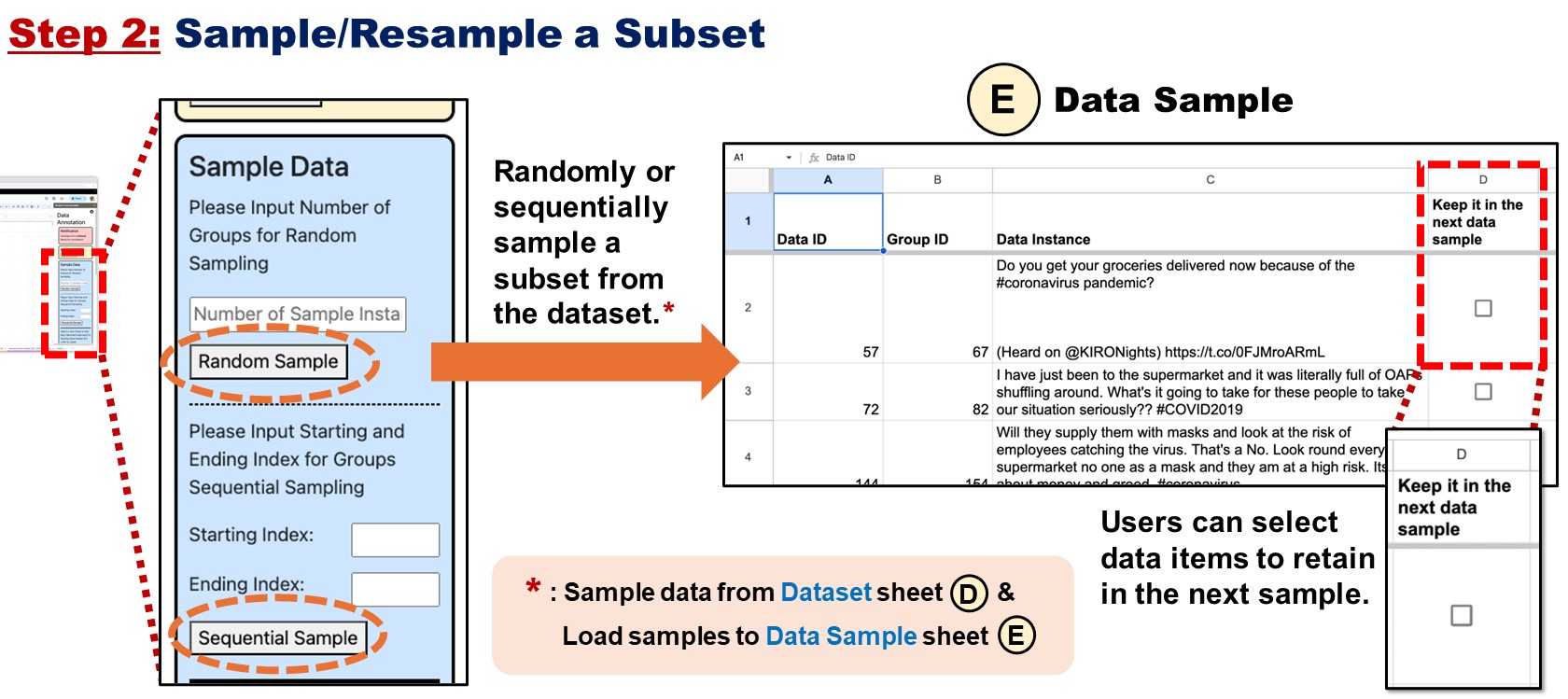}\Description{This is Step 2 described in Figure 1. Users can either random or sequential sample data from the Dataset tab to the Working Data Sample tab. In the Working Data Sample, users can check “Keep it in the next data sample” for data instances that users want to remain in the Working Data Sample tab during sampling.}
    \caption{The overview of step 2 of the data labeling process, sample or resample a subset.
The full dataset is often too large for the user to thoroughly review, so sampling a subset is necessary.
%Labeling only a subset, rather than the entire dataset, is necessary because 
%Additionally, labeling the entire dataset iteratively would be prohibitively expensive. 
In this step, the user can (1) randomly or (2) sequentially sample data from the Dataset (A) sheet.
}
    \label{fig:system-interface-step-2}
\end{figure*}

\begin{figure*}[t]
    \centering
    \includegraphics[width=0.99\linewidth]{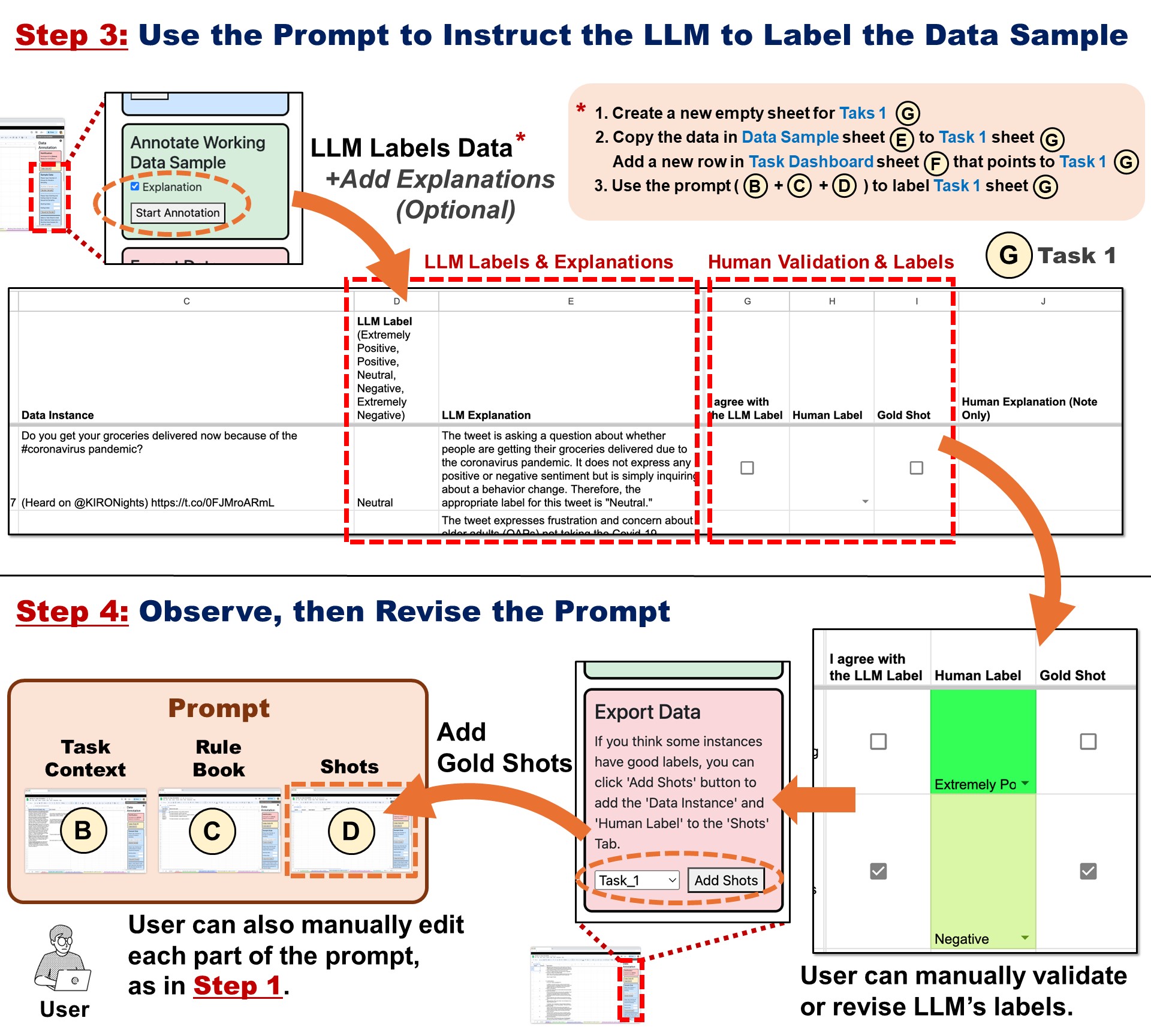}\Description{This is Step 3 and 4 described in Figure 1. After clicking Start Annotation, the results including LLM label and LLM explanation will be stored in a new tab. Users will review the data instances and LLM labels, by checking agree or providing their own labels. They also select the Gold Shot data instance to be added to Shots tab for later GPT to learn from. After verification, they can refine their prompt as Step 1 mentioned.}
    \caption{The overview of steps 3 and 4 of the data labeling process. After finalizing the prompt (Step 1) and sampling data instances (Step 2), in Step 3, the user clicks the ``Start Annotation'' button in the sidebar to annotate all instances in the Working Data Sample sheet. \system creates a new sheet, Task 1 (G), to store the data and labels of this labeling task, and also creates a new row for Task 1 in the Task Dashboard sheet. Then, in Step 4, the user can review the outcomes and refine the prompt accordingly (Step 1).
}
    \label{fig:system-interface-step-3-4-kenneth}
\end{figure*}

Users interact with \system to craft a prompt, use it to instruct the LLM in labeling data, review the results, and then revise the prompt through an iterative process. 
To demonstrate the users' workflow, we present a scenario where a user wants to employ \system to label a collection of tweets related to COVID with a 5-point sentiment scale, ranging from Very Negative (1) to Very Positive (5).
The goal is to analyze the sentiment of Twitter (now X) users toward COVID, with an emphasis on ensuring that the classification of each tweet reflects the user's own judgment.
In this case, the LLM's labels should align with the user's assessment of what is positive or negative, as well as the intensity of sentiment, rather than following an ``objective'' standard.
%In other words, the LLM's labels should align with the user's personal perception of the topic rather than adhering to an ``objective'' standard.\kenneth{This is not very accurate hmmm. Might need to edit later.}

%\begin{enumerate}
%    \item 
%\end{enumerate}

\begin{itemize}
   
\item 
\textbf{Step 0: Load and Index the Dataset.}
To begin using \system, the user opens a new Google Sheets document and activates the \system add-on. 
The system automatically sets up the necessary tabs, and the add-on interface appears on the right side of the spreadsheet (Figure~\ref{fig:system-interface-ui-kenneth}). 
The user then imports their data instances into the Dataset sheet, with the text of each tweet placed in the Data Instance column. 
The user must specify a Group ID for each instance. 
If the data are not sequential or grouped, they can assign unique Group IDs using Google Sheets' automatic numbering function.\footnote{\system is designed to accommodate single and grouped data instances within a Group ID. For tasks like sentiment analysis, each data instance is treated separately under its unique Group ID. For tasks that require contextual information, such as annotating text segments in an academic abstract (\eg, CODA-19~\cite{huang-etal-2020-coda}), \system can combine all data instances under the same Group ID into a single request to the LLM model. This flexibility allows the system to support different data instance formats based on user requirements.} 
Once the data is entered, the user clicks the ``Index Data ID'' button in the sidebar, and \system automatically assigns unique data IDs to each instance in the ``Data ID'' column.

\item 
\textbf{Step 1: Compose/Refine the Prompt (Figure~\ref{fig:step-1}).}
%Step 1: compose the promot using things. 
%Uses know a vague idea what they want and will keep revise that idea. But you need to write something down. When it comes to load data, spreadsheet is great!
This is the most critical step, where the user composes and refines prompts for the LLM to label the data. 
In \system, the prompt consists of three parts, each corresponding to a separate sheet: (1) Context, (2) Rule Book, and (3) Shots. 
Figure~\ref{fig:step-1} provides an overview of each sheet.
\begin{enumerate}

\item 
In the \textbf{Context} sheet, the user answers questions that describe the context of the data annotation task, such as the purpose of the annotation and the source of the data, to provide task-specific context for the LLM.

\item
In the \textbf{Rule Book} sheet, the user adds annotation labels along with their definitions. Providing content for both the Context and Rule Book sheets is mandatory, as the LLM requires this information in the prompt to function effectively.

\item
In the \textbf{Shots} sheet, the user adds data instances along with their corresponding gold labels, which serve as examples to help the LLM learn. While adding examples to the Shots sheet is optional during the first iteration---since the user may not yet have a well-defined gold standard for labeling---more examples can be identified as the user reviews data. These examples can be manually added or generated using \system's function (see Step 4).
\end{enumerate}
It is important to note that at the beginning of this labeling process, the user has only a vague idea of what they want to label and will continuously refine that idea. 
Each time the prompt is revised, it reflects an evolution of their understanding and approach to the labeling task.

\item 
\textbf{Step 2: Sample/Resample a Subset (Figure~\ref{fig:system-interface-step-2}).}
Next, the user employs \system to sample a subset of data for labeling. 
Labeling only a subset, rather than the entire dataset, is necessary because the full dataset is too large for the user to thoroughly review the LLM's results. 
Additionally, labeling the entire dataset iteratively would be prohibitively expensive. 
In this step, the user can (1) randomly or (2) sequentially sample data from the Dataset sheet into the Working Data Sample sheet:

\begin{itemize}

\item 
For a \textbf{Random Sample}, the user enters any whole number between 1 and the total number of group IDs in the dataset.
\system will then randomly select that number of groups and copy them into the Working Data Sample sheet. 

\item
In \textbf{Sequential Sample}, the user specifies a range of group IDs from the Dataset sheet, and \system will import the data instances from the selected range into the Working Data Sample sheet.
This feature allows users to process their data instances sequentially in batches, which is especially useful when the data instances have a sequential relationship, such as sentences within the same document.

%The purpose of this feature is to enable users to process their data instances sequentially in batches, making their work more manageable and easier to track.
%\kenneth{Mayeb say a few words on why we need this.}\steven{done.}

\end{itemize}

Once sampling begins, all previously existing data in the Working Data Sample sheet will be removed, except for instances marked as ``Keep it in the next data sample'' (Figure~\ref{fig:system-interface-step-2}). 
Only the data in the Working Data Sample sheet will be labeled by the LLM when the ``Start Annotation'' button is clicked in Step 3.

\item 
\textbf{Step 3: Use the Prompt to Instruct the LLM to Label the Data Sample (Figure~\ref{fig:system-interface-step-3-4-kenneth}).}
After finalizing the three prompt sheets---Context, Rule Book, and Shots---in Step 1 and sampling data instances in Step 2, the user clicks the ``Start Annotation'' button in the sidebar to annotate all instances in the Working Data Sample sheet (Figure~\ref{fig:system-interface-step-3-4-kenneth}).
\system creates a new sheet, Task 1 (Figure~\ref{fig:system-interface-step-3-4-kenneth}), to store the data and labels of this labeling task, and also creates a new row for Task 1 in the Task Dashboard sheet.

In the background, \system first combines the information in Context, Rule Book, and Shots sheets into a prompt (see Section~\ref{sec:implementation} for details).
%gathers the questions and answers from the Context sheet and feeds them into GPT-4 to generate an instruction prompt. 
%This prompt is then combined with the rules and provided gold shots to create the final annotation prompt. 
For each data group (\ie, data instances with the same Group ID), \system sends a request to the LLM using this prompt for annotation.
After receiving the LLM's output, the system parses the results and updates the Task 1 sheet with the annotated outcome for each instance. 
In our implementation, the LLM is always asked to provide explanations for its labels, though the user can decide whether to display these explanations in the annotation results.
%In our user study, we also explored the impact of showing the LLM's explanations to users.

%Step 3: Send it to LLM to label. System creat a new tab; you can navigate tasks using dashboard. Then you re

\item 
\textbf{Step 4: Observe, then Revise the Prompt (Figure~\ref{fig:system-interface-step-3-4-kenneth}).}
The labeling results are saved to the Task 1 sheet (Figure~\ref{fig:system-interface-step-3-4-kenneth}), where the user can manually verify the LLM's labels.
The user can review as many or as few data instances as they wish to develop a better understanding of the labeling task and the dataset. 
Based on this evolving understanding, they can refine the prompt by modifying the Context, Rule Book, and Shots sheets accordingly.

If the user disagrees with any of the labels, they can assign a new label to the data instance under the ``Human Label'' column. 
If the user identifies good examples, they can check the ``Gold Shot'' checkboxes. 
After selecting enough good examples, the user can click the ``Add Shots'' button in the sidebar to add these examples to the Shots sheet (Figure~\ref{fig:system-interface-step-3-4-kenneth}).
Like in other sheets, if the user wants certain data instances to be re-annotated in the next round, they can check the ``Keep it in the next data sample'' checkboxes. 
This will ensure that those instances are not removed during the next sampling process, allowing the user to observe whether the LLM's behavior changes over iterations.

\end{itemize}

When using \system, the user moves through Steps 1, 2, 3, and 4, and then returns to step 1 in an iterative process until they are satisfied with the LLM's labels.

%\subsubsection{Step 0: Initial Setups}

%\subsubsection{Step 1: Compose/Refine the Prompt}

%\subsubsection{Step 2: Sample/Resample a random subset}

%\subsubsection{Step 3: Use the Prompt to Instruct the LLM to Label the Data Sample}

%\subsubsection{Step 4: Observe, and Revise the Prompt}

\subsection{Implementation Details\label{sec:implementation}}
%\kenneth{TODO: Here we mention (1) what framework you used to implement Google Sheets add-on, (2) how do you convert Spreadsheet's content into a prompt, and (3) what LLM (which version exactly) you used and how did you send request (batch? or each data instance is one request?)--- Maybe talk about latency issue here a bit.}

\paragraph{Developing Google Sheets Add-On.}
%\kenneth{How do people built Google Sheets add-ons? Did we use an web server? Where do we store our data?}\steven{done}
\sloppy
\system utilized Google Sheets as its main platform, leveraging the convenience and functionality of its spreadsheet capability. The Google Sheets add-on was implemented in Google App Script, with Google Cloud Service serving as a back-end to store all action logging files. User-specific data, such as OpenAI information, was securely stored in user properties tied to individual email accounts, ensuring privacy protection. 

\paragraph{Converting a Spreadsheet's Content into a Prompt.}
Once users click on the ``Start Annotation'' button (Figure~\ref{fig:system-interface-step-3-4-kenneth}), \system will first collect all questions and answers from the ``Context'' tab and send a request to GPT-4o to generate an instructional prompt (Table~\ref{tab:instruction-prompt}). Next, \system will merge this generated prompt with rules and definitions from ``Rule Book'' and available gold standard labels from the ``Shots'' tab to create an annotation prompt (Table~\ref{tab:main-prompt} and Table~\ref{tab:main-multi-prompt}). Finally, \system will use this prompt to annotate all data instances. 

\paragraph{Interacting with the LLM through an API}
In this paper, we utilized OpenAI's \texttt{gpt-4o-2024-05-13} model for our study~\cite{openai2024gpt4o}.
%\kenneth{TODO: Add citation} \steven{done}
Technically, this LLM can be replaced by any other model that offers an API compatible with the ChatGPT-4 specification. 
In our implementation, we group all data instances with the same Group ID and send them in a single API request.

\section{User Study\label{sec:user-study}}
%\section{Comparative Study Procedure}
Our goal is to investigate how effective people are at prompt engineering when gold labels are absent, namely, ``prompting in the dark''. 
To study this, we conducted an in-lab user study in which participants used \system to perform a 5-point sentiment labeling task on a tweet dataset. 
This section overviews the details of this study.
This study has been approved by the IRB office of the authors' institute.

\subsection{Study Procedure}

%\subsection{In-lab Study}

%We conducted a 90-minute in-lab user study with participants using \system for an annotation task.

\subsubsection{Pilot Study\label{sec:pilot-study}}
Three participants were recruited through the authors' network for the pilot study. 
In the first pilot, we used CODA-19~\cite{huang-etal-2020-coda} as the data annotation task, where participants labeled text segments from academic abstracts into categories such as background, purpose, and findings.
We observed that the participant consistently agreed with nearly all the labels and did not suggest further refinements. 
This may have been due to the highly specialized nature of the abstracts, which made it difficult for a broader audience to fully understand and evaluate the labels. 
As a result, we decided to switch to a Twitter Sentiment task for the second pilot.
In this second pilot, we found that our guidelines were too flexible, leading to participant confusion and uncertainty about how to proceed. 
We made adjustments to provide more structure, such as requiring participants to complete at least four iterations, with each iteration involving the annotation of 10 out of 50 instances. 
After verification, participants were instructed to refine their rules and add gold standard labels.
Based on the results of the two pilot studies, we extended the study duration from 60 to 90 minutes to give participants enough time to learn the system and complete the tasks. Compensation was also adjusted to \$20. 
We tested these revised settings with the third participant and confirmed that the procedure worked effectively.
%Two participants were recruited via the authors' network for the pilot study. 
%\kenneth{TODO: say a few words about pilot study? what did we change after pilot study?}
%In the first pilot study, we used CODA-19~\cite{huang-etal-2020-coda} as the data annotation task, where participants annotated text segments from an academic abstract into background, purpose, method, finding, and others. We observed that the participant consistently agreed with almost all labels and did not suggest further refinement during the verification process. Additionally, while the LLM labels were different than the dataset gold labels, LLM labels and LLM explanations were logically sound. 
%Based on these findings, we decided to switch the annotation task to a Twitter Sentiment task for the second pilot. In this study, we found that our guidelines were too flexible, leading to participant confusion and uncertainty about how to proceed. To address that, we simplified and standardized the user study procedures. For example, we required them to do at least four iterations, with each iteration involving the annotation of 10 out of 50 instances. After verification, participants were instructed to refine their rules and add gold standard labels.

%Based on the results of two pilot studies, we extended the study duration from 60 minutes to 90 minutes to allow participants to have sufficient time to learn the system and complete the annotation tasks. The compensation was also adjusted to \$20.

\subsubsection{Participants Recruitment, Backgrounds, and Grouping}
%Recruitment
For our main study, we focused on recruiting individuals with reasonable familiarity with LLMs but relatively new to large-scale text data annotation. 
While \system is designed as an end-user prompting tool, in this study, we prioritized participants likely to represent the first wave of ``newcomers'' (as noted in our Design Goals in Section~\ref{sec:3-system-design}) entering LLM-powered data annotation. %\kenneth{TODO: Update the section ID} \steven{added}
This focus allowed us to avoid the need to teach participants the basics of LLMs, prompting, or text data annotation.
%For the main study, 
We recruited 20 participants from diverse educational backgrounds through the authors' networks, social media posts, and mailing lists within the authors' institute. 
The group included 1 Post-doctoral Researcher, 9 Ph.D. students, 9 Master's students, and 1 Undergraduate student.  
As part of the recruitment process, we specifically sought participants who met the criteria of possessing prior experience using LLMs. 
%\steven{added recruitment part}
%\steven{one participant was dropped because he did not attend the makeup session, should we mention that?}
Participants were compensated \$20 for their participation, and in our analysis, they are denoted as P1 to P20.
%\steven{Our system is designed for requesters or researchers who understand their task requirements. However, due to recruitment constraints, we could not involve participants with extensive data annotation experience. To address this, we chose a subjective task like Twitter sentiment analysis, where prior annotation expertise is less critical. This approach allows participants to rely on their own knowledge and judgment, simulating real-world scenarios where individuals guide LLMs in tasks that naturally depend on personal interpretation and expertise.}

%Backgrounds
%\kenneth{------------------KENNETH IS WORKING HERE---------------------------}

%\kenneth{------------------KENNETH IS WORKING HERE---------------------------}

%Grouping
Participants were randomly and evenly assigned to four groups based on two variables: 
% (1) whether or not they could view the LLM's explanations for its labels\footnote{\steven{Since two participants who had access to LLM explanations chose to turn them off, we grouped them with the no LLM explanation participants, resulting in 8 participants with access to LLM explanations and 12 participants without access.}}, and 
% (2) whether they had access to 50 instances per iteration or 10 instances per iteration.
(1) whether they had access to 50 instances per iteration or 10 instances per iteration, and
(2) whether or not they could view the LLM's explanations for its labels\footnote{Since one participant chose to disable the LLM explanations after the first iteration and another participant decided not to use the LLM explanations throughout the entire study, we grouped them with the no LLM explanation participants, resulting in 8 participants with access to LLM explanations and 12 participants without access.}.
Further details are provided in the study procedure section (Section~\ref{sec:study-procedure}). 

\paragraph{Survey on Participants' LLM Familiarity and Usage.}
To assess participants' familiarity with using LLMs, we conducted an optional post-study survey, offering an additional \$5 compensation for completion. 
(The full set of survey questions is provided in Table~\ref{tab:participants-llm-background-survey} in Appendix~\ref{sec:appendix=participant-background}.) %\kenneth{UPDATE REF}\steven{updated}
All participants responded. %\kenneth{UPDATE NUMBER}\steven{updated}
Most participants reported being familiar with LLMs, with an average familiarity score of 4.20 (SD=0.77) on a 5-point scale. %\kenneth{UPDATE NUMBER}\steven{updated}
15 participants had over one year of experience using LLMs, while 4 reported more than four months of experience, and 1 reported between one and three months. %\kenneth{UPDATE NUMBER} \steven{updated}
In terms of usage frequency, 16 participants used LLMs daily, 3 used them weekly, and 1 used them monthly. %\kenneth{UPDATE NUMBER} \steven{updated}
%Interaction durations varied: eight participants engaged with LLMs for more than 30 minutes, four for less than 5 minutes, three for 5-15 minutes, and four for 15-30 minutes. 
While most participants used LLMs for general tasks such as Q\&A, research, writing assistance, and programming/debugging, only 5 participants had experience using LLMs for data labeling. %\kenneth{UPDATE NUMBER}\steven{updated}
Participants rated their confidence in crafting prompts and their proficiency in interacting with LLMs similarly, with average scores of 3.75 (SD=0.85) and 3.85 (SD=0.88), respectively. %\kenneth{UPDATE NUMBER}\steven{updated}
%Many employed prompt engineering techniques, ranging from simple input adjustments to advanced methods like iterative refinement, system message editing, in-context learning, and chain-of-thought prompting to enhance LLM performance.
Overall, the participants represented individuals familiar with LLMs but relatively inexperienced with large-scale data annotation.

\subsubsection{Labeling Task, Scheme, and Data}
We selected the Coronavirus Tweet NLP Text Classification task, which categorizes tweets into five sentiment categories: Extremely Positive, Positive, Neutral, Negative, and Extremely Negative, using the dataset hosted on Kaggle.\footnote{Coronavirus tweets NLP - Text Classification: https://www.kaggle.com/datasets/datatattle/covid-19-nlp-text-classification/}
The dataset contains tweets from December 30, 2019, to September 7, 2020. 
For our study, we randomly sampled 1,060 tweets: 10 tweets were used for the tutorial task, 1,000 for the main study, and 50 for the final evaluation set (see Section~\ref{sec:study-procedure}).

This task was chosen, partially informed by our pilot study (Section~\ref{sec:pilot-study}), for several reasons. 
First, it strikes a balance in difficulty, being challenging enough to require iterative prompting efforts from LLMs, as a 5-class sentiment task is more complex than typical 2-class (positive, negative) or 3-class (positive, negative, neutral) sentiment classification tasks. 
Second, it avoids requiring specialized knowledge, ensuring a broad pool of potential participants. 
Tasks demanding domain-specific expertise would have significantly restricted recruitment; sentiment labeling for general COVID-related tweets is sufficiently accessible for this purpose. 
Finally, the task incorporates a subjective element, as it lacks universally agreed-upon gold labels. 
This aligns with our focus on ``prompting in the dark,'' where participants' understanding of the data, as well as labeling goals, evolve through iterations.
The subjective nature of the task allows participants to arrive at differing gold standards by the end of the process.
Considering these factors, we selected this task for our study.

%We chose this task during our pilot study (Section~\ref{sec:pilot-study}), primarily for its accessibility and its somewhat subjective nature: 
%sentiment analysis does not require specialized expertise; 
%the evaluation often depends on personal judgment.
%\kenneth{TODO Kenneth: This is not very accurate. need to revise later.}
%\steven{In contrast to tasks with strictly defined objectives, such as CODA-19~\cite{huang-etal-2020-coda}, which require extensive domain knowledge, subjective tasks like Twitter sentiment analysis do not have universally correct answers, relying instead on personal judgment for evaluation. 
%In this study, we encouraged participants to guide the LLM to align with their individual standards rather than steering it toward a fixed standard grounded in domain expertise}
%This \steven{task} allowed participants to guide the LLM according to their own interpretation of sentiment, particularly in deciding what qualifies as ``extremely'' positive or negative.

\subsubsection{Study Procedure\label{sec:study-procedure}}
For our main user study, most sessions were conducted remotely via Zoom or Microsoft Teams, with each session lasting between 87 and 127 minutes. 
Participants who attended in person used one of the author's laptops, while remote participants used their own computers. 
Since \system was a Google Add-on in a development version, it was installed on one of the author's laptops. 
Remote participants were given control of this laptop to conduct the experiment. 
Each session was recorded, capturing the screen, audio, and video for further analysis.

The study followed these steps:

\begin{enumerate}

\item 
\textbf{Onboarding:}
Participants were first introduced to the study's objectives and procedures, and their informed consent was obtained.

\item 
\textbf{Tutorial Task:}
Participants were then presented with a tutorial on the system's workflow and features, either delivered by one of the authors or via a prerecorded video, depending on their preference. 
Afterward, they completed a short tutorial task, identical to the main study task but involving only 10 data instances, to ensure their understanding of the system.

\item 
\textbf{Main Study:}
Participants were then asked to use \system to iteratively compose a prompt to label the sentiment of COVID-related tweets in alignment with their personal judgment of sentiment scores. 
Each participant was asked to complete at least four iterations (\ie, going through Steps 1 to 4 four or more times).
Participants were free to do additional iterations beyond the required four; on average, participants completed 4.75 iterations.
%\kenneth{update numbers}\steven{done}

Depending on their assigned group, participants used \system to annotate either 50 or 10 instances per iteration and then review the results.
For participants working with 50 instances per iteration, we advised that it was not necessary to manually verify all labels, as that would take too much time. 
% Participants in the group with access to LLM explanations could manually turn off the explanations if they felt that reading them was too time-consuming; only a few participants chose to do so.
Participants in the group with access to LLM explanations were explicitly informed that they
could manually turn on the explanations if they felt that they needed reasoning for each label. 
Most participants turned on the LLM explanations in the first iteration;
however, one participant chose to disable the LLM explanations after the first iteration and another participant decided not to use the LLM explanations throughout the entire study.
% , but a few participants chose to disable them after the first iteration. 
% One participant opted not to use the LLM explanations throughout the entire study.
%\kenneth{Is this true? How does this work?}\steven{the default is off, they can choose to turn on or keep it off.}\kenneth{hmmm how did we test the effect in such case then? Did they turn it on very often??? Did we always ask LLM to provide explanations?}\steven{All participants turned on the LLM explanation in the first iteration and use LLM explanation. and three participants decided to turn off the LLM explanation. }\kenneth{how about implementation? Did we always ask for LLMs to give us explanations in our prompt, just some participants do not have acees to it?}\steven{yes, the LLM explanation always in the raw output. Our LLM explanation checkbox is used to display or not display the LLM explanation. The label outputs remain consistent for all participants. }\steven{my bad, one participant did not turn on the LLM explanation the whole time.}

%After reviewing the LLM-generated labels, participants were encouraged to refine their Context, Rule Book, and Shots sheets if they gained new insights into the task, or to add gold shots.
%All participants were required to complete at least four iterations (i.e., going through Steps 1 to 4 four times). 

\item 
\textbf{Manual Labeling of the Gold Set:}
Upon completing the iterative process, participants manually labeled 50 tweets based on their own sentiment judgments. 
These labels reflected the participants' understanding of the data and the final labeling they aimed to achieve by the end of the study session. 
These manually labeled tweets were used as an evaluation dataset to assess the performance of the participants' prompts; they were not used to train or fine-tune any AI models.

%Upon completing the iterative process, participants manually labeled 50 tweets based on their own sentiment judgments. 
%These manually labeled tweets served as the evaluation dataset to assess the performance of their prompt.
%\kenneth{TODO Kenneth: (1) Say it captured at last understanding of users. (2) WE do not use it for any training or fine-tuning!!!}

\item 
\textbf{Post-Study Survey and Feedback Collection:}
At the end of the session, participants completed a questionnaire to rate the system's effectiveness, performance, and accessibility. They were also asked the following questions:
(1) Without this tool, how would you typically approach prompt engineering?
(2) How does your prompt engineering process compare before and after using this tool?
(3) Did the system help you complete the tasks more efficiently? If yes, please explain how.
(4) What features did you find most useful?
(5) Would you be interested in using this annotation system in your regular work or study? If no, please explain why.
(6) Do you have any suggestions for making the system more suitable for your needs?

Upon completing the questionnaire, we verbally asked participants to provide some last comments about the workflow, labeling task, and our system.

\end{enumerate}

\section{Findings}
In this section, we organize our findings under each research question (RQs) mentioned in Section~\ref{sec:rq}.

\subsection{RQ 1-1: How effective are people at prompt engineering in the ``prompting in the dark'' scenario?}
\begin{figure*}[t]
    \centering
    \begin{subfigure}{0.48\textwidth}
    \includegraphics[width=\linewidth]{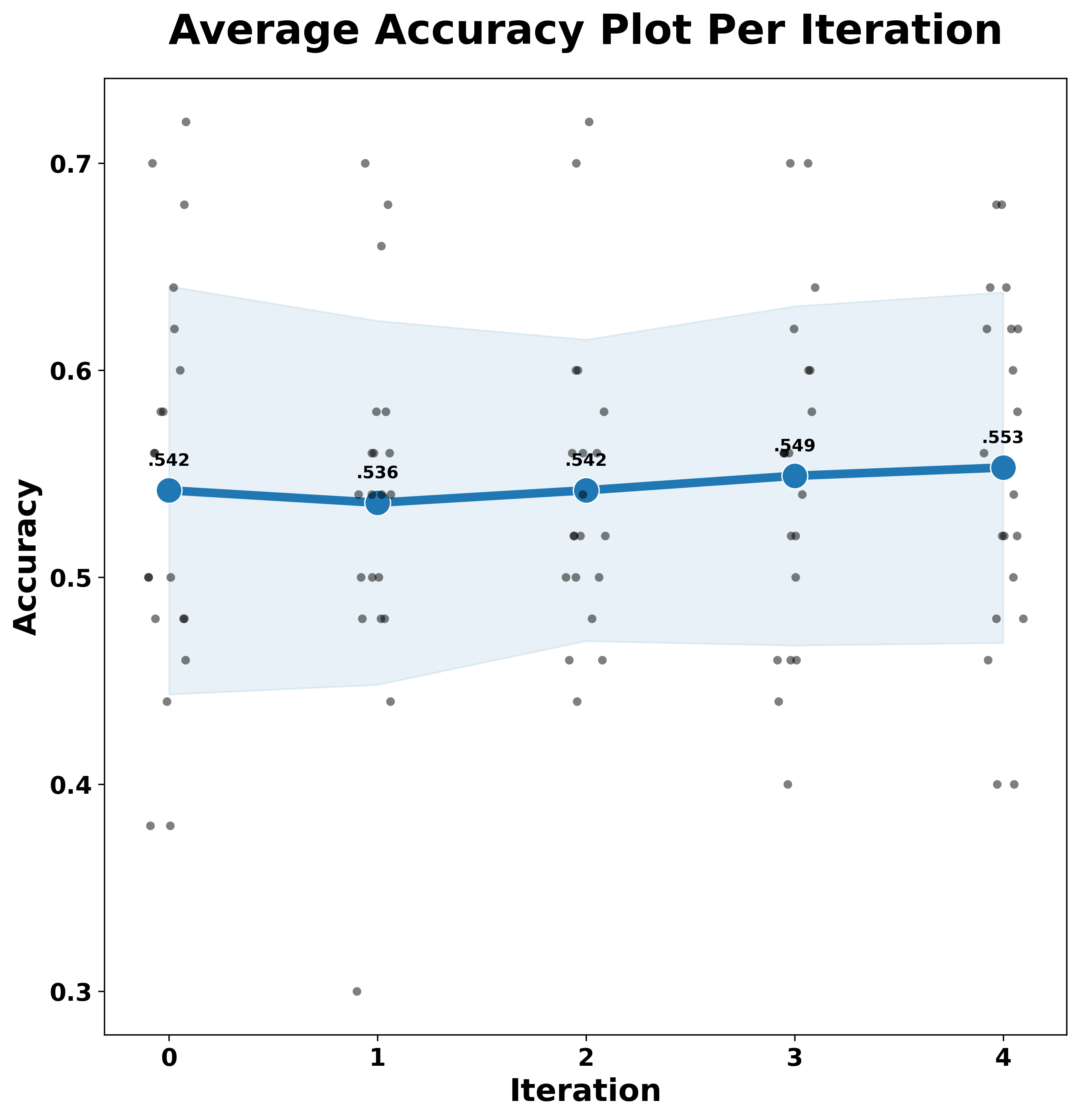}\Description{This is an average accuracy plots subfigure of the participants prompt performance. The subfigure includes shaded regions representing the standard deviation and scatter points indicating individual participants’ data.}
    \caption{The average ACC of the first prompt and the four revisions.}
    \label{fig:average-acc}
  \end{subfigure}
  \hfill
  \begin{subfigure}{0.48\textwidth}
    \includegraphics[width=\linewidth]{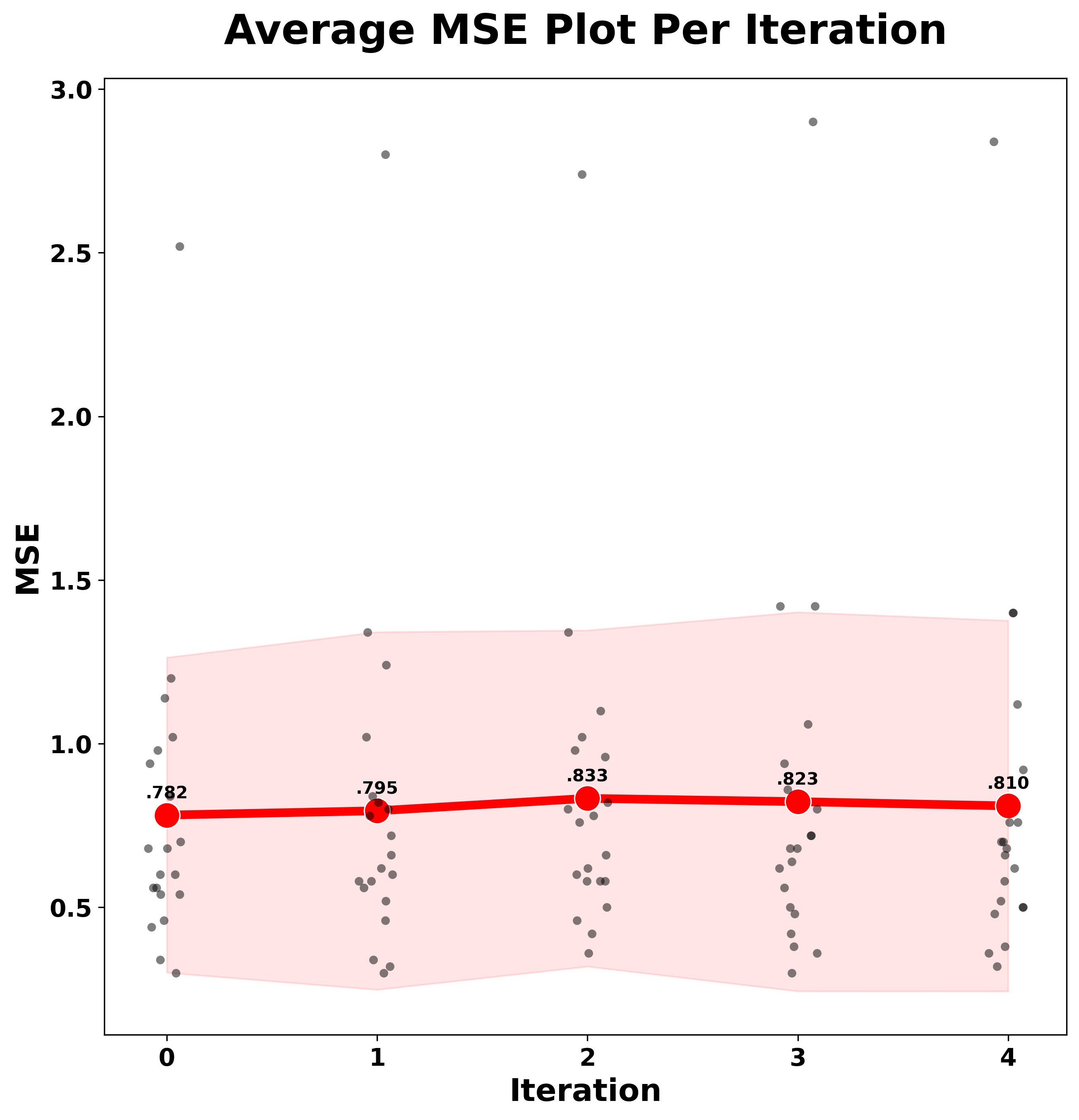}\Description{This is a MSE accuracy plots subfigure of the participants prompt performance. The subfigure includes shaded regions representing the standard deviation and scatter points indicating individual participants’ data.}
    \caption{The average MSE of the first prompt and the four revisions.}
    \label{fig:average-mse}
  \end{subfigure}
  \caption{The average ACC and MSE of the first prompt and the four subsequent revisions. These results show that prompting in the dark is not particularly effective. 
Average labeling accuracy only slightly improved after four iterations; the average MSE fluctuated, ultimately increasing only marginally by the fourth revision.}
  \label{fig:average-acc-mse-performance}
\end{figure*}

\begin{table*}[t]
\centering
%\small
\begin{tabular}{lcccc}
 & \multicolumn{2}{c}{\textbf{ACC$\uparrow$}} & \multicolumn{2}{c}{\textbf{MSE$\downarrow$}} \\ \hline
\multicolumn{1}{c}{} & \textbf{Avg.(SD)} & \textbf{\begin{tabular}[c]{@{}c@{}}\%Participant\\ Improved\\ Over Initial\end{tabular}} & \textbf{Avg.(SD)} & \textbf{\begin{tabular}[c]{@{}c@{}}\%Participant\\ Improved\\ Over Initial\end{tabular}} \\ \hline
\textbf{Initial Prompt} & .542 (.099) & - & .782 (.482) & - \\ \hline
\textbf{1st Revision} & .536 (.088) & 35\% & .795 (.546) & 40\% \\
\textbf{2nd Revision} & .542 (.073) & 40\% & .833 (.514) & 45\% \\
\textbf{3rd Revision} & \underline{\textbf{.549 (.082)}} & 50\% & .823 (.579) & 50\% \\
\textbf{4th Revision} & \underline{\textbf{.553 (.085)}} & 45\% & .810 (.566) & 40\% \\ \hline
\textbf{\begin{tabular}[c]{@{}l@{}}End of Session\\ (Avg \#Iter = 4.75)\end{tabular}} & \underline{\textbf{.546 (.084)}} & 45\% & .815 (.489) & 45\%
\end{tabular}
\caption{The average ACC and MSE of the first prompt, the four subsequent revisions, and at the end of the session. Improvements over the initial prompt are bolded and underlined. }
\label{tab:overall-ACC-MSE}
\end{table*}

\begin{figure*}[t]
    \centering
    \includegraphics[width=0.95\linewidth]{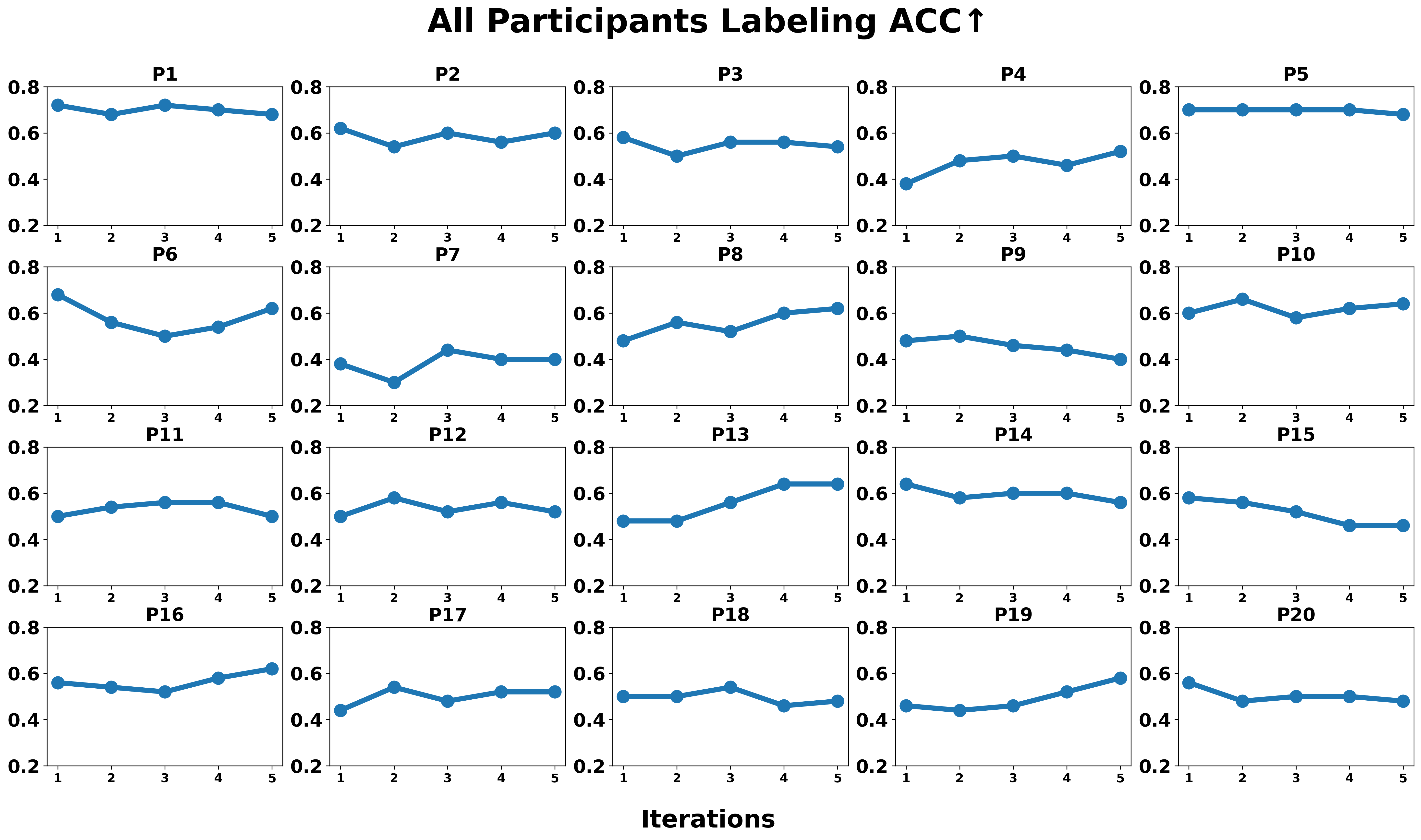}\Description{It contains 20 subfigures of the average ACC of each participant.}
    \caption{Accuracy Plots for all participants. The results show that the process is highly unreliable. 
Labeling accuracy improved for 9 participants after four iterations, declined for 10, and remained unchanged for 1. }
    \label{fig:individual-pure-acc}
\end{figure*}

\begin{figure*}[t]
    \centering
    \includegraphics[width=0.95\linewidth]{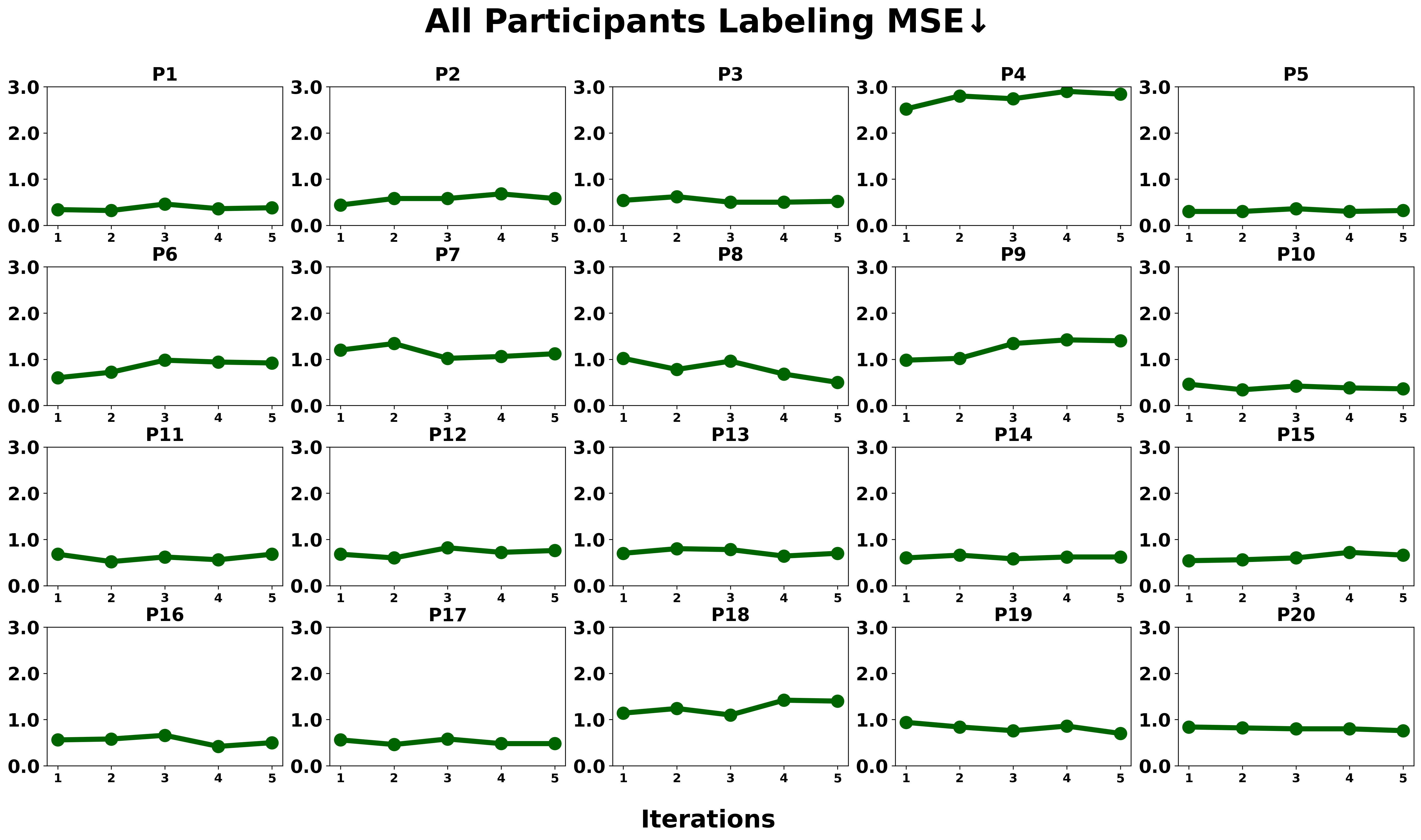}\Description{It contains 20 subfigures of the average MSE of each participant.}
    \caption{MSE Plots for all participants.
    The results show that the process is highly unreliable. 
MSE improved (\ie, decreased) for 8 participants, worsened for 10, and stayed the same for 2.}
    \label{fig:individual-pure-mse}
\end{figure*}

%\kenneth{NO ONE SHOULD DO THIS. THIS IS HARDDD.}

We evaluated each prompt iteration by comparing the labels generated by the LLM (GPT-4o) to the gold labels participants manually annotated for a set of 50 tweets at the end of their study session.
Performance was measured using accuracy (ACC) and Mean Square Error (MSE), two commonly used metrics in sentiment analysis~\cite{saxena2022introduction}.
%\kenneth{Why do we cite this here?} \steven{this describe that ACC and MSE are two evaluation matrices for analyzing sentiment task}
In the main study, each participant provided at least five prompts (with some offering more), and our evaluation focuses primarily on these five prompts, which were shared across all participants.
The ACC and MSE were calculated by comparing the annotation results from participant-provided prompts with the manually generated gold-standard labels. Each label was treated as a distinct category, and accuracy was then calculated accordingly.

\subsubsection{5-Point sentiment rating is a highly subjective task.}
%\kenneth{I kinda feel we don't need a table for kappa}

%The premise of our study is that users bring their own personal perspectives and judgments to even seemingly identical labeling tasks. 
%Before analyzing the results, we first validate this assumption in our study.
%We calculated both Cohen's kappa and Kendall's correlation between participants' manual ratings, collected at the end of our study and the ``gold-standard'' labels provided by the dataset, as well as the value between participants.
The premise of our study is that users bring their own personal perspectives and judgments to seemingly identical labeling tasks. 
To validate this assumption, we calculated both Cohen's kappa
%, as well as Spearman's and Kendall's correlations 
between participants' manual ratings---the labels collected at the end of the study---and the ``gold-standard'' labels from the dataset, as well as the agreement between participants themselves.
%\kenneth{TODO Steven: Kappa is pretty harsh. Can you additionally try Spearman or Spearman, Kendall correlation?}\steven{sure, Spearman: participant vs dataset is 0.343(SD=0.121), between participants is 0.530(SD=0.078). Kendall: participant vs dataset is 0.299(SD=0.107), and between participants is 0.474(SD=0.071) These two correlation scores are slightly better than Kappa but still not very good.}
%\paragraph{Inter-annotator agreement (Cohen's kappa)} 
Participants' labels show a poor alignment with the labels from the original dataset, with an average Kappa of 0.114 (SD=0.070). 
%Spearman's of 0.343 (SD=0.121), and 
%Kendall's of 0.299 (SD=0.107).
%Interestingly, 
The average Kappa value between participants was only slightly higher, with an average Kappa of 0.249 (SD=0.059).
%Spearman's of 0.530(SD=0.078), and 
%Kendall's of 0.474(SD=0.071).
%\kenneth{is this average between all the participant pairs?}\steven{yes}
%each individual participant and the rest of the participants was only slightly  higher at 0.249 (SD=0.059), suggesting participants were more consistent with each other than with the original dataset labels.
The low Kappa score indicated the Twitter Sentiment task we used in our study was a highly subjective task. 
Namely, each participant's gold labels were highly influenced by their personal interpretations and preferences.

\subsubsection{Prompting in the dark is ineffective}
Table~\ref{tab:overall-ACC-MSE} and Figure~\ref{fig:average-acc-mse-performance} show the average ACC and MSE for the first prompt and the four subsequent revisions.
%\kenneth{TODO: Add reference to the Table}\steven{done}
Our analysis, as captured in \system, indicates that prompting in the dark is not particularly effective. 
Among the 20 participants, average labeling accuracy (where higher is better) only slightly improved from 0.542 to 0.553 after four iterations (Figure~\ref{fig:average-acc}). 
Some participants went through more than four iterations. 
The average labeling accuracy at the end of their sessions---the final iteration for all participants---was improved to 0.546.
%\kenneth{Add one sentence about the end-of-session ACC performance}\steven{done}
Meanwhile, the average MSE (where lower is better) fluctuated, ultimately increasing from 0.782 to 0.810 by the fourth revision (Figure~\ref{fig:average-mse}). 
The average MSE for the end-of-session increased to 0.815.
%\kenneth{Add one sentence about the end-of-session MSE performance}\steven{done}
It is important to note that, since this is a 5-scale rating task, ACC is a more harsh metric, awarding credit only for exact matches, while MSE considers the distance between the predicted rating and the user-specified rating.

%\steven{significant increasing trend by using linear mixed-effect model}

\subsubsection{Prompting in the dark is unreliable.}
To further illustrate how the process unfolded for each participant, we present individual ACC and MSE charts in Figure~\ref{fig:individual-pure-acc} and Figure~\ref{fig:individual-pure-mse}. 
The results show that the process is highly unreliable. 
Labeling accuracy improved for 9 participants after four iterations, declined for 10, and remained unchanged for 1. 
Similarly, MSE improved (\ie, decreased) for 8 participants, worsened for 10, and stayed the same for 2. 
Overall, the practice of ``prompting in the dark''---iterating prompts without reference to gold labels---proved unreliable, with over half of the participants experiencing a decline in performance by the end of the study.

%--------------- dead kitten ---------------
\begin{comment}

The average labeling accuracy had a slight increase (Figure~\ref{fig:average-acc}), rising from 54.20\% to 55.30\% in the final refined prompt compared to the initial prompt. Meanwhile, the average MSE (Figure~\ref{fig:average-mse}) fluctuated and ultimately increased from 0.782 to 0.810 by the last prompt.

Figure~\ref{fig:individual-pure-acc} and Figure~\ref{fig:individual-pure-mse} illustrate the performance of participant-guided LLM for each iteration. 
Labeling accuracy improved for 9 participants after four iterations, decreased for 10 participants, and remained unchanged for 1 participant.
Labeling MSE improved (dropping in MSE) for 8 participants, decreased for 10 participants, and remained unchanged for 2 participants.

Overall, the prompt refined during the user study was \textbf{highly unreliable} as over half of the participants experienced a decline in their performance at the end of the study.

\end{comment}

\subsection{RQ 1-2: How does sample size affect human performance in prompt engineering?\label{sec:rq-1-2}}
%\usepackage{booktabs}
\begin{table*}[t]
\centering
\footnotesize
\begin{tabular}{@{}lcccccccc@{}}
 & \multicolumn{4}{c}{\textbf{50 Samples/Round (N=10)}} & \multicolumn{4}{c}{\textbf{10 Samples/Round (N=10)}} \\ \cmidrule{2-9} 
 & \multicolumn{2}{c}{\textbf{ACC$\uparrow$}} & \multicolumn{2}{c}{\textbf{MSE$\downarrow$}} & \multicolumn{2}{c}{\textbf{ACC$\uparrow$}} & \multicolumn{2}{c}{\textbf{MSE$\downarrow$}} \\ \midrule
\multicolumn{1}{c}{} & \textbf{Avg.(SD)} & \textbf{\begin{tabular}[c]{@{}c@{}}\%Participant\\ Improved\\ Over Initial\end{tabular}} & \textbf{Avg.(SD)} & \textbf{\begin{tabular}[c]{@{}c@{}}\%Participant\\ Improved\\ Over Initial\end{tabular}} & \textbf{Avg.(SD)} & \textbf{\begin{tabular}[c]{@{}c@{}}\%Participant\\ Improved\\ Over Initial\end{tabular}} & \textbf{Avg.(SD)} & \textbf{\begin{tabular}[c]{@{}c@{}}\%Participant\\ Improved\\ Over Initial\end{tabular}} \\ \midrule
\textbf{Initial Prompt} & .520 (.075) & - & .742 (.242)  & - & .564 (.117) & - & .822 (.654) & - \\ \midrule
\textbf{1st Revision} & \underline{\textbf{.530 (.094)}} & 50\% & \underline{\textbf{.720 (.285)}} & 40\% & .542 (.086) & 20\% & .870 (.733) & 40\% \\
\textbf{2nd Revision} & \underline{\textbf{.528 (.050)}} & 50\% & .780 (.268) & 50\% & .556 (.091) & 30\% & .886 (.692) & 40\% \\
\textbf{3rd Revision} & \underline{\textbf{.546 (.083)}} & 70\% & \underline{\textbf{.722 (.308)}} & 60\% & .552 (.085) & 30\% & .924 (.768) & 40\% \\
\textbf{4th Revision} & \underline{\textbf{.536 (.095)}} & 60\% & \underline{\textbf{.730 (.310)}} & 40\% & \underline{\textbf{.570 (.074)}} & 30\% & .890 (.753) & 40\% \\ \midrule
\textbf{\begin{tabular}[c]{@{}l@{}}End of Session\\ (Avg \#Iter=4.75)\end{tabular}} & \underline{\textbf{.536 (.095)}} & 60\% & \underline{\textbf{.736 (.303)}} & 40\% & .556 (.075) & 30\% & .894 (.632) & 50\%
\end{tabular}
\caption{Comparison of participants who reviewed 50 instances per iteration versus those who reviewed 10 instances per iteration. Reviewing 50 instances per iteration resulted in more frequent and consistent improvements compared to reviewing 10 instances. Improvements over the initial prompt are bolded and underlined.}
\label{tab:table-50-example-kenneth}
\end{table*}

\begin{figure*}[t]
        \centering
        \includegraphics[width=0.98\linewidth]{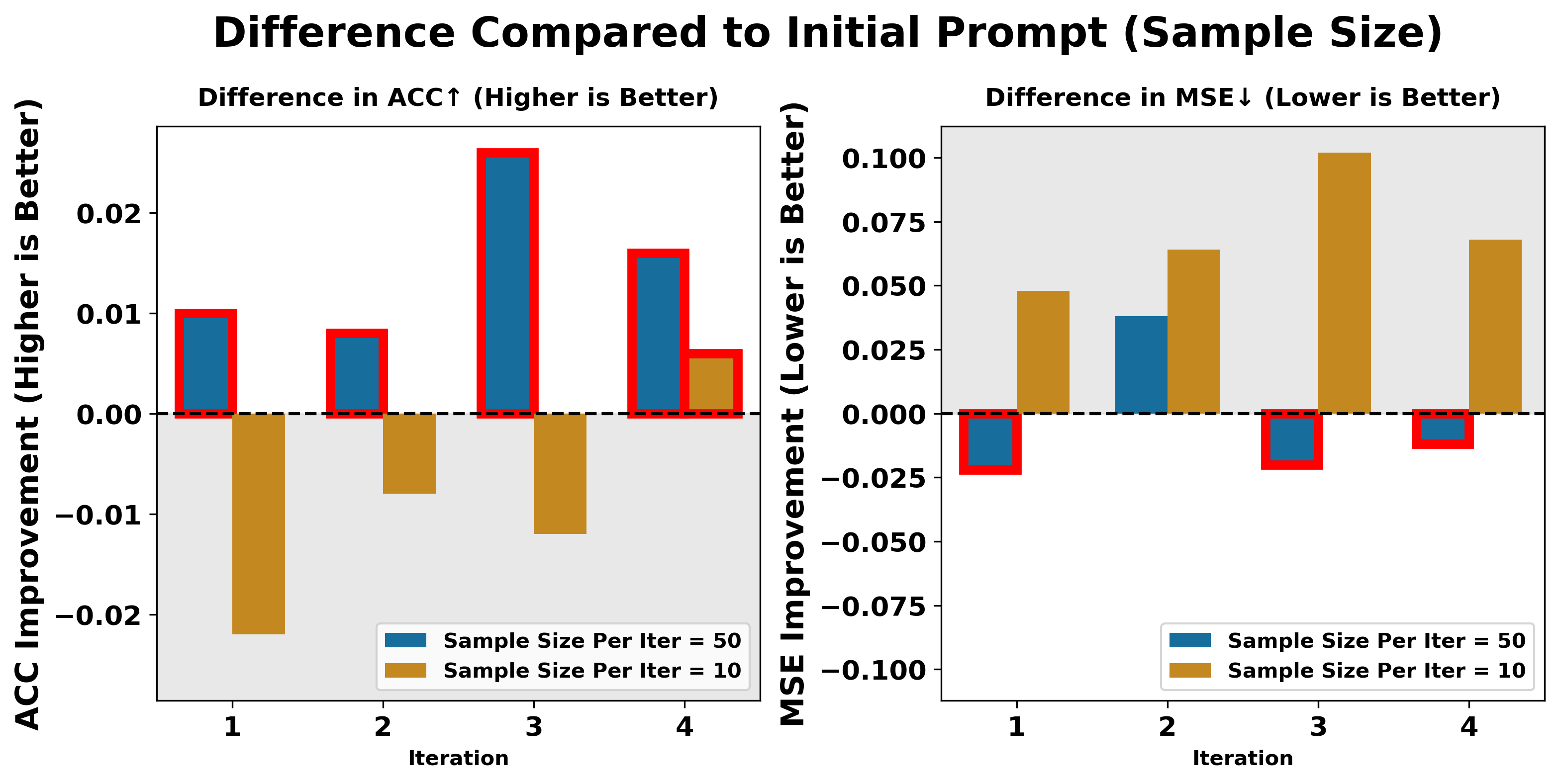}\Description{It contains 2 subplots illustrating the effect of sample size on ACC and MSE across iterations. The bars represent performance changes relative to the initial prompt, with improvements highlighted by a thick red border.}
  \caption{Comparison of improvement (difference) in ACC and MSE over the initial prompt between the 50-instance and 10-instance groups. Bars with red borders indicate positive improvements over the initial prompt's outcome. Reviewing 50 instances per iteration results in more consistent improvements in ACC compared to reviewing 10 instances. }
%.\kenneth{TODO Steven: (1) The fonts need to be bigger, especially the legend label. They are way too small. (2) The title should say ``50 Samples/Round'' vs. ``10 Samples/Round'' instead of w/ w/o. (3) MAYBE use Red for 50 and }
  \label{fig:average-50-instance-acc-mse-performance}
\end{figure*}

\subsubsection{Reviewing 50 instances per iteration leads to more frequent and consistent improvements compared to reviewing 10 instances}

%\kenneth{TODO Steven: Can you talk to Zixin about how to do t-test (or any significance test) in this case? In particular (1) how to do it between iterations and (2) how to do it for the "differences" instead of for absolute value. I kinda feel like we care about the diff than the absolute value. 50 vs 10 group, the 10 group had a better starting acc/mse but that's not relevant to our system but rather just by chance. }

%Participants who reviewed 50 instances were designated as \textbf{50Y}, and those who reviewed 10 instances were labeled as \textbf{50N}.
%For 50Y participants, 6 out of 10 participants demonstrated an improvement in their labeling accuracy, whereas 3 out of 10 participants in the 50N group improved. 
%However, participants under 50Y demonstrated a slightly better change in MSE compared to 50N, as 50Y had 2 fewer individuals with a decline in MSE performance. 
Table~\ref{tab:table-50-example-kenneth} and Figure~\ref{fig:average-50-instance-acc-mse-performance} present a comparison of participants who reviewed 50 instances per iteration against those who reviewed 10 instances per iteration. The detailed breakdown is shown in Figure~\ref{fig:average-acc-mse-llm-instances-performance-new}.
%\kenneth{TODO Steven: Update the figure and table references.}\steven{done}
In terms of accuracy, at the end of the session (\ie, four or more revisions),
6 out of 10 participants in the 50-instance group showed improvement, while only 3 out of 10 participants in the 10-instance group improved.
%\kenneth{Is this AFTER 4 REVISIONs or AT THE END OF SESSION?}\steven{4 revision and at the end have the same number of improved partcipant}
On average, every iteration in the 50-instance group resulted in better accuracy compared to the initial prompt, though the improvement was not strictly increasing with each iteration.

For MSE, participants in the 50-instance group improved across three iterations, while those in the 10-instance group showed no improvement over the initial prompt in any round.

We also note that participants in the 10-instance group began with higher initial performance, but this was before they viewed the labeling results and occurred by chance, unrelated to the experimental conditions. 
Our analysis focuses on performance differences between iterations across both groups.

\paragraph{Significant Tests.}
We conducted eight linear mixed-effects models to examine the effect of iteration across four different conditions. 
The dependent variables were ACC and MSE, and participants were treated as random effects.
%Under conditions where participants \textbf{did not have access to LLM explanations,} we observed a significant increasing trend in accuracy with each iteration ($\beta$=0.013, p-value=0.009**). 
%Conversely, 
In the condition where participants \textbf{reviewed only 10 instances per iteration}, we found a significant increasing trend in MSE as the iterations progressed ($\beta$=0.019, p-value=0.043*).

%\kenneth{Bigger is better!}

\subsection{RQ 1-3: How does displaying LLM explanations impact human performance in prompt engineering?\label{sec:llm-explanation-result}}
\begin{table*}[t]
\centering
\footnotesize
\begin{tabular}{@{}lcccccccc@{}}
 & \multicolumn{4}{c}{\textbf{LLM Explanations Shown (N=8)}} & \multicolumn{4}{c}{\textbf{No LLM Explanations Shown (N=12)}} \\ \cmidrule{2-9} 
 & \multicolumn{2}{c}{\textbf{ACC$\uparrow$}} & \multicolumn{2}{c}{\textbf{MSE$\downarrow$}} & \multicolumn{2}{c}{\textbf{ACC$\uparrow$}} & \multicolumn{2}{c}{\textbf{MSE$\downarrow$}} \\ \midrule
\multicolumn{1}{c}{} & \textbf{Avg. (SD)} & \textbf{\begin{tabular}[c]{@{}c@{}}\%Participant\\ Improved\\ Over Initial\end{tabular}} & \textbf{Avg. (SD)} & \textbf{\begin{tabular}[c]{@{}c@{}}\%Participant\\ Improved\\ Over Initial\end{tabular}} & \textbf{Avg. (SD)} & \textbf{\begin{tabular}[c]{@{}c@{}}\%Participant\\ Improved\\ Over Initial\end{tabular}} & \textbf{Avg.(SD)} & \textbf{\begin{tabular}[c]{@{}c@{}}\%Participant\\ Improved\\ Over Initial\end{tabular}} \\ \midrule
\textbf{Initial Prompt} & .590 (.063) & - & .608 (.259) & - & .510 (.107) & - & .898 (.566) & - \\ \midrule
\textbf{1st Revision} & .558 (.074) & 12.5\% & .633 (.292)  & 37.5\% & \underline{\textbf{.522 (.096)}} & 50\% & .903 (.655) & 41.7\% \\
\textbf{2nd Revision} & .568 (.070)  & 12.5\% & .640 (.221) & 50\% & \underline{\textbf{.525 (.072)}} & 58.3\% & .962 (.616) & 41.7\% \\
\textbf{3rd Revision} & .555 (.082) & 25\% & .660 (.349) & 50\% & \underline{\textbf{.545 (.085)}} & 66.7\% & .932 (.685)  & 50\% \\
\textbf{4th Revision} & .563 (.084) & 25\% & .645 (.333) & 50\% & \underline{\textbf{.547 (.088)}}  & 58.3\% & .920 (.671) & 33.3\% \\ \midrule
\textbf{\begin{tabular}[c]{@{}l@{}}End of Session\\ (Avg \#Iter=4.75)\end{tabular}} & .563 (.087) & 25\% & .680 (.311) & 50\% & \underline{\textbf{.535 (.083)}} & 58.3\% & .905 (.574) & 41.7\%
\end{tabular}
\caption{Comparison of ACC and MSE between participants with and without access to LLM explanations during the labeling process. Participants without access to LLM explanations showed improvement in accuracy over multiple revisions, while those with access did not exhibit the same level of improvement. Improvements over the initial prompt are bolded and underlined.}
\label{tab:results-llm-new}
\end{table*}

\begin{figure*}[t]
    \centering
    \includegraphics[width=0.98\linewidth]{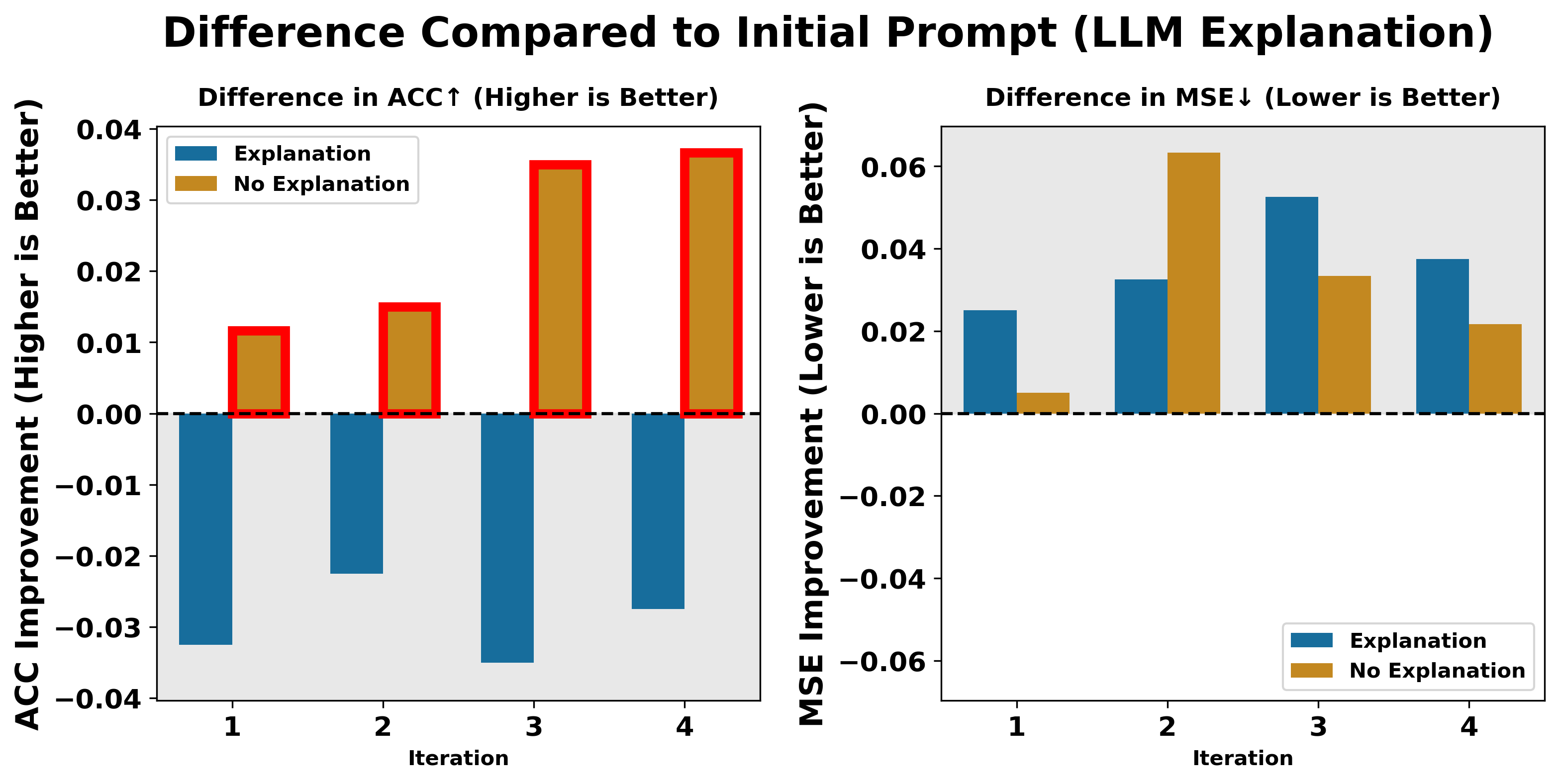}\Description{It contains 2 subplots illustrating the effect of LLM explanation on ACC and MSE across iterations. The bars represent performance changes relative to the initial prompt, with improvements highlighted by a thick red border.}
  \caption{Comparison of improvement (difference) in ACC and MSE over the initial prompt between the Explanation and No-Explanation groups. Bars with red borders indicate positive improvements over the initial prompt's outcome. Participants without access to LLM explanations improved their accuracy over multiple revisions, while those with access did not.}
  \label{fig:diff-bar-average-acc-mse-llm-explanation-performance-new}
\end{figure*}

%new added explnation and 50 instance figures

%% end adding

\subsubsection{Participants without access to LLM explanations improved their accuracy over multiple revisions, while those with access did not.}
Table~\ref{tab:results-llm-new} and Figure~\ref{fig:diff-bar-average-acc-mse-llm-explanation-performance-new} show a comparison between participants with and without access to the LLM's explanations during the labeling process. The detailed breakdown is shown in Figure~\ref{fig:average-acc-mse-llm-explanation-performance-new}.
%\kenneth{TODO Steven: Update Table references} \steven{done}
%After four revisions, 
At the end of the session (\ie, four or more revisions),
7 out of 12 participants without access to explanations improved their accuracy, while only 2 out of 8 participants with access to explanations showed improvement.
%\kenneth{Is this after four revisions or at the end of the session?} \steven{both after 4th and at end session}
On average, the group without access saw improved accuracy over the initial prompt with each of the four revisions, whereas the group with access to explanations experienced a decline in accuracy across all revisions.

In terms of MSE, there was no significant difference between the two groups in the number of individuals who showed improvement. 
In fact, both groups saw an increase in MSE during the revision process.

As with the results related to data sample size (Section~\ref{sec:rq-1-2}), we are aware that participants with access to LLM explanations started with higher initial performance; this initial advantage occurred before the participants reviewed the labeling results and explanations.
Our analysis focuses on performance changes across iterations rather than the absolute performances.

\paragraph{Significant Tests.}
We conducted eight linear mixed-effects models to examine the effect of iteration across four different conditions. 
The dependent variables were ACC and MSE, and participants were treated as random effects.
Under conditions where participants \textbf{did not have access to LLM explanations,} we observed a significant increasing trend in accuracy with each iteration ($\beta$=0.010, p-value=0.027*).

%Conversely, 
%In the condition where participants \textbf{reviewed only 10 instances per iteration}, we found a significant increasing trend in MSE as the iterations progressed ($\beta$=0.019, p-value=0.043*). 

%Participants with access to the LLM explanation were labeled as \textbf{LY}, while those without access were labeled as \textbf{LN}. 
%Table~\ref{fig:average-acc-mse-llm-performance} show the number of participants whose labeling accuracy and MSE improved, declined, or remained unchanged before and after using \system. Among 10 participants under LN, 7 participants improved their labeling accuracy, while only 2 participants under LY had improvement. 
%For labeling MSE, there was no difference between LY and LN participants in the number of individuals who showed improvement.

\subsubsection{Showing LLM explanations reduced labeling variation}
We observed that providing LLM explanations to participants led to more consistent labeling, as participants' labels became more similar to each other. 
Those with access to the LLM explanations had a higher Cohen's Kappa (0.333, SD=0.039), as well as higher Spearman (0.556, SD=0.053) and Kendall (0.504, SD=0.049) correlations compared to the group without access, whose Kappa was 0.193 (SD=0.047), Spearman 0.492 (SD=0.094), and Kendall 0.433 (SD=0.084).
%\kenneth{TODO Steven: Update the numbers.}
Additionally, Table~\ref{tab:results-llm-new} shows that the standard deviations (SD) of both ACC and MSE were systematically lower in the group with access to LLM explanations.
%\kenneth{TODO Steven: Update the reference.}\steven{done}

This suggests that \textbf{rather than users tailoring the LLM's behavior to their individual preferences, the LLM---through its explanations---encouraged users to align with its behavior}.

%\subsubsection{LLM explanations guides users, not users guide LLMs. (variation is bigger!)}
%We calculated Kappa under different conditions. 
%We found that participants who explored more data instances had the highest Kappa value of 0.135 (SD=0.073). 
%Participants who had access to the LLM explanation achieved the highest average inter-condition Kappa value of 0.333 (SD=0.039).

\begin{figure*}[t]
    \centering
    \begin{subfigure}[b]{0.49\linewidth}
        \centering
        \includegraphics[width=\linewidth]{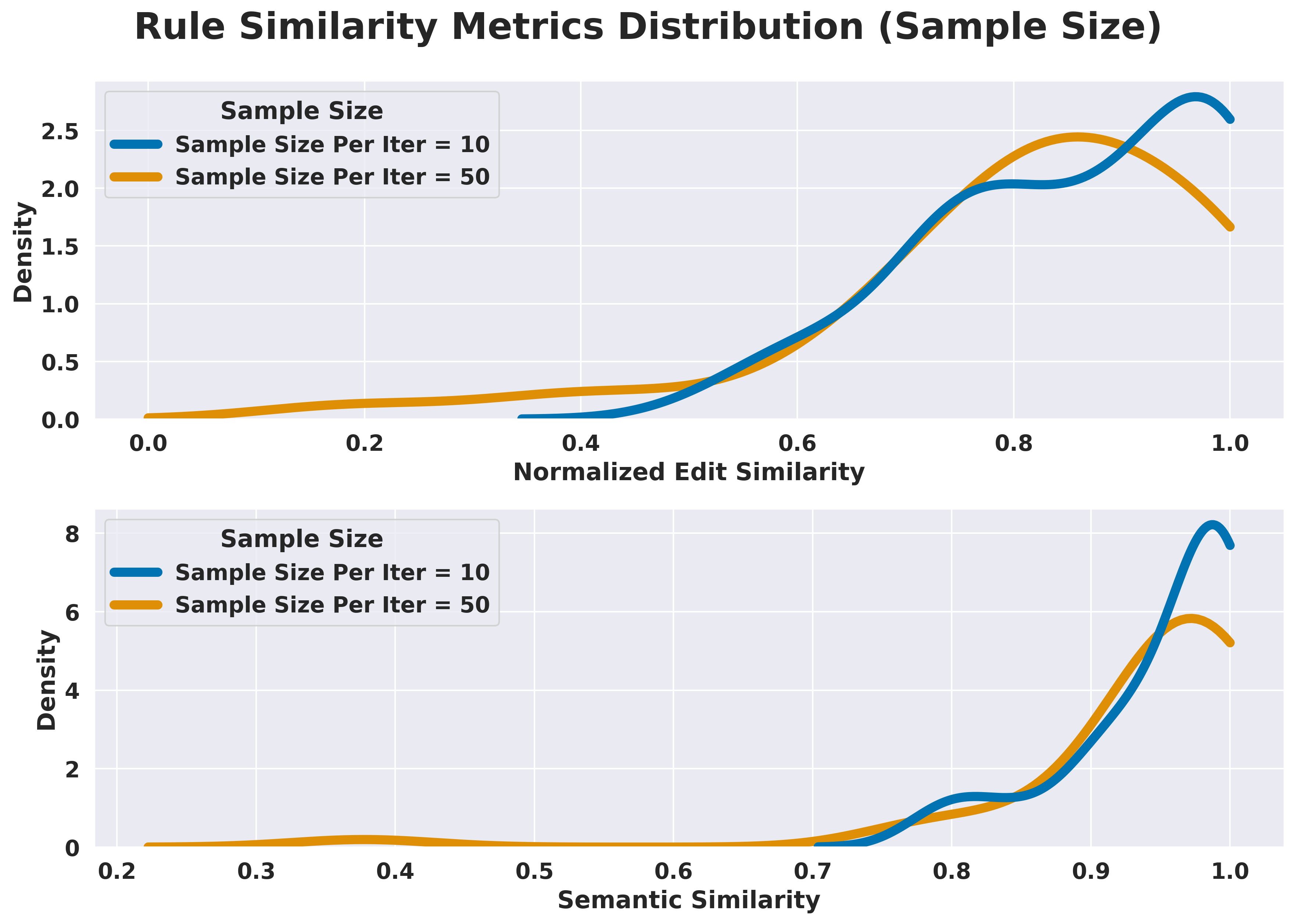}\Description{This is a subplot presenting rule similarity metrics for data sample and explanation group. Each subplot includes two density graphs representing normalized edit similarity and semantic similarity. For the data sample group, participants reviewing 50 instances per iteration tend to exhibit lower similarity within their own rules.}
        \caption{Data Sample Similarity Metrics}
        \label{fig:data-similarity}
    \end{subfigure}
    % \hfill
    \begin{subfigure}[b]{0.49\linewidth}
        \centering
        \includegraphics[width=\linewidth]{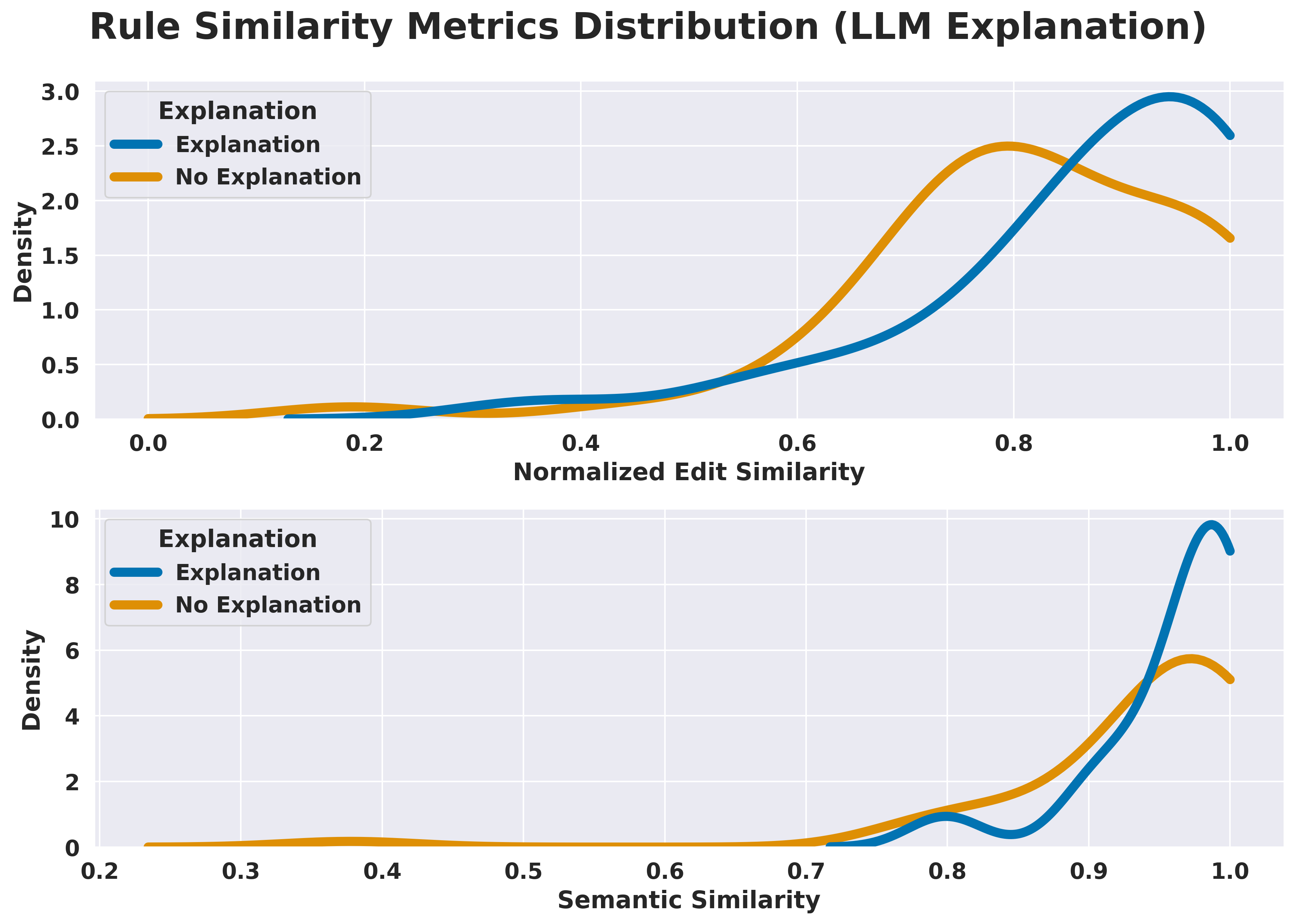}\Description{This is a subplot presenting rule similarity metrics for data sample and explanation group. Each subplot includes two density graphs representing normalized edit similarity and semantic similarity. In the explanation group, participants without access to LLM explanations generally show lower similarity within their own created rules.}
        \caption{Explanation Similarity Metrics}
        \label{fig:explanation-similarity}
    \end{subfigure}
    \caption{Comparison of Rule Similarity Metrics for Data Samples and Explanations. Overall, participants in the 50-instance group or those without access to LLM explanations were more likely to modify their rules.}
    \label{fig:rule-similarity-metrics}
\end{figure*}

\subsection{Additional Analysis: How Does Each Variable Influence Rule Editing?} %\steven{Rule Cos Sim}

Building on the impact of the two variables on prompting outcomes (RQ 1-2 and 1-3), a key follow-up question is \textit{why} these variables affect the outcomes differently. 
To explore this, we examined how the sample size per iteration and the presentation of LLM explanations influence how users edit the labeling rules---one of the main components of the final prompt---in \system.
For each prompt collected, we compiled all rules written by participants in the Rule Book sheet into a single string. 
We then calculated the sentence-level similarity between prompts from consecutive iterations for each session (\eg, between the initial prompt and iteration 1, iteration 1 and iteration 2, and so on).
Using \citet{huang2023ting}'s analysis method, we measured similarity in two ways:
(1) \textbf{Normalized Edit Similarity}: Calculated as $(1-Normalized~Edit~Distance)$,
where higher scores indicate greater similarity~\cite{yujian2007normalized, luozhouyang_python_similarity}.
(2) \textbf{Semantic Similarity}: Measured as the cosine similarity between semantic representations generated with Sentence-BERT~\cite{reimers-2019-sentence-bert}.
Each participant yielded eight similarity scores (2 similarity metrics $\times$ 4 pairs).
We used Kernel Density Estimation (KDE) to visualize the distribution of similarity between rules across consecutive iterations, as shown in Figure~\ref{fig:rule-similarity-metrics}. 
A similarity score of 1.0 indicates no changes, while 0.0 represents substantial modifications. 
Each chart compares participants in the 10-instance group to the 50-instance group or those with versus without access to LLM explanations.

\subsubsection{Larger rule changes were linked to better prompting outcomes.}
Our analysis revealed an interesting pattern: 
conditions that had better prompting outcomes---showing more data items, or not displaying LLM explanations---tended to have \textit{lower} similarity between labeling rules across consecutive iterations. 
In Figure~\ref{fig:rule-similarity-metrics}, the KDE curves for the 50-instance group (Figure~\ref{fig:rule-similarity-metrics}a) and the no-explanation group (Figure~\ref{fig:rule-similarity-metrics}b) skew further left compared to their counterpart conditions, regardless of the similarity metric. 
This indicates that participants in these settings---when seeing more data items, or when not having access to LLM explanations---made \textit{larger} changes to the rule books.
% Using the Kolmogorov-Smirnov (KS) test, only one significant difference was found in the normalized edit similarity between the group with LLM explanation access and the group without (p-value=0.031*). Specifically, participants without access to LLM explanations made significantly more frequent and substantial revisions to their rules than participants who had access to LLM explanations.
Using the Kolmogorov-Smirnov (KS) test, only one significant difference was found in the normalized edit similarity between the group with LLM explanation access and the group without (p-value=0.031*). Specifically, \textbf{participants without access to LLM explanations made significantly more frequent and substantial revisions to their rules} than participants who had access to LLM explanations.
This finding suggests an intriguing implication for human-LLM interaction: 
proactive and frequent revisions during iterative prompt refinement lead to better outcomes compared to making fewer revisions.
Encouraging meaningful revisions in prompting-in-the-dark scenarios---where no gold labels are available to guide or ``reward'' users---presents an interesting challenge for HCI research.

\subsection{RQ 2: Can automatic prompt optimization tools like DSPy improve human performance in ``prompting in the dark'' scenarios?}

\begin{table*}[t]
\centering
\small
\begin{tabular}{lccccccccccc}
 & \multicolumn{3}{c}{\multirow{2}{*}{\textbf{Human}}} & \multicolumn{8}{c}{\textbf{DSPy}} \\ \cline{5-12} 
 & \multicolumn{3}{c}{} & \multicolumn{2}{c}{\textbf{\begin{tabular}[c]{@{}c@{}}Simple\\ Prompt\end{tabular}}} & \multicolumn{2}{c}{\textbf{\begin{tabular}[c]{@{}c@{}}Bootstrap-\\ FewShots\end{tabular}}} & \multicolumn{2}{c}{\textbf{COPRO}} & \multicolumn{2}{c}{\textbf{MIPRO}} \\ \hline
\multicolumn{1}{c}{} & \textbf{\begin{tabular}[c]{@{}c@{}}Avg.\\ \#Shot\end{tabular}} & \textbf{ACC$\uparrow$} & \textbf{MSE$\downarrow$} & \textbf{ACC$\uparrow$} & \textbf{MSE$\downarrow$} & \textbf{ACC$\uparrow$} & \textbf{MSE$\downarrow$} & \textbf{ACC$\uparrow$} & \textbf{MSE$\downarrow$} & \textbf{ACC$\uparrow$} & \textbf{MSE$\downarrow$} \\ \hline
\textbf{Initial} & 0.00 & .542 & .782 & - & - & - & - & - & - & - & - \\ \hline
\textbf{1st Revision} & 2.52 & .536 & .795 & .533 & .864 & .526 & .915 & \underline{\textbf{.565}} & .822 & .527  & .973 \\
\textbf{2nd Revision} & 4.80 & .542 & .833 & .533 & .864 & .534 & .934 & .538 & .873 & .536 & .862 \\
\textbf{3rd Revision} & 7.29 & \underline{\textbf{.549}} & .823 & .533 & .864 & \underline{\textbf{.550}} & .850 & \underline{\textbf{.544}} & .801 & \underline{\textbf{.547}} & .873 \\
\textbf{4th Revision} & 10.04 & \underline{\textbf{.553}} & .810 & .533 & .864 & .526 & .889 & \underline{\textbf{.554}} & .809 & .536 & .874 \\ \hline
\textbf{\begin{tabular}[c]{@{}l@{}}End of Session\\ (Avg \#Iter=4.75)\end{tabular}} & 11.14 & \underline{\textbf{.546}} & .815 & .533 & .864 & .535 & .873 & .528 & .817 & .528 & .868
\end{tabular}
\caption{Comparison of ACC and MSE between users' original prompts and prompts improved by four DSPy approaches. DSPy showed limited effectiveness in enhancing ACC or MSE, potentially due to the small number of gold shots. Improvements over the initial prompt are bolded and underlined.}
\label{tab:dspy-results}
\end{table*}

\begin{figure*}[t]
    \centering
    \begin{subfigure}{0.48\textwidth}
        \includegraphics[width=\linewidth]{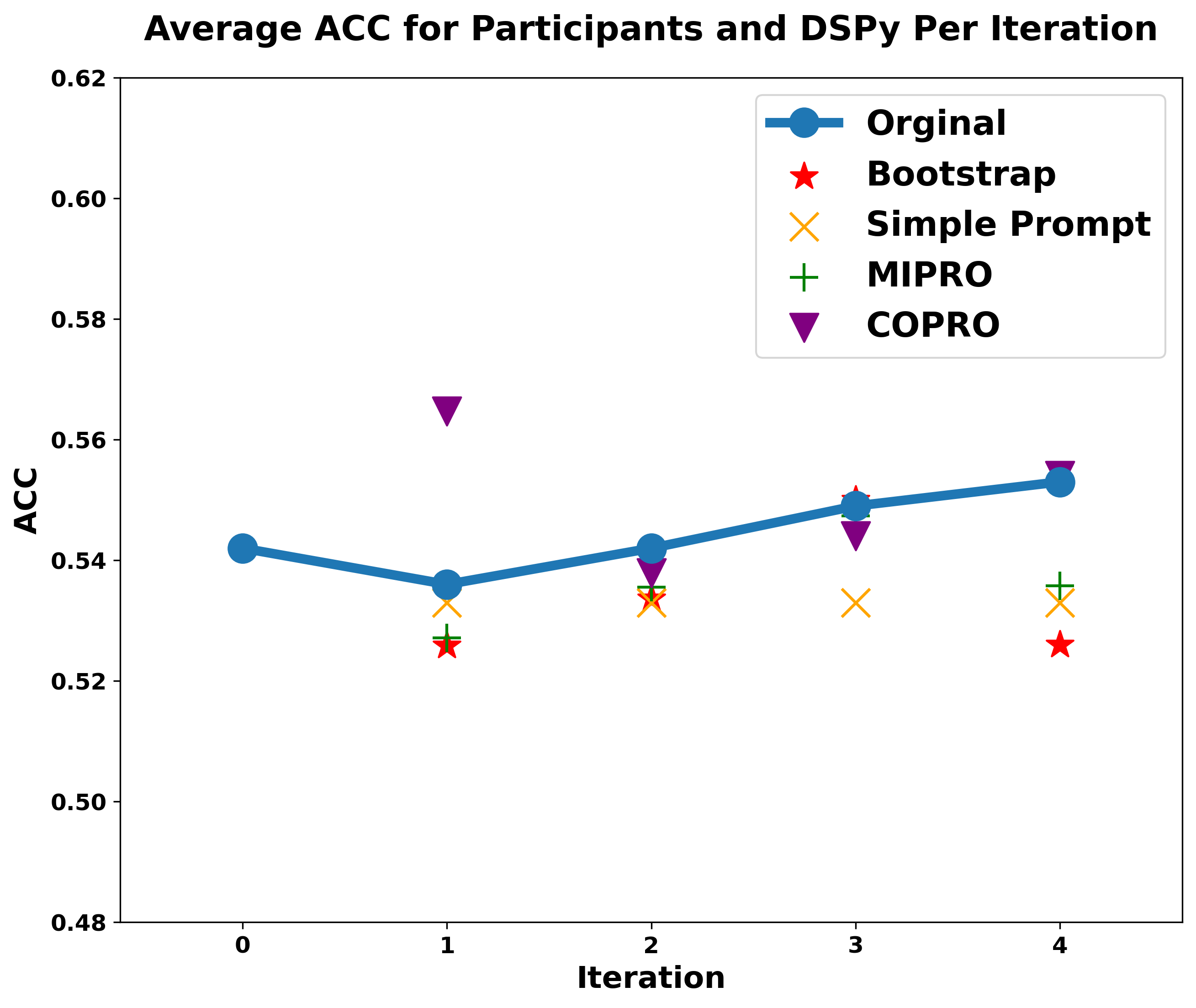}\Description{This is an average accuracy plots subfigure of the participants prompt performance. There are scatter points for each DSPy algorithm from 1 to 4 iterations.}
        \caption{Average ACC of participants' prompts (Original) compared to prompts improved by DSPy's four approaches across each iteration.}
        \label{fig:dspy-plot-acc}
    \end{subfigure}
    \hfill
    \begin{subfigure}{0.48\textwidth}
        \includegraphics[width=\linewidth]{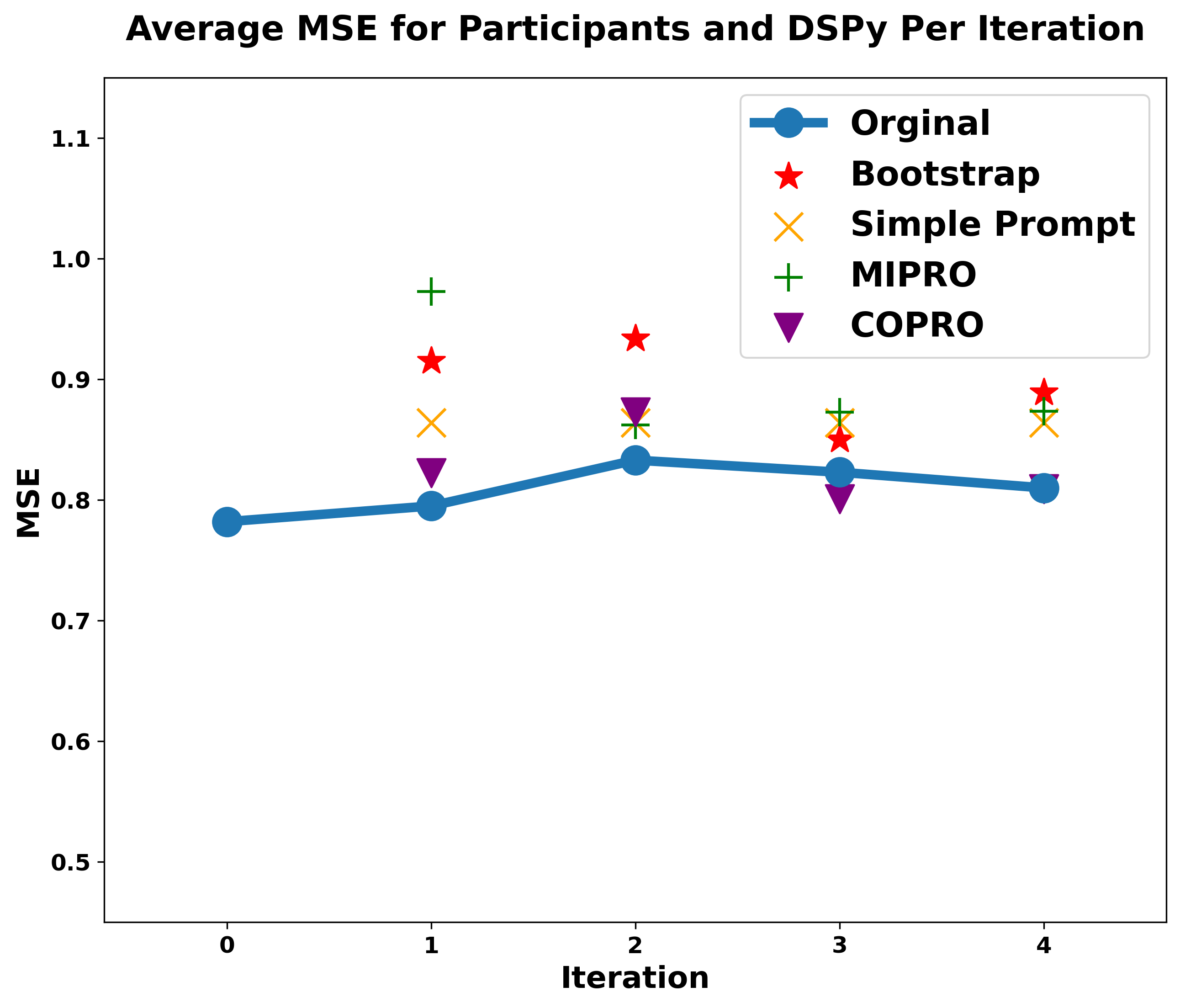}\Description{This is a MSE accuracy plots subfigure of the participants prompt performance. There are scatter points for each DSPy algorithm from 1 to 4 iterations.}
        \caption{Average MSE of participants' prompts (Original) compared to prompts improved by DSPy's four approaches across each iteration.}
        \label{fig:dspy-plot-mse}
    \end{subfigure}
    
    \caption{Average performance comparison between participants' prompts (Original) and those improved by DSPy's four approaches. DSPy was not effective in enhancing ACC or MSE, potentially due to the small number of gold shots.}
    %\kenneth{The space between these two subfigures should be wider. Captions stitch together....}\steven{done}
    \label{fig:dspy-four-settings-charts}
\end{figure*}

\begin{figure*}[t]
    \centering
    \includegraphics[width=0.92\linewidth]{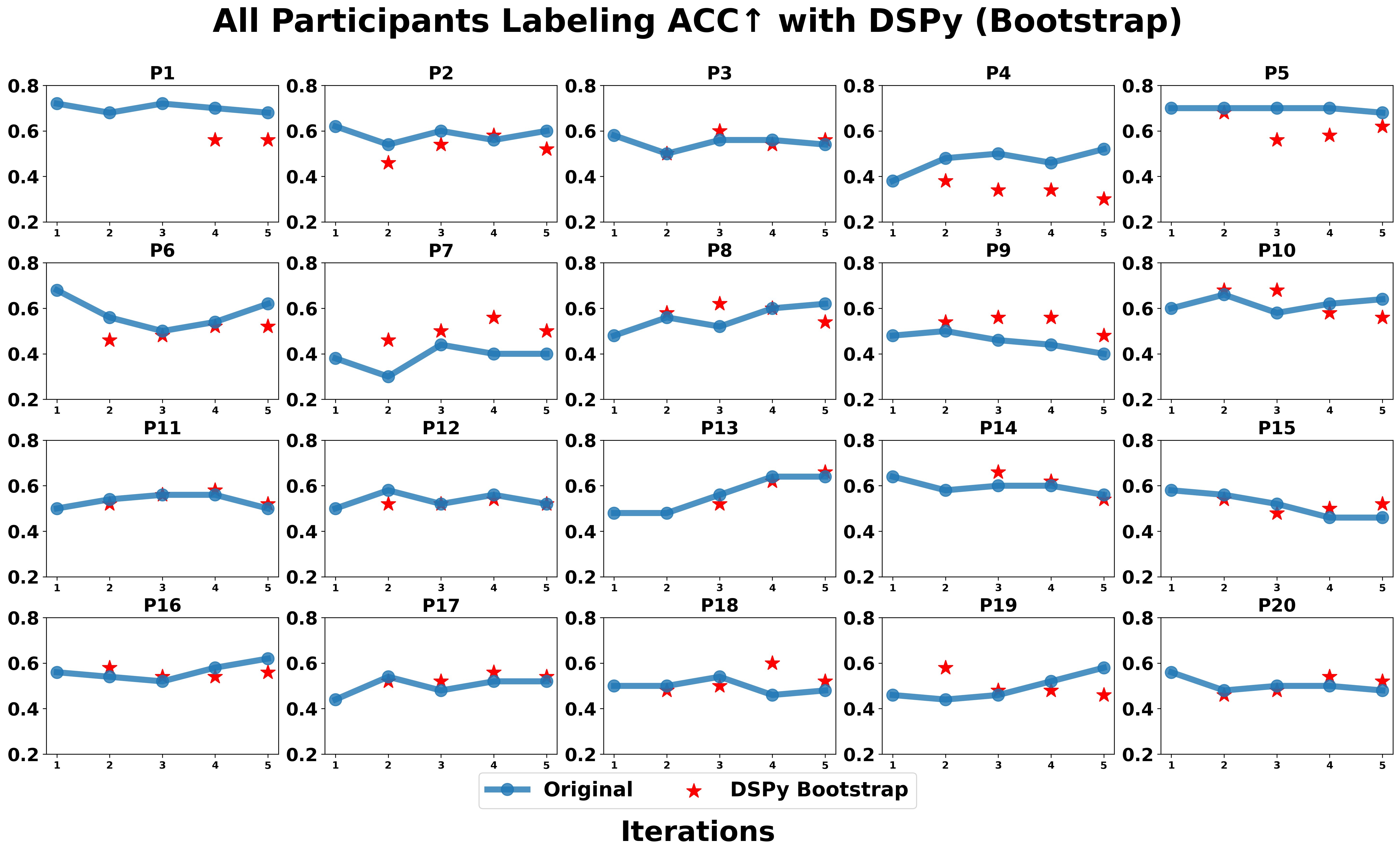}\Description{It contains 20 subfigures of average ACC of each participant plus DSPy Bootstrap scatter.}
    \caption{ACC of all participants compared to DSPy's BootstrapFewShots approach in each iteration. DSPy was not reliable in providing consistent improvements. (Some DSPy dots are missing because participants did not provide examples required for generating augmented samples in those iterations.)}
    \label{fig:acc-participant-plus-dspy}
\end{figure*}

%\steven{done. Make the plot a little transparent to ensure the DSPy dots covered by plot line more visible}

\begin{figure*}
    \centering
    \includegraphics[width=0.92\linewidth]{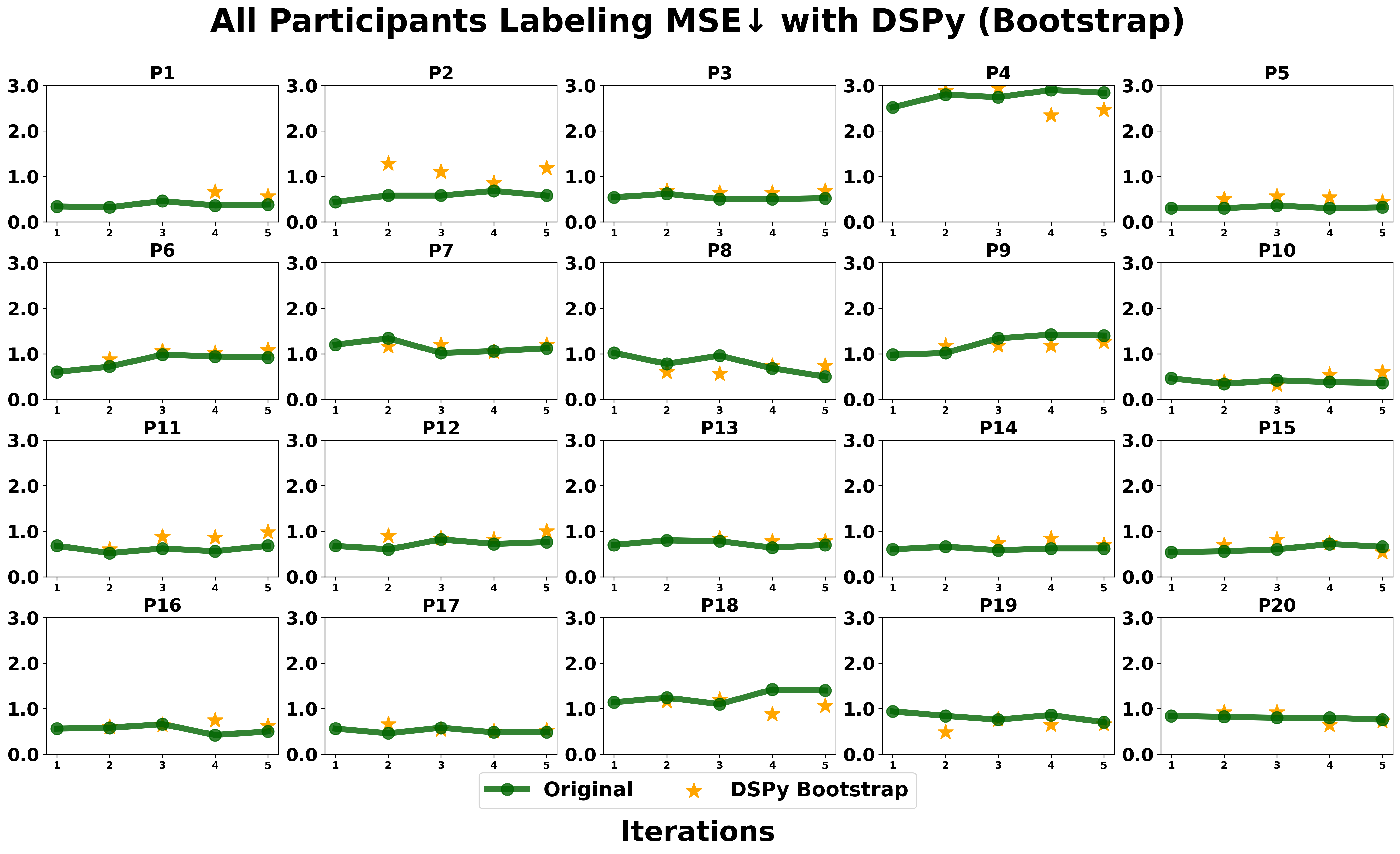}\Description{It contains 20 subfigures of average MSE of each participant plus DSPy Bootstrap scatter.}
    \caption{MSE of all participants compared to DSPy's BootstrapFewShots approach in each iteration. DSPy was not reliable in providing consistent improvements. (Some DSPy dots are missing because participants did not provide examples required for generating augmented samples in those iterations.)}
    \label{fig:mse-participant-plus-dspy}
\end{figure*}

%\steven{done}\steven{todo: missing dots. done}

The previous section demonstrated that only a few settings were effective in helping participants improve prompt performance. 
This raises the question of how automatic prompt optimization tools might assist with refining prompts.
In our study, we explored DSPy~\cite{khattab2023dspy}, a framework designed to algorithmically optimize LLM prompts, to enhance the prompt at each stage of revision by participants. 
DSPy is particularly effective at working with small sets of labeled data and abstract, generic initial prompts, making it well-suited for the ``prompting in the dark'' scenario.

\paragraph{Study Setups.}

%Specifically, 
We experimented with the following four approaches offered by DSPy:

\begin{itemize}
\item 
\textbf{Simple Prompt (Baseline):}
This approach uses the abstract prompts constructed by DSPy and employs DSPy's simplest teleprompter, BootstrapFewShots, to generate optimized examples based on all the few-shot examples labeled by participants throughout the study session. 
It is a simple method that does not account for differences between iterations or the user's initial prompt, making it a baseline approach for using DSPy.
\item 
\textbf{BootstrapFewShots:}
This approach uses the task context (from Context sheet in \system), label definitions (Rule Book sheet), and few-shot examples (Shots sheet) provided by participants in each iteration, and applies DSPy's simplest teleprompter, BootstrapFewShots, for optimization. 
The BootstrapFewShots teleprompter automatically generates optimized examples to be included in the user-defined prompt based on the provided few-shot examples.
BootstrapFewShots is recommended when only a small amount of labeled data is available, such as 10 examples.\footnote{The recommended amount of data is based on DSPy's documentation: https://dspy-docs.vercel.app/docs/building-blocks/optimizers}
\item 
\textbf{COPRO:}
This approach is identical to the BootstrapFewShots setup but uses the COPRO teleprompter instead. 
The COPRO teleprompter focuses on optimizing the prompt instructions while keeping the few-shot examples constant. 
This enables the generation of more refined prompt instructions, even when labeled examples are limited or absent.
\item 
\textbf{MIPRO:}
This approach is identical to the BootstrapFewShots setup but uses the MIPRO teleprompter instead. 
MIPRO combines the features of both COPRO and BootstrapFewShots, refining the prompt instructions while also generating optimized examples using the provided few-shot data. 
MIPRO is recommended when a slightly larger amount of labeled data is available, such as 300 examples or more.
\end{itemize}

All the prompts were optimized with the goal of maximizing accuracy (ACC).

%\kenneth{Not really!}
%\subsubsection{Using DSPy}

\subsubsection{DSPy was not effective in improving ACC or MSE, possibly due to the small number of gold shots}
%Table~\ref{tab:dspy-results}
As shown in Table~\ref{tab:dspy-results} and Figure~\ref{fig:dspy-four-settings-charts}, none of the DSPy algorithms consistently improved ACC or MSE.
This may be attributed to the limited number of gold labels generated in our study, which was insufficient for DSPy to operate effectively, especially given the difficulty of a 5-class classification task. 
Table~\ref{tab:dspy-results} details the average number of gold shots used for DSPy in each revision.

\subsubsection{DSPy was not reliable in delivering consistent improvements}
Figure~\ref{fig:acc-participant-plus-dspy} and Figure~\ref{fig:mse-participant-plus-dspy} display the performance of DSPy's BootstrapFewShots (red dots) alongside human performance (blue line, which is the same as in Figure~\ref{fig:average-acc-mse-performance} for RQ1.)
These results indicate that DSPy's performance was unreliable across participants, as none of the DSPy algorithms consistently improved accuracy or MSE. 
We chose to plot the results for BootstrapFewShots because there was no significant difference across the four DSPy approaches we tested, and BootstrapFewShots was specifically recommended by DSPy's documentation when working with only 10 examples.
Some DSPy dots were missing because participants did not provide examples prior to those iterations, which was necessary for the DSPy algorithm to generate new augmented samples for prompt optimization. 

\section{User Feedback}
In addition to addressing the main research questions, a post-study survey (Appendex~\ref{sec:post-question-survey}) consisted of twenty-two questions, including seven Likert scale ratings and fifteen free-text responses from participants provided valuable insights on both ``prompting in the dark'' practices and our system.
We summarize these insights in this section.

%\kenneth{I re-organized this section. Please take a look.}

\subsection{Two Variables Impacting Participant Ratings\label{sec:two-var-on-rating}}
% \alan{Two variables impacting participant ratings?}
Figure~\ref{fig:user-rating} displayed the seven Likert scale rating responses by participants. The seven survey questions can be categorized into seven different categories. 
Appendix~\ref{app:two-var-on-rating} shows
the survey questions and the accompanying categories were rated on a seven-point Likert scale. 

\begin{comment}

listed below:

\begin{itemize}
    \item 
    \textbf{(Q1) Understandable}: The annotation task was easy to understand.

    \item
    \textbf{(Q2) Ease of Use}: The annotation tool is easy to use.

    \item
    \textbf{(Q4) Intuitive System}: The interface of the annotation system is intuitive.

    \item
    \textbf{(Q5) Performance Satisfaction}: How satisfied are you with the performance of the system?

    \item
    \textbf{(Q6) Prompt Improvement}: This tool was helpful in improving my prompt. 
    
    \item
    \textbf{(Q7) Process Efficiency}: Using this tool made the process of prompt engineering more efficient.

    \item
    \textbf{(Q19) Task Completion}: I completed the annotation tasks efficiently.

\end{itemize}
    
\end{comment}

%Figure~\ref{fig:user-rating} shows the participant's rating across different conditions.
\subsubsection{Participants reviewing 10 instances reported higher satisfaction ratings.}
Figure~\ref{fig:sample-size-rating} compares participants who reviewed 10 instances per iteration with those who reviewed 50.
Both groups provided similar ratings for system ease of use, system intuitiveness, and efficiency in processing prompt engineering, with comparable variation.
However, participants who reviewed 10 instances found the annotation tasks more difficult to understand compared to those who reviewed 50. Comparatively, participants who reviewed 10 instances reported higher levels of satisfaction with their performance, a stronger sense of prompt improvement, and better task completion rating. This could be attributed to their minimal modifications to the rule.
% We performed a KS test on all rating categories and no significant difference was found between two groups, indicating that the observed difference did not reach statistical significance.
It is noteworthy that we performed a Kolmogorov-Smirnov (KS) test on all rating categories, and no significant difference was found between the two groups, indicating that the observed difference did not reach statistical significance.

% \steven{Only the performance ratings from participants showed a significant difference between two instances groups based on the t-test (p-value=0.031)}

\subsubsection{Participants without LLM explanations rated the system as more intuitive, effective, and satisfying.}
\begin{figure*}
    \centering
    \begin{subfigure}[t]{0.48\textwidth}
        \includegraphics[width=\linewidth]{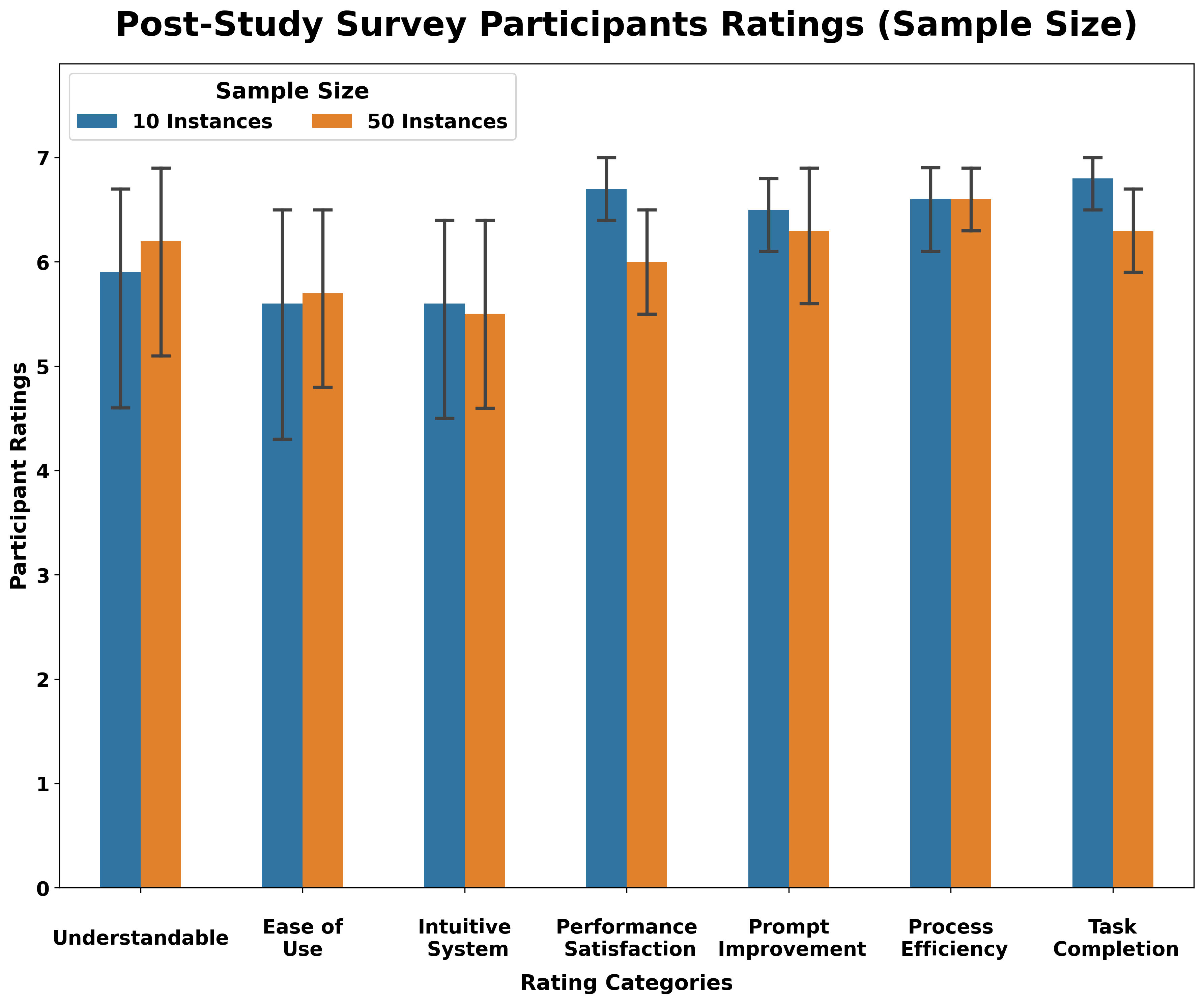}\Description{This subplot is for the data sample group, based on participants’ post-survey responses. Each subplot features bar charts comparing two settings: 50 instances vs. 10 instances. Each bar is accompanied by a confidence interval displayed at the top.}
        \caption{Participants' ratings of the system and the annotation task, comparing those who accessed 10 instances per iteration to those who accessed 50 instances.}
        \label{fig:sample-size-rating}
    \end{subfigure}
    \hfill
    \begin{subfigure}[t]{0.48\textwidth}
        \includegraphics[width=\linewidth]{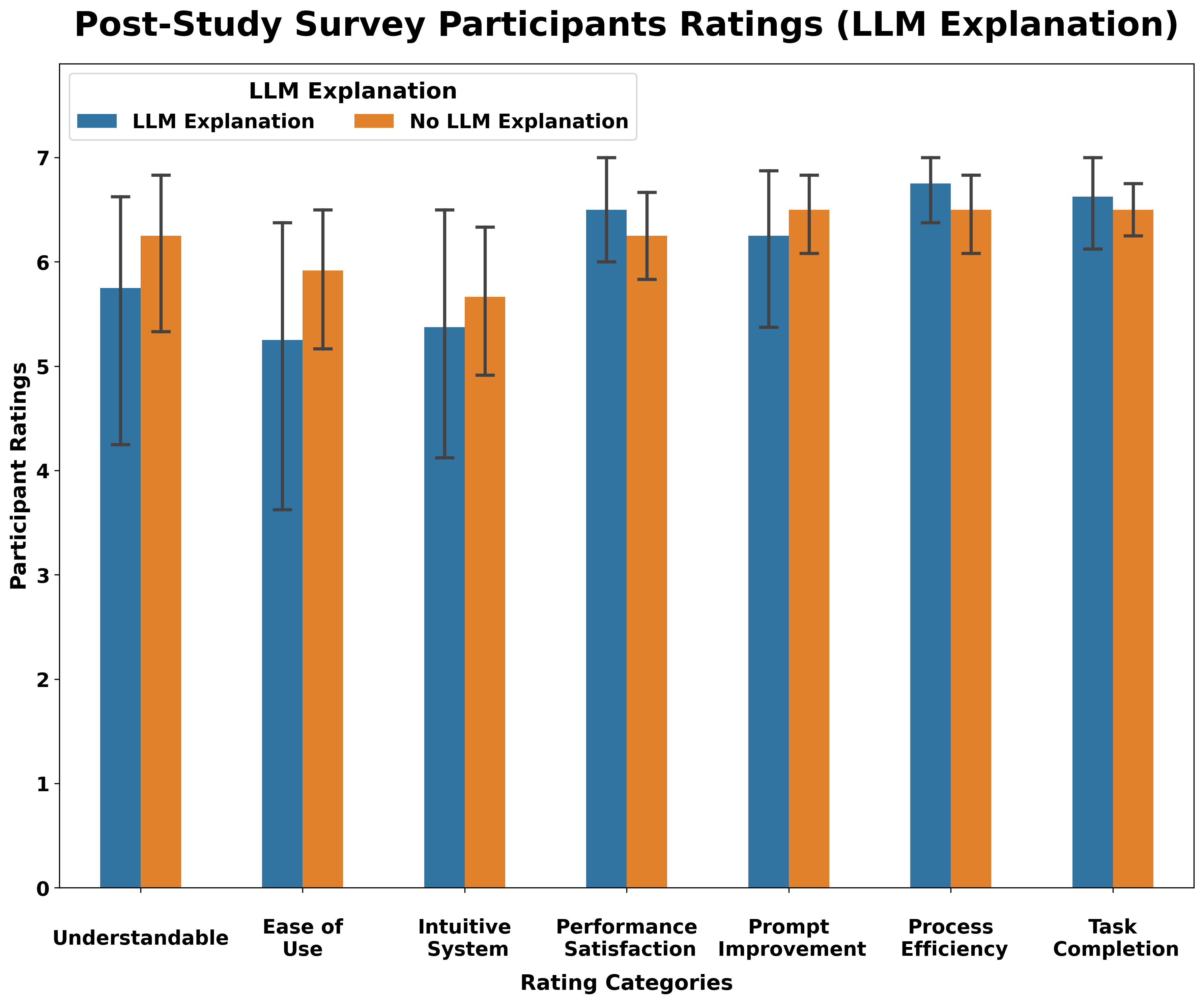}\Description{This subplot is for the explanation group, based on participants’ post-survey responses. Each subplot features bar charts comparing two settings: no explanation vs. explanation. Each bar is accompanied by a confidence interval displayed at the top.}
        \caption{Participants' ratings of the system and the annotation task, comparing those who utilized LLM explanations to those who did not.}
        \label{fig:explanation-rating}
    \end{subfigure}
    \caption{Participants' rating of the system and the annotation task. Each rating category refers to one question in the post-study survey. 
    \textbf{(Q1) Understandable}: The annotation task was easy to understand; 
    \textbf{(Q2) Ease of Use}: The annotation tool is easy to use;
    \textbf{(Q4) Intuitive System}: The interface of the annotation system is intuitive;
    \textbf{(Q5) Performance Satisfaction}: How satisfied are you with the performance of the system? 
    \textbf{(Q6) Prompt Improvement}: This tool was helpful in improving my prompt; 
    \textbf{(Q7) Process Efficiency}: Using this tool made the process of prompt engineering more efficient; 
    \textbf{(Q19) Task Completion}: I completed the annotation tasks efficiently.}
    \label{fig:user-rating}
\end{figure*}

Figure~\ref{fig:explanation-rating} shows the comparison of ratings between participants with and without LLM explanations.
Participants with LLM explanations found the annotation tasks more challenging, rated the system as less intuitive and harder to use, and viewed it as less effective in improving prompts, also with greater variation in their ratings. In contrast, participants without LLM explanations expressed higher level of satisfaction with the system performance, believing the tool improved prompt engineering efficiency and task completion effectiveness.
% We conducted a KS test on all ratings from participants and no significant difference was found between the two groups, suggesting that the observed difference were not statistically significant.
Notably, we conducted a Kolmogorov-Smirnov (KS) test on all participants' ratings, and no significant difference was found between the two groups, suggesting that the observed differences were not statistically significant.

%\steven{added user rating.}\kenneth{(1) Font for axis titles and the title of the figure are too big, (2) Make the figure wider so that the x-axis labels do not need to rotate (use newline for x-labels if possible), (3) this is kinda extra: can we break it down in two ways (a) with/without explanations and (b) smaller/bigger sample size.}\steven{done}

\subsection{Is \system Useful?}
\subsubsection{Participants considered \system helpful and efficient.}
In the post-study survey, we asked participants to rate the (Q5) performance satisfaction, (Q6) helpfulness of the tool, and (Q7) its efficiency on a seven-point Likert scale.
% The detailed questions asked are shown in Appendex~\ref{sec:post-question-survey}.
%\kenneth{TODO Steven: Update refernece}
Participants expressed high satisfaction, with an average rating of 6.350 (SD=0.745), and found the system helpful for improving prompts (6.400, SD=0.883) and making prompt engineering more efficient (6.600, SD=0.598).
%\kenneth{TODO Steven: Add numbers---- Are there really high???}
P4 noted, ``\textit{I really like this tool instead of traditional prompt engineering on ChatGPT and Copilot.}''
%\kenneth{TODO Steven: Add a few more examples for other participants.}
P15 mentioned, ``\textit{I would be interested in using this annotation system in my regular work or study, because I really like the idea [of] improving annotation performance by considering iteration annotation process between human and the GPT.}''
% A participant (P18) wrote in the questionnaire, ``I like how the system's UI has been designed and programmed.''

%\kenneth{Not sure how to phrase this...}
\subsubsection{\system is easy to use but less intuitive and with a steep learning curve.}
We asked participants to rate whether ``\textit{(Q1) The annotation task was easy to understand},'' ``\textit{(Q2) The annotation tool is easy to use},'' and ``\textit{(Q4) The interface of the annotation system is intuitive}'' on a seven-point Likert scale from ``Strongly Disagree'' (1) to ``Strongly Agree''(7). 
The average score for the ``easy to understand annotation task'' was 5.812, for the ``easy to use annotation tool'' was 5.375, and for the ``system is intuitive'' was 5.250. Suggesting that while the task and tool itself are not hard to understand and use, learning to properly use the tool can be harder for participants and required some learning.
For example, it was noted that the need to switch between tabs during the task can cause confusion.
% One participant mentioned that the system workflow was unclear, as he had to switch between different tabs, causing confusion and disruption.

\subsubsection{Participants found the Shots and Rule Book useful.}
%\paragraph{Having the flexibility to structure tasks freely and make adjustments on the go—whether modifying the Gold Shots or the Rule Book—reduces the burden of the traditional, iterative labeling process.}
We asked participants, in a free-text format, ``\textit{(Q10) What features did you find most useful?}'' Fourteen participants specifically mentioned that `Gold Shots' were particularly valuable, as they provided explicit examples to guide LLMs. 
Additionally, six participants highlighted the usefulness of the Rule Book. 
These two features stood out among the responses, demonstrating their importance in enhancing the user experience.
Participants noted that the flexibility to structure tasks freely and make on-the-fly adjustments---such as modifying the Gold Shots or Rule Book---eases the burden of the traditional iterative labeling process.

% We asked participants, ``What features did you find most useful?'' in free text form and 14 participants mentioned that ``Gold Shots'' were useful and could explicitly provide examples to guide LLMs.
% % 14 participants (P1-P6, P9, P11, P13, P14, P15, P16, P18, P20) thought the ``Gold Shot'' feature was useful as it could explicitly provide examples to guide LLMs. 
% On the other hand, availability of the Rule book was considered useful for 6 participants. 
%\steven{there are some participant like both.}

%\kenneth{Not sure about this...}

\subsubsection{Dilemma of showing LLM explanations.}
%\paragraph{Participants' desire to include LLM explanations.} 
Although our study found that providing LLM explanations can sometimes lead participants to generate labels more aligned with those produced by the LLM, participants still expressed a strong desire to have them included. P8 explicitly recommended incorporating LLM explanations, noting that participants were interested in understanding the reasoning behind potential discrepancies between their own labels and those generated by the LLM. P15 also emphasized the value of these explanations, stating, ``\textit{The explanation from GPT gave me some insights to modify my rules,''} and, ``\textit{I think GPT's explanation of the tweets is very helpful and it may help me to improve the accuracy of human annotation.''}

\begin{comment}

\subsubsection{What can be improved in \system?}
%\paragraph{Workflow can be confusing and interface not user friendly}
Some participants faced challenges when learning the system, and it can take them a long time to get comfortable with it. P1 said, ``\textit{The system is not logically clear for me because the system needs to jump in between different tabs, which is different than a normal workflow.}'' 
Some participants became confused with the system, even after several iterations. For instance, a few participants forgot that to proceed to the next iteration, they needed to sample the data and click ‘Start Annotation.’ Research team members need to remind participants of the workflow continuously. Additionally, the nature of the task requires moving from one tab to another, which adds difficulty for participants to navigate the interface and causes confusion.

\end{comment}

\subsection{About ``Prompting in the Dark''}

%\kenneth{Not sure about this too}
\subsubsection{Prompting in the dark without any tool is common.}
%\alan{Iterative, trial and error prompting is common without the help of tools.}}
We also asked participants, ``\textit{Without this tool, how would you typically approach prompt engineering?}''
We found that many participants commonly rely on iterative, trial and error strategies. Specifically, they start with prompts from scratch, test them on data points, adjust based on incorrect labels, and re-test until they are satisfied with the results.
% Responses varied: P0 emphasized giving LLMs as much context as possible\steven{this is P0, which is not included in our actual user study}, while others (P1, P3, P4) described a general refining process---starting prompts from scratch, testing on data points, adjusting based on incorrect labels, and re-testing until satisfied.
%\kenneth{TODO Steven: Add a few more examples for other participants.}
%\alan{trial and error, iterative process, personify}

P1 said, `\textit{I need to start with a prompt from scratch; then I will test it on real data points; I will observe those wrongly labeled data points and adjust my prompt accordingly. After the adjustment, I will rerun the testing on the real data points. The whole process is trail-and-error, which is really time-consuming and labor-consuming.}''
P3 stated, ``\textit{Give an initial prompt, if the answer is not meeting expectation, then change the prompt.}''
% P5[Keep trying different prompts to see if the responses I get satisfies my expectations.]
P6 reflected, ``\textit{Normally, if I do not get the desired output from the LLM, I will try to give more  specific instruction maybe some examples.}''
% P14[Try prompting with ChatGPT, if ChatGPT cannot provide a good answer, just rephrase the prompt and ask the question again.]
P17 said, ``\textit{I re-write my prompts several times (3-5 times) until I got an output that I like.}''

\begin{comment}

We also want to emphasize that, although not common, some participants employ a personification method by asking LLMs to take on a specific role or personality and make decisions based on that role. For example, P12 stated, ``\textit{I will first assign a role to GPT like ``Supposing you are an expert in coding, ...''. Then I will ask it to follow my instructions.}''
% Personify, P103 [Make llm assume that it is not an AI and act as a specific person who is involved in that specific activity. By Giving as much context as possible.], P12 [I will first assign a role to GPT like ``Supposing you are an expert in coding, ...''. Then I will ask it to follow my instructions.]
P13 said, ``\textit{I would narrate the incident or situation environment and then give prompt asking specific and questions clearly.}''

\end{comment}

% \paragraph{\system gave participants new insights into prompt engineering.}
% Interestingly,
% our survey also showed that using \system gave participants new insights into prompt engineering. 
% For example, P1 observed that ``adding or deleting a few words can change the overall output,'' and P2 noted, ``I can compare how many labels are correctly labeled before and after modifying prompts.''
%\kenneth{TODO Steven: Add a few more examples for other participants.--- This one is interesting. Say a few more if possible?}

\subsubsection{Prompting in the dark is hard, as participants lacked confidence in their labels.}
%\paragraph{Participants lacked confidence in labeling}
Without a comprehensive understanding of the entire dataset, participants found it challenging to generate suitable labels.
P19 mentioned, ``\textit{I am not confident about the label}''.
P12 pointed out that, ``\textit{When I need to express sentiment, I tend to be more reserved and avoid extremes. So, when labeling data, I usually prefer to choose negative/[positive] rather than extremely negative/[extremely positive]}''

\subsection{Users' Suggested Features}

\subsubsection{More automated supports for rule creation.}
Participants expressed concerns about creating rules that effectively suit the labeling task at hand. As a result, support for Rule Book creation is a welcome addition.
P2 remarked, ``\textit{It was hard to set the right rules,}'' while P3 suggested providing initialized instructions for labels and rules to ease the process. Additionally, participants (P1, P3) proposed that new rules could be automatically generated based on Gold Shots, existing rule books, and human explanations, streamlining the rule creation process.

%\kenneth{Can we make subsubsection title a complete sentence?}
%\alan{something like this?}
\subsubsection{Shorter LLM explanations for easier consumption.}

Although LLM-generated explanations received positive feedback from participants, there was concern about the length of these explanations. Many felt that the explanations were too long and could be difficult to consume. P1 recommended, ``\textit{It would be better if the LLM explanation could be shorter.}'' emphasizing the need for more concise outputs to improve user experience.

% P9 pointed out, ``the `Start Annotation' button is hard to find.''.

% \paragraph{Simplified Interfaces and Workflows.}
% Some participants suggested hiding uncommon features (P17) or displaying the function descriptions only when hovering over them (P20).

%\paragraph{Other suggestions}\steven{Not related to the annotation system}
% (P1)If there's a graphical UI, the system will be more accessible to laypersons without the HCI background. Currently, the system is too flexible to get lost in the interactions.
%(P3) More tutorials on the data annotation task.

%\subsection{Participant In-Lab Annotation Process Analysis}

%\subsection{Participant Self-Reported Response Analysis}

\section{Discussion}
\subsection{Design Suggestions}

Our study shows that ``prompting in the dark''---where users gradually refine their expectations and understanding of data characteristics while iteratively prompting LLMs---is indeed a challenging task. 
Using \system, we quantitatively assessed the difficulty of this process and the unreliability of the progress. 
Many strategies we tested proved ineffective, and the few that showed promise offered only weak signals. 
So, what should be designed in response? 
From the study, we derived insights that we believe are valuable for shaping future systems intended to support iterative LLM-powered labeling, which we summarize into three design suggestions.

\begin{itemize}

\item 
\textbf{Gold Labels as a Necessary Evil.}
%-Need some good gold labels even the users change their mind, it's still very very useful to give guide
The key premise of ``prompting in the dark'' is to avoid assuming a stabilized gold standard from the outset.
This practice embraces the potential for the so-called ``gold standard'' in data labeling to evolve and remain dynamic, realizing this through the interactive nature of LLMs. 
Our system, \system, embodies this practice, using a spreadsheet format.
That being said, throughout our user studies, many participants faced struggles, frustrations, and errors due to the lack of direction or guidance during the exploratory labeling process. 
%The elephant in the room is that much of this lack of guidance was caused by the absence of sufficient gold labels.
The elephant in the room---as highlighted by decades of research on the importance of gold labels in crowdsourced data annotation pipelines (Section~\ref{sec:related-work-gold-label})---is that much of this lack of guidance stemmed from the absence of sufficient gold labels. %\kenneth{TODO: Refer to Related Work}\steven{done}
To clarify, we mean ``sufficient'' as in 100 to 200 labeled text items before starting to use the system, not just the 10 examples users provided via the ``add gold shot'' function.
%we could not use them to calculate numbers like similarity, accuracy, or error rates to provide users with meaningful signals. 
%We also could not use gold labels to track the direction of LLMs' behavioral shifts. 
With no gold labels or fewer than 10, we could not calculate metrics like similarity, accuracy, or error rates to give users meaningful feedback, nor could we track LLMs' behavioral shifts.
These challenges arose as part of our effort to ensure that no substantial labeled items would anchor users' exploration, allowing them to freely evolve their goals.
Our first design suggestion is that this cost might be \textit{too high}.
Based on our observations, having some form of guidance in the ``prompting in the dark'' scenario is likely worth it. 
We propose that users manually label at least a small set of data (\eg, 50-100 items), and that the system be intentionally designed to minimize the impact of these labels on constraining exploration. 
For example, the system could hide the labeled items from the user (even the ones they just labeled), show only signals such as accuracy, or periodically invite users to relabel data.
One example worth highlighting is PromptAid~\cite{mishra2023promptaid}, which dedicates most of its interface to higher-level signals, abstracting away from low-level data details.
Manually labeling some ``gold'' data to start with, when carefully managed, should be considered a reasonable trade-off to balance reliability in the process and the freedom to explore.

\item
\textbf{Automated Support to Reduce Distractions.}
%-Need some support----typo check, rule induction, prompt optimization, and maybe some color/viz aid
Our second design suggestion is to incorporate automation to help reduce users' distractions during the ``prompting in the dark'' process. 
Even with some gold labels---as we suggested---iterating on prompts with an LLM and reviewing tens or hundreds of text items and labels remains exhausting. 
In our study, only a few participants were able to steadily improve labeling accuracy over time.
We believe the root of this mental strain stems from at least two causes.
First, the behavior of LLMs is a mystery. 
Even with carefully crafted prompts, users are still navigating an intangible and unpredictable fog, trying to guess how the LLM will respond.
Second, real-world data is inherently noisy. 
Even when using a strong classifier, users must still contend with numerous edge cases and ambiguous linguistic nuances. 
These challenges are inherently difficult and clash with established human-AI design guidelines~\cite{amershi2019guidelines}. 
%For instance, 
In the ``prompting in the dark'' scenario, there is no efficient way to recover from incorrect labels (Guideline G9 in~\cite{amershi2019guidelines}), LLMs can not reliably explain their labeling decisions (Guideline G11), users lack mechanisms to provide granular feedback to LLMs (Guideline G15), and the consequences of each revision are unclear (Guideline G16).
Given the complexity of these challenges, systems should aim to adhere to all other feasible design guidelines to mitigate user strain.
%These challenges are difficult by nature.
Throughout our study, %while many of the struggles were due to the inherent complexity of LLMs and data, 
we observed that some confusion arose from the workflow or tools. 
Issues such as how to sample a subset from the full dataset, or how to add gold short from validated data, added unnecessary complexity.
Our recommendation is to design systems that reduce these unnecessary distractions. 
For example, the system could default to randomly sampling a subset for the user, or automatically add validated gold labels into the prompt and manage the number of shots in the background.

\item
\textbf{Design to Prevent Overreliance on LLMs.}
Our final and most critical design suggestion is to create systems that reduce the human tendency to overly trust and accept outputs from AI, particularly LLMs. 
Overreliance on AI is a well-documented phenomenon where people are influenced by AI decisions and often accept them without verification, even when they are incorrect~\cite{vasconcelos2023explanations}.
Overreliance is particularly problematic in exploratory labeling because the core idea---central to ``prompting in the dark''---is that users should own the exploration process and have strong control over which direction to pursue next. 
Users choose to prompt in the dark because they do not want to be guided by someone else's light.
However, introducing AI, such as LLMs, can lead users to fixate on or anchor to the LLMs' suggestions.
If this results in everyone's labels becoming more similar, it undermines the very spirit of exploratory labeling.
In our study, participants who saw LLMs' explanations tended to agree more with the LLMs' labels, leading to higher inner-annotator agreement among those exposed to LLM explanations compared to participants who were not (Section~\ref{sec:llm-explanation-result}). 
%This is yet another instance of automation bias, which is particularly harmful because it goes against the principles of exploratory labeling.
To address this, we advocate designing systems that actively mitigate overreliance on AI.
While explainable AI has been proposed as a solution to overreliance by helping users better judge when to trust AI predictions, empirical evidence shows mixed results. 
%As noted in Related Work , 
Explanations do not always reduce cognitive effort for verifying AI predictions, leading users to ignore them (Section~\ref{sec:rq}). %\kenneth{UPDATE Section REF}
A cost-benefit framework interprets that explanations are less effective when the cognitive cost of verification outweighs the perceived benefits~\cite{vasconcelos2023explanations}.
Text data labeling often falls into this category, as reading and verifying text-based explanations is cognitively expensive. 
This leaves scaffolds, such as cognitive forcing functions, as a viable solution. 
These functions require users to engage more thoughtfully with AI explanations and have been shown to effectively reduce overreliance~\cite{buccinca2021trust}.
%\kenneth{TODO: Cite ``To trust or to think: cognitive forcing functions can reduce overreliance on AI in AI-assisted decision-making''}

\end{itemize}

%\kenneth{Interact with literature in XAI:}

%\input{Sections/2-6-explaination-data-annotation}

% 2- how sensitive was the performance to rule, example changes - R2 (NOT DONE)
% There are related literature on sensitivity on prompting. 1. Few-shot is important (NLP). 2. DSPy study, we know prompt engineering in some cases can improve performance. 3. More recent study, LLM is not sensitive. Prompt is the only way to manage the LLM. We cannot measure it on this scale. 

\subsection{Limitations}
%-We only try annotation task and one annotation task
%-We only try DSPy
%-Admittedly, the system's workflow is a bit complicated and some reliability issues might come from that--- BUT really there's no other way to make it more doable I guess. 
%-Users only try 1-2 hours
%-Some limitation or design constraints come from spreadsheet, e.g., resample data is hard, can not label image, etc
We acknowledge several limitations in our work. 
First, we focused solely on the ``prompting in the dark'' practice for data labeling tasks, while many other scenarios, such as text generation or conversational agents, are also widely used when working with LLMs.
Second, for RQ2, we only studied one automatic prompt optimization technique, DSPy. 
%While DSPy is one of the most effective and widely used approaches, 
It is possible that other methods might achieve better performance. 
Third, we recognize that, aside from the inherent challenges of the open-ended and exploratory nature of ``prompting in the dark,'' the workflow of our system is somewhat complex. 
This complexity may have contributed to some of the reliability issues observed in our study, though we believe similar challenges would apply to any system supporting newly emerged user practices. 
Fourth, since \system is a Google Sheets Add-on, it inherits limitations and design constraints typical of spreadsheets, 
%such as the difficulty of frequently moving rows between sheets and 
such as limited support for images or videos.
Fifth, it was difficult to track how thoroughly participants reviewed samples. 
Some may have directly modified rules or added gold shots without careful consideration.
%\steven{Fifth, tracking the number of samples reviewed by participants is challenging. It is possible that some participants did not thoroughly review the samples and instead directly modified rules and added gold shots without careful consideration. }
%\steven{
Sixth, we lack a clear understanding %or benchmark 
for how sensitive LLMs are to prompt revisions. 
Tools like DSPy~\cite{khattab2023dspy} and related experiments show that prompt revisions can systematically influence LLM prediction performance, but the degree of sensitivity remains unclear.
%, with recent evidence suggesting that larger models may be less affected by prompt variations~\cite{zhuo2024prosa}. 
This limited understanding of LLM behavior impacts the generalizability of our findings.
%even though DSPy has demonstrated that optimizing prompts can enhance performance in certain cases, it remains unclear which specific modifications or aspects of these optimizations are most sensitive to performance improvements.
%This ambiguity stems from inherent limitation in the underlying LLMs. A recent study found that larger models are less sensitive to prompt variation~\cite{zhuo2024prosa}. Additionally, the limited number of participants in our study makes it more challenging to accurately access sensitivity in our experimental settings. 
%}
Finally, our user study, though already 1-2 hours long, only captures a single session's worth of behavior. 
A longitudinal study that observes how users interact with \system over months or even years might offer different insights.

\section{Conclusion and Future Work}
This paper studies a scenario in LLM-powered data labeling called ``prompting in the dark,'' where users iteratively prompt LLMs to label data without relying on a stabilized, pre-labeled gold-standard dataset for benchmarking. 
%In this scenario, users' understanding of the data and their desired labeling scheme evolves through their interactions with the LLM and its outputs. 
We developed \system, a Google Sheets add-on that allows users to compose, revise, and iteratively label data within spreadsheets. 
Our user study revealed that prompting in the dark was highly ineffective and unreliable, and automated prompt optimization tools like DSPy struggled when few gold labels were available. 
We concluded the paper with a set of design recommendations. 
%starting with a small set of gold labels is a reasonable compromise, and systems should be designed to minimize unnecessary distractions as well as to mitigate overreliance on LLMs' predictions. 
Based on these insights, our next step is to explore automation technologies, such as automatic rule creation, to further support users. %this exploratory, iterative data labeling process. 
Simultaneously, we will investigate methods to track system performance under evolving standards with minimal labeling effort. 
Additionally, we aim to better capture users' understanding and use this information to monitor and improve LLMs, fostering more effective human-AI collaboration. 
%At the meantime, will explore methods to track system performance under evolving standards with less labeling efforts. Furthermore, we will investigate how to better capture human's understanding and monitor LLMs with the captured information to better assist human-AI collaboration.
Finally, we plan to enhance and deploy \system for more users
%the system's implementation and workflow in the next version, using the improved \system for a longer-term deployment study 
to observe how people prompt LLMs to label data they truly care about in real-world scenarios.

\begin{acks}
We thank the anonymous reviewers for their insightful feedback. 
We also extend our gratitude to all the participants in our study. 
This work was partially supported by a seed grant from the Institute for Computational and Data Sciences (ICDS) at Pennsylvania State University and by the Alfred P. Sloan Foundation (Grant Number: 2024-22721).

%We would like to express our sincere gratitude to the Institute for Computational and Data Sciences (ICDS) at the Pennsylvania State University for providing the seed grant that supported this work. We also thank the reviewers for their valuable insights and constructive feedback. Finally, we would like to extend our appreciation to all the participants who contributed to our experiment.
\end{acks}

\bibliographystyle{Style/ACM-Reference-Format}
\bibliography{Bibtex/sample-base,Bibtex/software,Bibtex/main}

%TC:ignore
\appendix
% \section{\system User Interface}
% \begin{figure}[tbh]
%     \centering
%     \includegraphics[width=0.85\linewidth]{Figures/Interfaces/System-Interface-5.jpeg}
%     \caption{The Task Dashboard tab systematically tracks all iteration records. This includes Task ID, a hyperlink to the specific task tab, timestamp, the prompt used in that iteration, and the total labeling cost.}
%     \label{fig:system-interface-5}
% \end{figure}

% \begin{figure}
%     \centering
%     \includegraphics[width=0.35\linewidth]{Figures/Interfaces/Remove-Clear-figure.jpg}
%     \caption{Remove Unselected Data Instance and Clear Data Instance Features}
%     \label{fig:remove-clear}
% \end{figure}

% \section{Participant Overall ACC and MSE}
% \input{Tables/single-condition-t-test}
% \input{Tables/mixed-condition-t-test}

% \input{Tables/dspy-t-test/dspy-t-test-detail-conditions-under-each-optimizer}

\section{\system Task Context Questions}\label{sec:context-question-appendix}
\begin{table*}[h]
\footnotesize
\centering
\begin{tabular}{@{}lll@{}}
\toprule
\textbf{\#} & \textbf{Questionnaire Question} \\ \midrule
Q1 &  \begin{tabular}[c]{@{}l@{}}What is the purpose of annotating this data? Common answers include gaining insight about something, wanting to compare something, \\wanting to create prompts, etc. Try to give us more details about your higher-level goal.\end{tabular} \\ \midrule
Q2 & \begin{tabular}[c]{@{}l@{}}How do you want to use the annotated data? Common answers include further analysis, training an AI model, presenting it to people, \\or using it in some downstream tasks inside some computer system. Try to give us more details about the use cases of the annotated data. \end{tabular} \\ \midrule
Q3 & \begin{tabular}[c]{@{}l@{}}What are these data? Please tell us more about the source and the characteristics of the data. For example, ``These are real-world product \\reviews written by Amazon users. We obtained this dataset by downloading it from Kaggle.'' or ``This is the transcript of interviews of our \\participants. Each interview is about 30 minutes long. The interview is about their experience in creative writing.'' or ``These are tweets \\posted on Twitter between Jan 2024 to March 2024.'' \end{tabular} \\ \midrule
Q4 & \begin{tabular}[c]{@{}l@{}}What is the size of each data instance (each row)? For example, ``Each instance is a tweet.'', ``Each instance is one Amazon product review.'', \\or ``Each instance is one sentence from the interview transcript.'' \end{tabular} \\ \midrule
Q5 & \begin{tabular}[c]{@{}l@{}}Is there anything particular you want us to mention in the prompt? We will add all the context you mentioned in this tab to the prompt \\for LLMs. Please mention anything you want the LLMs to be aware of. \end{tabular} 
\\ \bottomrule
\end{tabular}
\caption{Questions for \system Task Context.}
\label{tab:task-sheet-questions}
% \vspace{-5mm}
\end{table*}

% \begin{table*}
% \small
% \centering
% \begin{tabular}{@{}lll@{}}
% \toprule
% \textbf{\#} & \textbf{Aspect} & \textbf{Questionnaire Question} \\ \midrule
% Q1 & Reach-Out & \begin{tabular}[c]{@{}l@{}}If I have this question, I would reach out to other people, such as friends, family members, colleagues,\\ or experts, to get help. As compared to finding the answers by looking up information by myself with\\ a computer or a smartphone.\\(1) Strongly Disagree  (2) Disagree  (3) Neutral  (4) Agree  (5) Strongly Agree \end{tabular} \\ \midrule

% \\ \bottomrule
% \end{tabular}
% \caption{Questionnaire questions used for \system.}
% \label{tab:study-survey}
% \end{table*}

\section{Post-Study Survey for the user study}\label{sec:post-question-survey}
Table~\ref{tab:study-survey} shows the questions for the post-study survey.
\begin{table*}
\footnotesize
\centering
\begin{tabular}{@{}lll@{}}
\toprule
\textbf{\#} & \textbf{Aspect} & \textbf{Post-Study Survey Question} \\ \midrule
Q1 & Task Experience & \begin{tabular}[c]{@{}l@{}}The annotation task was easy to understand. \\(1) Strongly Disagree  (2) Disagree  (3) Somewhat Disagree (4) Neutral  (5) Somewhat Agree (6) Agree (7) Strongly Agree \end{tabular} \\ \midrule
Q2 & Task Experience & \begin{tabular}[c]{@{}l@{}}The annotation tool is easy to use.\\(1) Strongly Disagree  (2) Disagree  (3) Somewhat Disagree (4) Neutral  (5) Somewhat Agree (6) Agree (7) Strongly Agree \end{tabular} \\ \midrule
Q3 & Task Experience & \begin{tabular}[c]{@{}l@{}}Did you encounter any difficulties while using the system? If yes, please describe the difficulties. \end{tabular} \\ \midrule
Q4 & Task Experience & \begin{tabular}[c]{@{}l@{}}The interface of the annotation system was intuitive. \\(1) Strongly Disagree  (2) Disagree  (3) Somewhat Disagree (4) Neutral  (5) Somewhat Agree (6) Agree (7) Strongly Agree \end{tabular} \\ \midrule
Q5 & Task Experience & \begin{tabular}[c]{@{}l@{}}How satisfied are you with the performance of the system? \\(1) Strongly Disagree  (2) Disagree  (3) Somewhat Disagree (4) Neutral  (5) Somewhat Agree (6) Agree (7) Strongly Agree \end{tabular} \\ \midrule
Q6 & Practice Prompting & \begin{tabular}[c]{@{}l@{}}This tool was helpful in improving my prompt. \\(1) Strongly Disagree  (2) Disagree  (3) Somewhat Disagree (4) Neutral  (5) Somewhat Agree (6) Agree (7) Strongly Agree \end{tabular} \\ \midrule
Q7 & Practice Prompting & \begin{tabular}[c]{@{}l@{}}Using this tool made the process of prompt engineering more efficient. \\(1) Strongly Disagree  (2) Disagree  (3) Somewhat Disagree (4) Neutral  (5) Somewhat Agree (6) Agree (7) Strongly Agree \end{tabular} \\ \midrule
Q8 & Practice Prompting & \begin{tabular}[c]{@{}l@{}}Without this tool, how would you typically approach prompt engineering? \end{tabular} \\ \midrule
Q9 & Practice Prompting & \begin{tabular}[c]{@{}l@{}}How would you compare your prompt engineering process before and after using this tool? \end{tabular} \\ \midrule
Q10 & System Usability & \begin{tabular}[c]{@{}l@{}}What features did you find most useful? \end{tabular} \\ \midrule
Q11 & System Usability & \begin{tabular}[c]{@{}l@{}}What features did you find least useful or problematic? \end{tabular} \\ \midrule
Q12 & System Usability & \begin{tabular}[c]{@{}l@{}}Did you feel the need for any additional features or improvements? If yes, please describe them.\end{tabular} \\ \midrule
Q13 & Overall Feedback & \begin{tabular}[c]{@{}l@{}}What did you like most about the annotation system?\end{tabular} \\ \midrule
Q14 & Overall Feedback & \begin{tabular}[c]{@{}l@{}}What did you like least about the annotation system?\end{tabular} \\ \midrule
Q15 & Overall Feedback & \begin{tabular}[c]{@{}l@{}}Any additional comments or suggestions?\end{tabular} \\ \midrule
Q16 & User Interaction and Behavior & \begin{tabular}[c]{@{}l@{}}Did you find the system responsive to your actions? If yes, please describe them.\end{tabular} \\ \midrule
Q17 & User Interaction and Behavior & \begin{tabular}[c]{@{}l@{}}Were there any delays or lags while performing the tasks? If yes, please describe them\end{tabular} \\ \midrule
Q18 & User Interaction and Behavior & \begin{tabular}[c]{@{}l@{}}Did you use any help or support features provided by the system? If yes, were they helpful?\end{tabular} \\ \midrule
Q19 & Efficiency and Effectiveness & \begin{tabular}[c]{@{}l@{}}I completed the annotation tasks efficiently. \\(1) Strongly Disagree  (2) Disagree  (3) Somewhat Disagree (4) Neutral  (5) Somewhat Agree (6) Agree (7) Strongly Agree \end{tabular} \\ \midrule
Q20 & Efficiency and Effectiveness & \begin{tabular}[c]{@{}l@{}}Did the system help you complete the tasks more efficiently? If yes, please explain how.\end{tabular} \\ \midrule
Q21 & Future Use & \begin{tabular}[c]{@{}l@{}}Would you be interested in using this annotation system in your regular work or study? If no, please explain why.\end{tabular} \\ \midrule
Q22 & Future Use & \begin{tabular}[c]{@{}l@{}}Do you have any suggestions for making the system more suitable for your needs?\end{tabular}
\\ \bottomrule
\end{tabular}
\caption{Post-Study Survey questions used for \system. The survey is consisted of twenty-two questions, including seven Likert scale ratings and fifteen free-text responses.}
\label{tab:study-survey}
% \vspace{-5mm}
\end{table*}

% \begin{table*}
% \small
% \centering
% \begin{tabular}{@{}lll@{}}
% \toprule
% \textbf{\#} & \textbf{Aspect} & \textbf{Questionnaire Question} \\ \midrule
% Q1 & Reach-Out & \begin{tabular}[c]{@{}l@{}}If I have this question, I would reach out to other people, such as friends, family members, colleagues,\\ or experts, to get help. As compared to finding the answers by looking up information by myself with\\ a computer or a smartphone.\\(1) Strongly Disagree  (2) Disagree  (3) Neutral  (4) Agree  (5) Strongly Agree \end{tabular} \\ \midrule

% \\ \bottomrule
% \end{tabular}
% \caption{Questionnaire questions used for \system.}
% \label{tab:study-survey}
% \end{table*}

\section{Prompt}
Table~\ref{tab:instruction-prompt} shows the prompt \system used for generating instructional prompt.
\begin{table*}[h]
\centering
    \begin{tabular}{@{}p{\textwidth}@{}}
        \hline
        \textbf{Instructional Prompt Creation Prompt} \\
        Here are questions and corresponding answers for a task description.\\ \\

        ```\\
        \textbf{Question:} [What is the purpose of annotating this data? Common answers include gaining insight about something, wanting to compare something, wanting to create prompts, etc. Try to give us more details about your higher-level goal.] \textbf{Answer:} [Answer 1...] \\
        \textbf{Question:} [How do you want to use the annotated data? Common answers include further analysis, training an AI model, presenting it to people, or using it in some downstream tasks inside some computer system. Try to give us more details about the use cases of the annotated data.] \textbf{Answer:} [Answer 2...] \\
        \textbf{Question:} [What are these data? Please tell us more about the source and the characteristics of the data. For example, ``These are real-world product reviews written by Amazon users. We obtained this dataset by downloading it from Kaggle.'' or ``This is the transcript of interviews of our participants. Each interview is about 30 minutes long. The interview is about their experience in creative writing.'' or ``These are tweets posted on Twitter between Jan 2024 to March 2024.''] \textbf{Answer:} [Answer 3...] \\
        \textbf{Question:} [What is the size of each data instance (each row)? For example, ``Each instance is a tweet.'', ``Each instance is one Amazon product review.'', or ``Each instance is one sentence from the interview transcript.''] \textbf{Answer:} [Answer 4...] \\
        \textbf{Question:} [Is there anything particular you want us to mention in the prompt? We will add all the context you mentioned in this tab to the prompt for LLMs. Please mention anything you want the LLMs to be aware of.] \textbf{Answer:} [Answer 5...] \\
        \textbf{Question:} [Is it a single-class or multi-class labeling task? [required]] \textbf{Answer:} [single-class]
        ......\\
        '''\\
        \\
        Based on task questions and answers, help me generate a concrete DETAILED task instruction.\\
        Provide Instruction ONLY!\\
        DO NOT ADD ANY ADDITIONAL INFORMATION NOT INCLUDE IN THE PREVIOUS Q and A!!!\\
        This Instruction is generated for LLM!\\
        \hline
    \end{tabular}
    \caption{Prompt to generate instructional prompts in \system. Participants are required to answer the first five questions to provide context for the task. The last question and its answer were intentionally fixed, as our study focuses on a single-class labeling task. In the future, participants will be allowed to answer this question.}
    \label{tab:instruction-prompt}
\end{table*}

%\steven{added instructional prompt generation prompt}

Table~\ref{tab:main-prompt} and Table~\ref{tab:main-multi-prompt} show the prompts \system used for the annotation process.
\begin{table*}[h]
\centering
    \begin{tabular}{@{}p{\textwidth}@{}}
        \hline
        \textbf{Prompt for Annotation Process} \\
        \{\textbf{\$Instructional\_Prompt}\}\\ \\
        Please ensure each label adheres to its following rules and regulations.\\
        \\
        Below are the descriptions of various labels. Please assign the most appropriate label to each description provided.\\
        Label `\{\textbf{\$label\_1}\}': Assign this label if the tweet meets any of the following criteria:\\
        \{\textbf{\$rule\_1}\}\\ 
        Label `\{\textbf{\$label\_2}\}': Assign this label if the tweet meets any of the following criteria:\\
        \{\textbf{\$rule\_2}\}\\ 
        Label `\{\textbf{\$label\_3}\}': Assign this label if the tweet meets any of the following criteria:\\
        \{\textbf{\$rule\_3}\}\\ 
        Label `\{\textbf{\$label\_4}\}': Assign this label if the tweet meets any of the following criteria:\\
        \{\textbf{\$rule\_4}\}\\
        Label `\{\textbf{\$label\_5}\}': Assign this label if the tweet meets any of the following criteria:\\
        \{\textbf{\$rule\_5}\}\\
        ......\\ \\
        Please refer to the following Shots (Examples for LLMs to Learn) for annotation tasks, where each instance is corresponded with a label.\\
        Example:```{\textbf{\$instance\_1}}''' $=>$ Label:```{\textbf{\$label\_x}}'''\\
        Example:```{\textbf{\$instance\_2}}''' $=>$ Label:```{\textbf{\$label\_x}}'''\\
        Example:```{\textbf{\$instance\_3}}''' $=>$ Label:```{\textbf{\$label\_x}}'''\\
        ......\\ \\
        Output Format\\
        Your output should consist of two sections: ANSWER and EXPLANATION.\\
        ANSWER: Label: []\\
        EXPLANATION: Provide a brief explanation for your label choice.\\
        The following is the data instance need to be annotated:\\
        data-instance: \{\textbf{\$data\_instance}\}\\
        \hline
    \end{tabular}
    \caption{Prompt for the annotation process in \system for groups which have only one data instance.}
    \label{tab:main-prompt}
\end{table*}

%\steven{added prompt for single instance like Twitter sentiment task}

\begin{table*}[h]
\centering
    \begin{tabular}{@{}p{\textwidth}@{}}
        \hline
        \textbf{Prompt for Annotation Process} \\
        \{\textbf{\$Instructional\_Prompt}\}\\ \\
        Please ensure each label adheres to its following rules and regulations.\\
        \\
        Below are the descriptions of various labels. Please assign the most appropriate label to each description provided.\\
        Label `\{\textbf{\$label\_1}\}': Assign this label if the tweet meets any of the following criteria:\\
        \{\textbf{\$rule\_1}\}\\ 
        Label `\{\textbf{\$label\_2}\}': Assign this label if the tweet meets any of the following criteria:\\
        \{\textbf{\$rule\_2}\}\\ 
        Label `\{\textbf{\$label\_3}\}': Assign this label if the tweet meets any of the following criteria:\\
        \{\textbf{\$rule\_3}\}\\ 
        Label `\{\textbf{\$label\_4}\}': Assign this label if the tweet meets any of the following criteria:\\
        \{\textbf{\$rule\_4}\}\\
        Label `\{\textbf{\$label\_5}\}': Assign this label if the tweet meets any of the following criteria:\\
        \{\textbf{\$rule\_5}\}\\
        ......\\ \\
        Please refer to the following Shots (Examples for LLMs to Learn) for annotation tasks, where each instance is corresponded with a label.\\
        Example:```{\textbf{\$instance\_1}}''' $=>$ Label:```{\textbf{\$label\_x}}'''\\
        Example:```{\textbf{\$instance\_2}}''' $=>$ Label:```{\textbf{\$label\_x}}'''\\
        Example:```{\textbf{\$instance\_3}}''' $=>$ Label:```{\textbf{\$label\_x}}'''\\
        ......\\ \\
        Output Format\\
        For each labeled data instance, your output should consist of two sections: ANSWER and EXPLANATION, with the data instance id. Each label fragment should be divided by ``======''\\
        ANSWER: Label: []\\
        EXPLANATION: Provide a brief explanation for your label choice.\\
        The following are data instances from a group that need to be annotated:\\
        data-instance-1: \{\textbf{\$data\_instance\_1}\}\\
        data-instance-2: \{\textbf{\$data\_instance\_2}\}\\
        data-instance-3: \{\textbf{\$data\_instance\_3}\}\\
        ......\\
        \hline
    \end{tabular}
    \caption{Prompt for the annotation process in \system for groups which have multiple data instances.}
    \label{tab:main-multi-prompt}
\end{table*}

%\steven{added prompt for tasks require context like CODA-19}

\section{Supplemental System Figures}
\begin{figure*}
    \centering
    \includegraphics[width=0.90\linewidth]{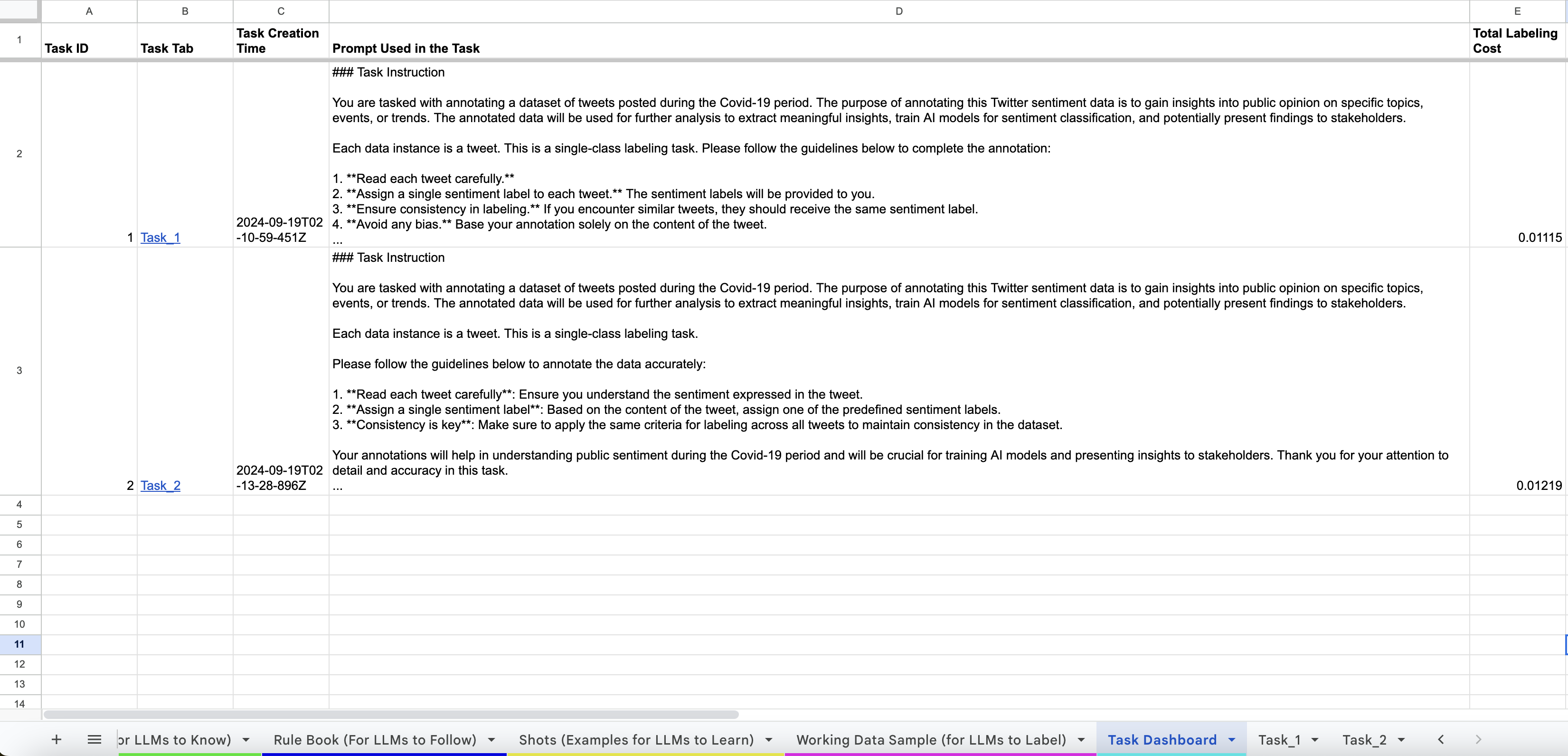}\Description{The task dashboard for PromptingSheet. This spreadsheet records key details, including the Task ID, Task Tab, Task Creation Time, Prompt Used for the Task, and Total Label Cost.}
    \caption{The Task Dashboard tab provides an overview of all labeling tasks. This spreadsheet records key details, including the Task ID, Task Tab, Task Creation Time, Prompt Used for the Task, and Total Label Cost. Users can click the hyperlink in the Task Tab column to access the corresponding task tab for more detailed information.}
    \label{fig:task-dashboard-new}
\end{figure*}

Figure~\ref{fig:task-dashboard-new} shows the task dashboard layout in \system.

%\steven{response to task dashboard question}

\section{Supplemental ACC and MSE Figure}
\begin{figure*}
    \centering
    \begin{subfigure}{0.48\textwidth}
     \includegraphics[width=\linewidth]{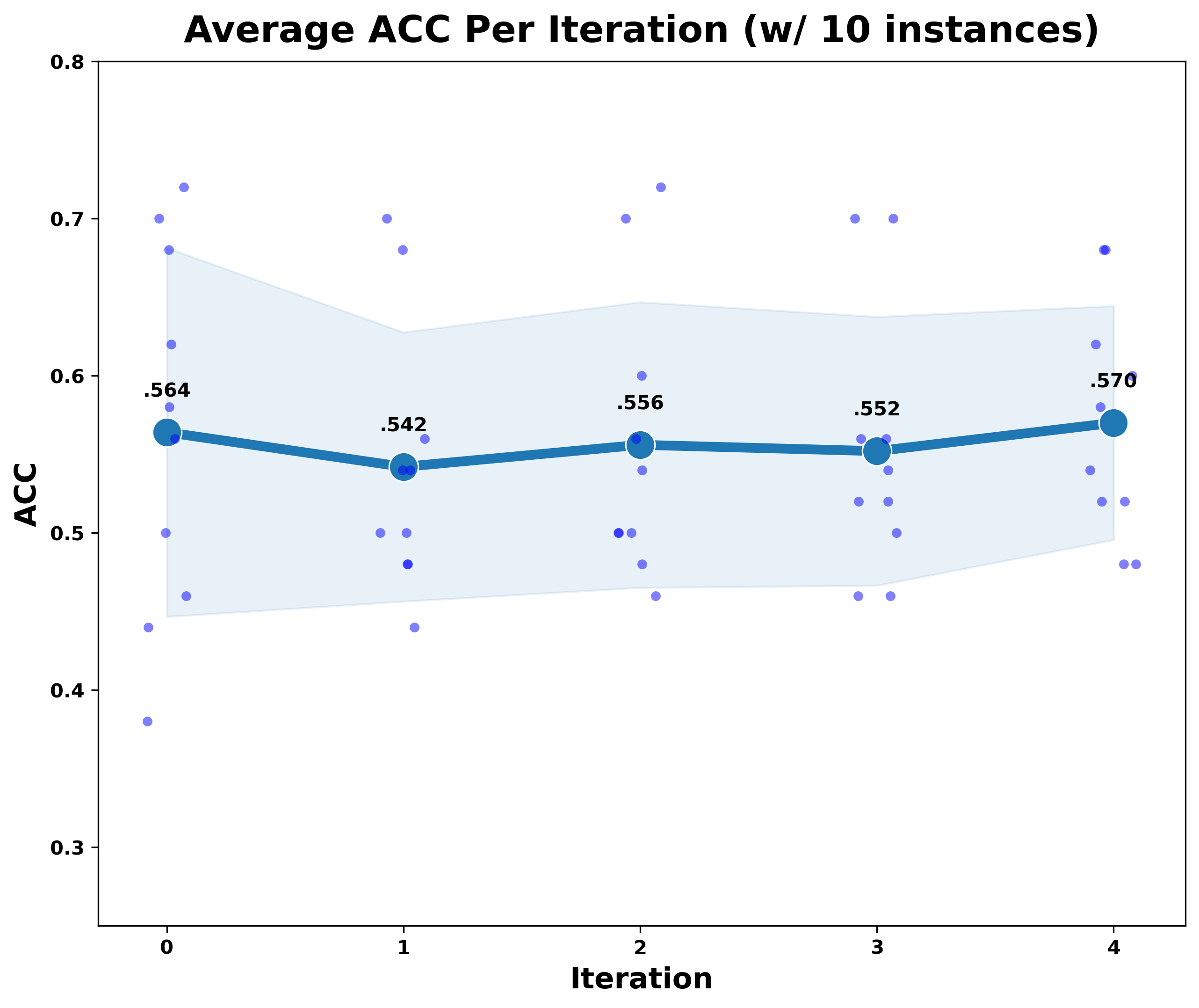}\Description{This subplot shows the average ACC for participants reviewing 10 instances. The plot includes shaded regions representing the standard deviation and scatter points representing the data of individual participants.}
    \caption{ACC with 10 instances.}
    \label{fig:new-lineplot-acc-10}
  \end{subfigure}
  \hfill
  \begin{subfigure}{0.48\textwidth}
    \includegraphics[width=\linewidth]{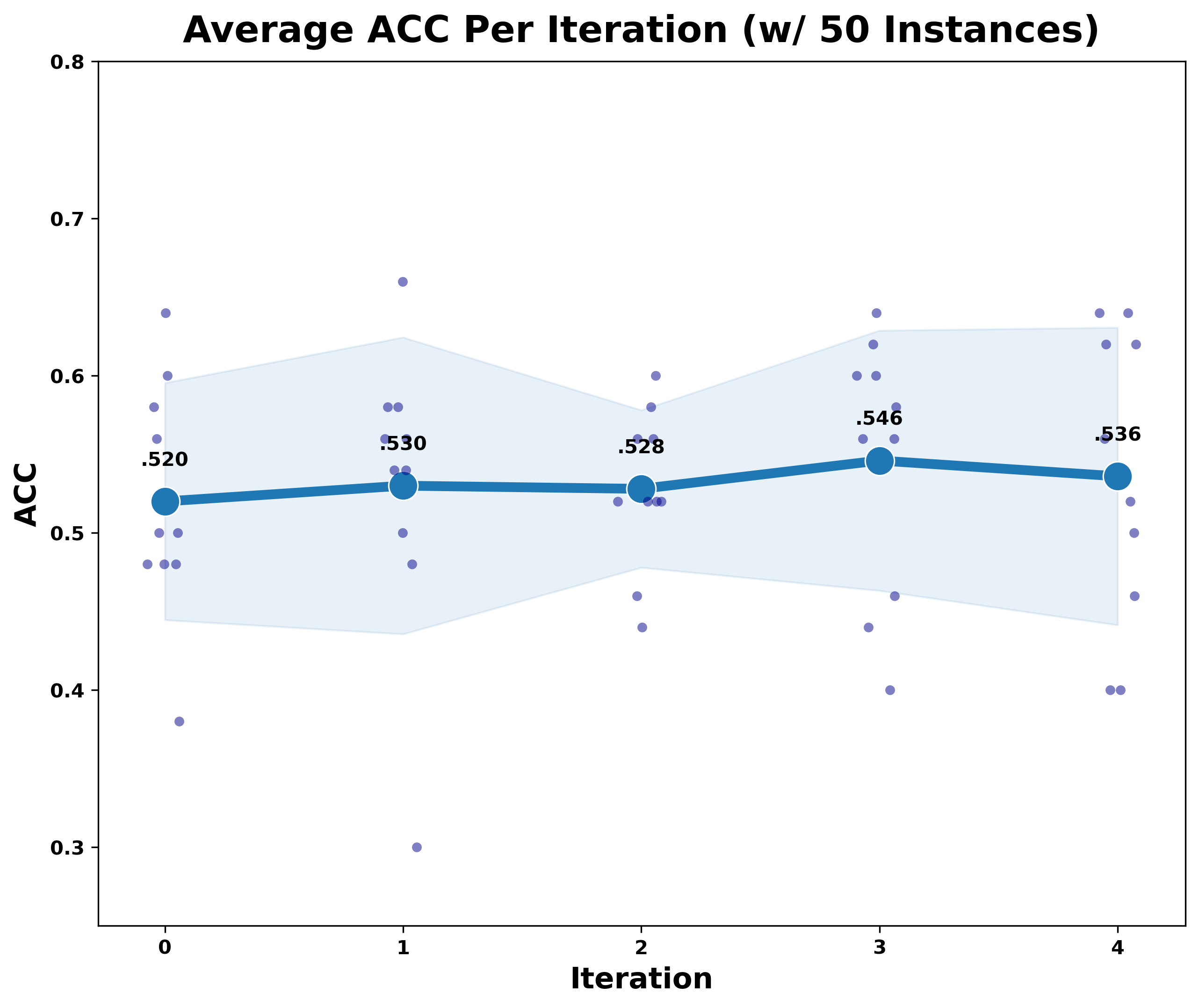}\Description{This subplot shows the average ACC for participants reviewing 50 instances. The plot includes shaded regions representing the standard deviation and scatter points representing the data of individual participants.}
    \caption{ACC with 50 instances.}
    \label{fig:new-lineplot-acc-50}
  \end{subfigure}
  \begin{subfigure}{0.48\textwidth}
     \includegraphics[width=\linewidth]{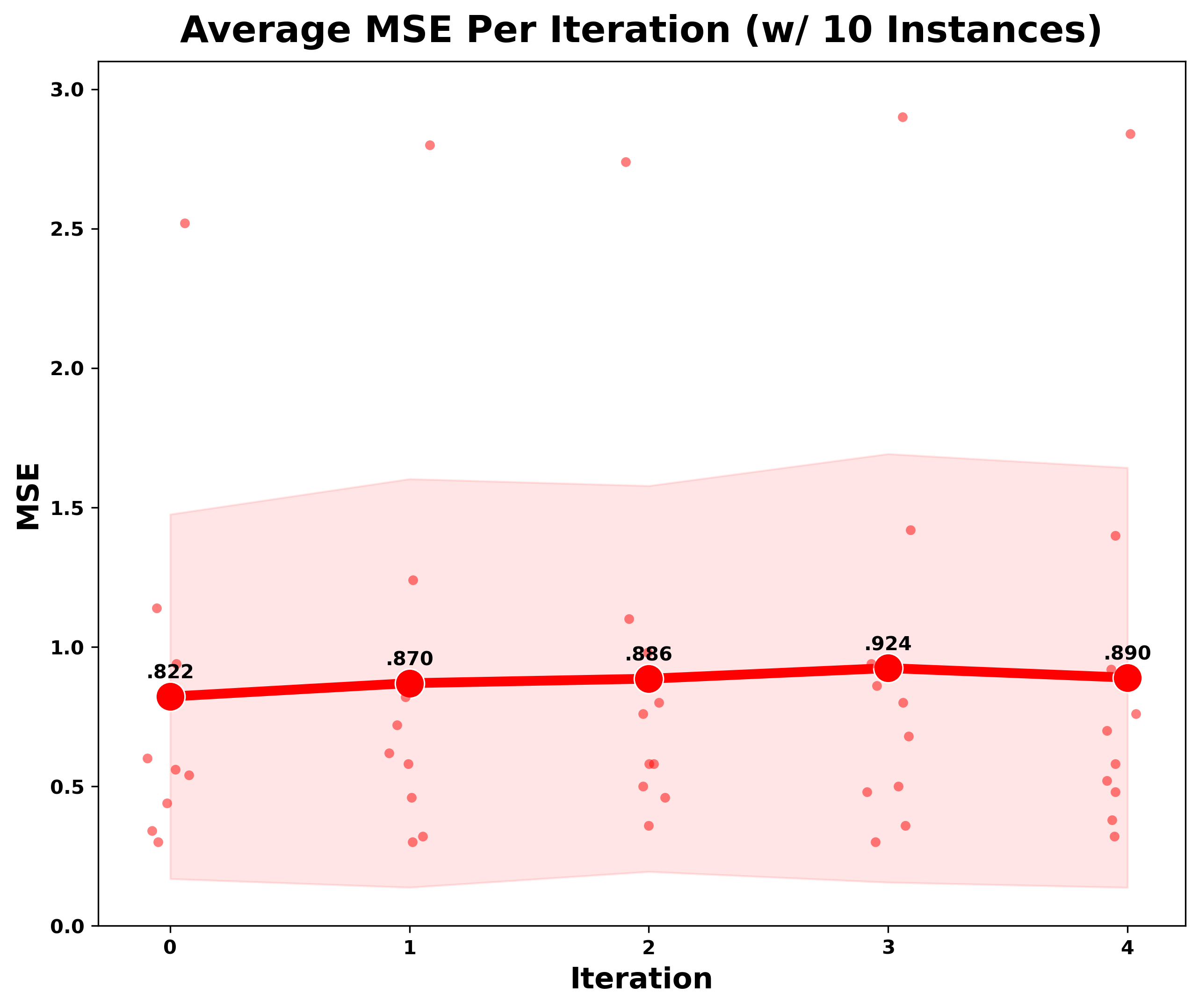}\Description{This subplot shows the average MSE for participants reviewing 10 instances. The plot includes shaded regions representing the standard deviation and scatter points representing the data of individual participants.}
    \caption{MSE with 10 instances.}
    \label{fig:new-lineplot-mse-10}
  \end{subfigure}
  \hfill
  \begin{subfigure}{0.48\textwidth}
    \includegraphics[width=\linewidth]{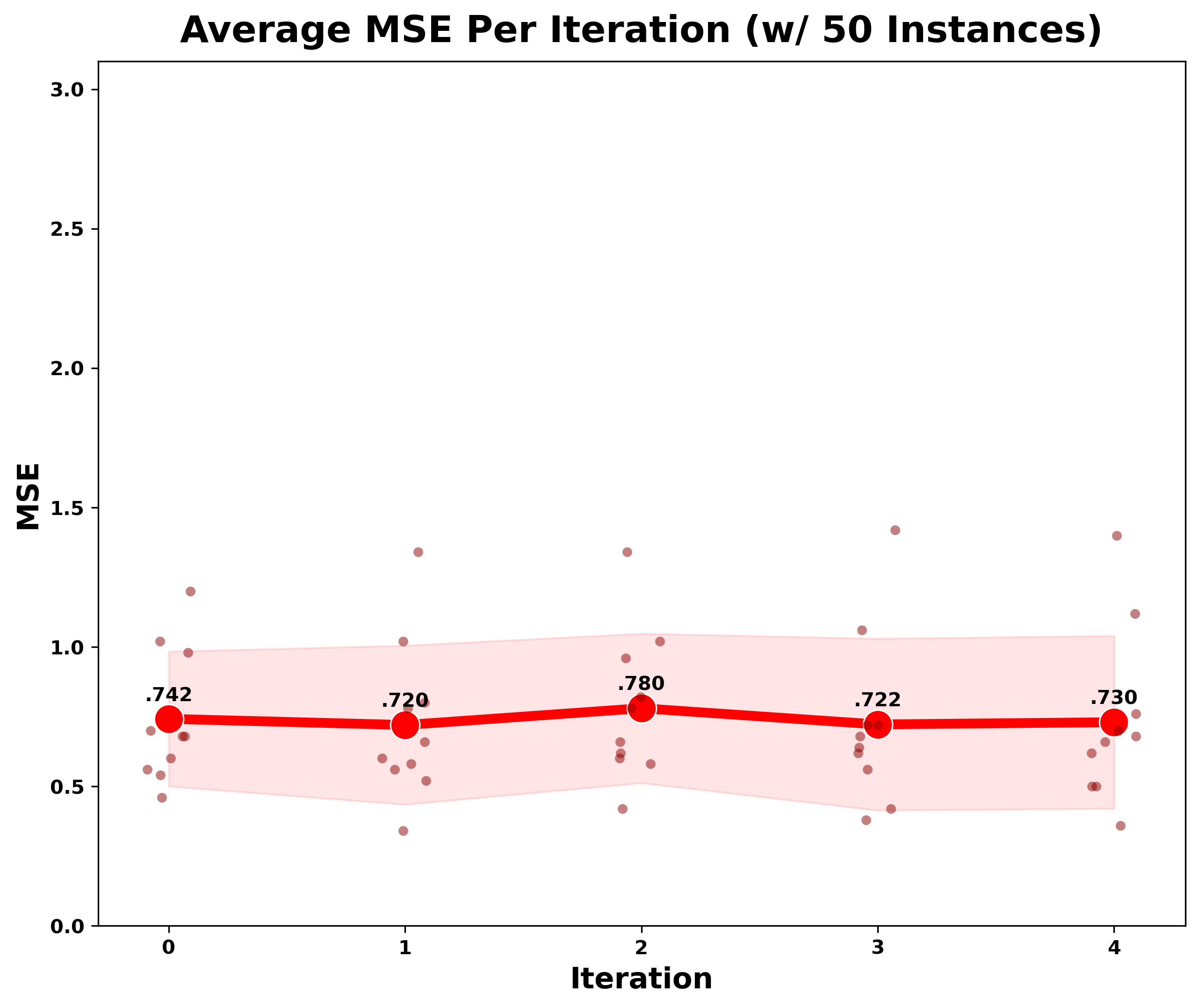}\Description{This subplot shows the average MSE for participants reviewing 50 instances. The plot includes shaded regions representing the standard deviation and scatter points representing the data of individual participants.}
    \caption{MSE with 50 instances.}
    \label{fig:new-lineplot-mse-50}
  \end{subfigure}
  \caption{The average ACC and MSE for participants reviewing 10 or 50 instances per iteration are presented in four subfigures, comparing ACC and MSE between the two conditions.}
  \label{fig:average-acc-mse-llm-instances-performance-new}
\end{figure*}

\begin{figure*}
    \centering
    \begin{subfigure}{0.48\textwidth}
     \includegraphics[width=\linewidth]{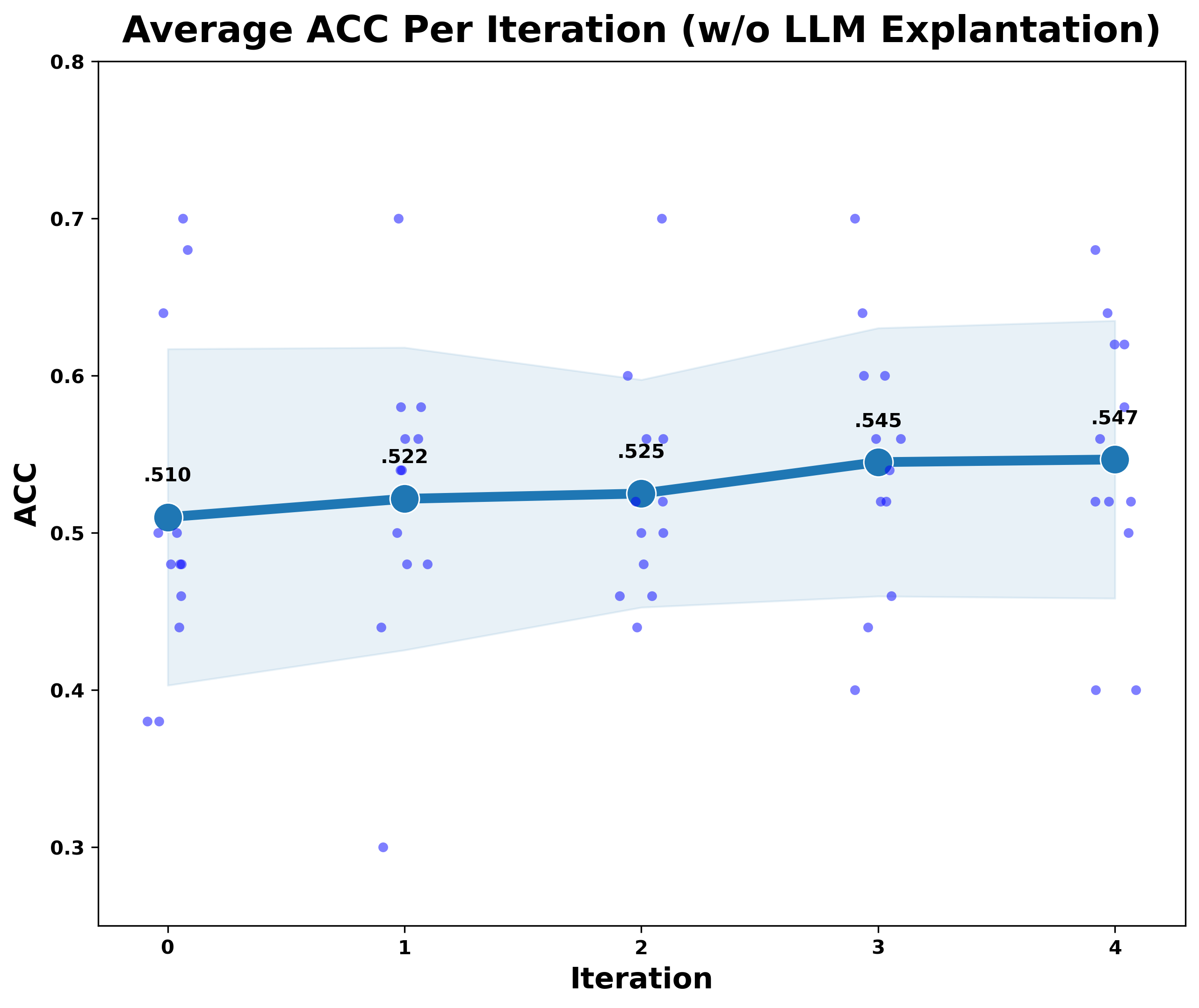}\Description{This subplot shows the average ACC for participants without access to LLM explanation. The plot includes shaded regions representing the standard deviation and scatter points representing the data of individual participants.}
    \caption{ACC without LLM Explanation.}
    \label{fig:new-lineplot-acc-no-exp}
  \end{subfigure}
  \hfill
  \begin{subfigure}{0.48\textwidth}
    \includegraphics[width=\linewidth]{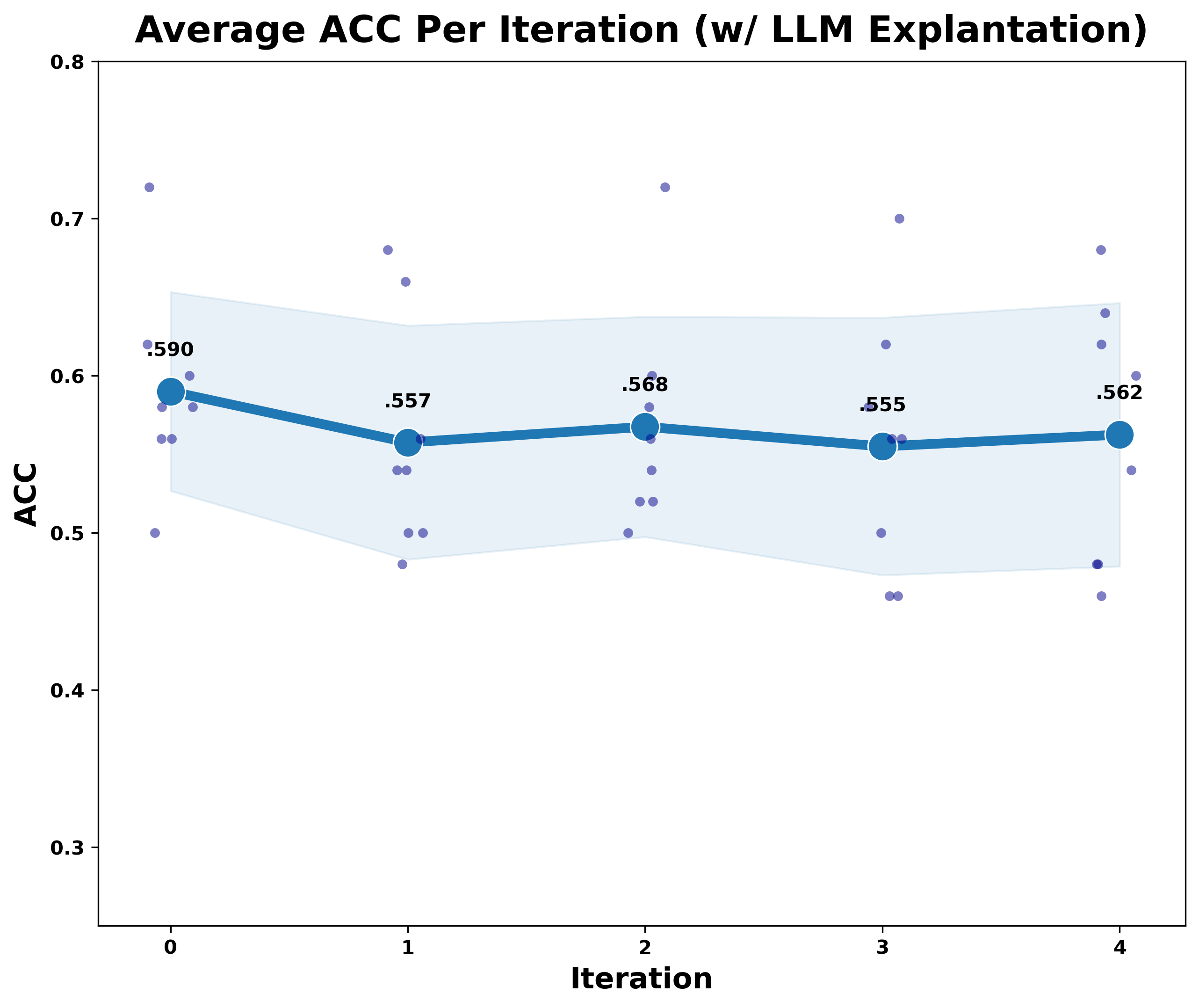}\Description{This subplot shows the average ACC for participants with access to LLM explanation. The plot includes shaded regions representing the standard deviation and scatter points representing the data of individual participants.}
    \caption{ACC with LLM Explanation.}
    \label{fig:new-lineplot-acc-exp}
  \end{subfigure}
  \begin{subfigure}{0.48\textwidth}
     \includegraphics[width=\linewidth]{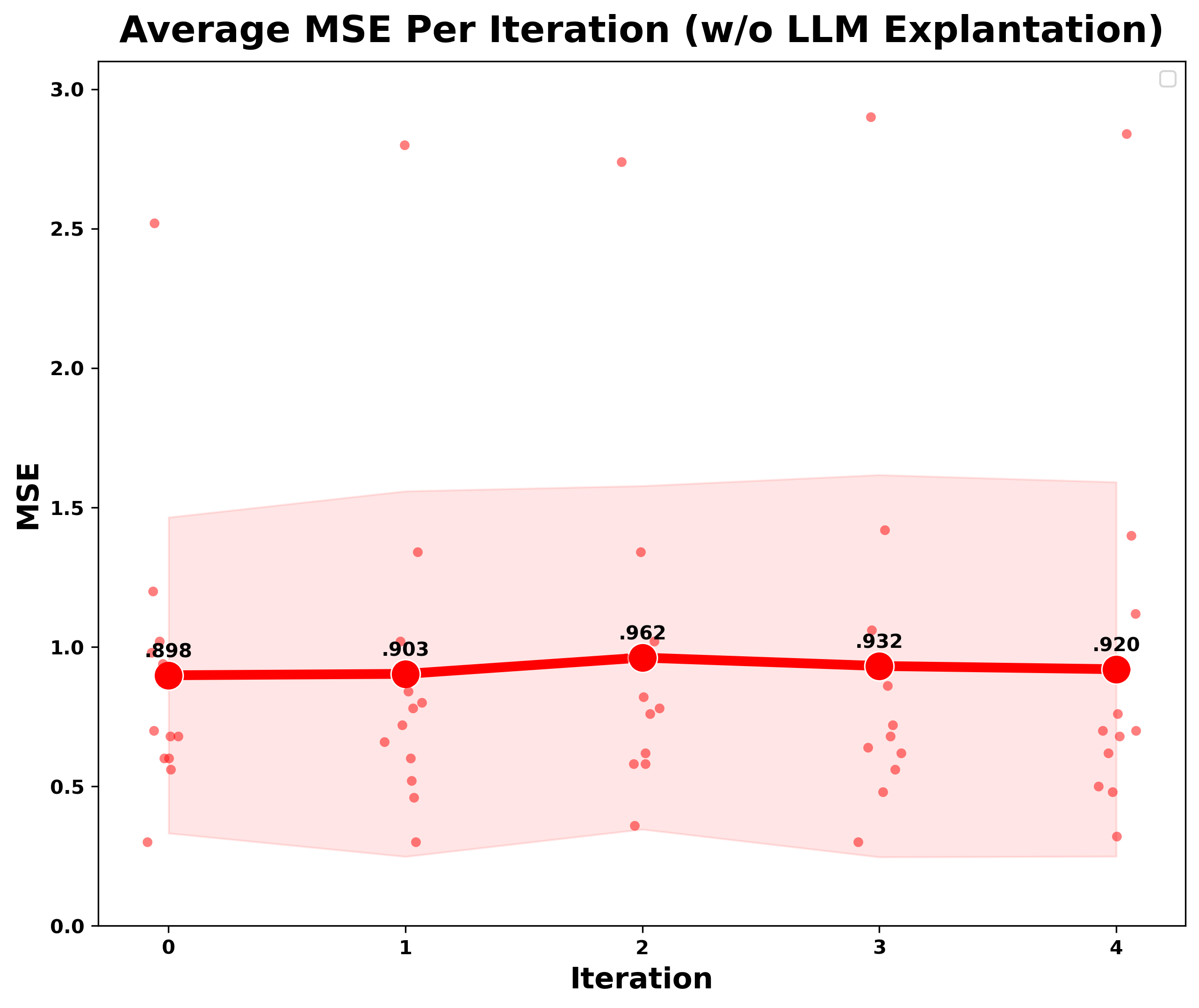}\Description{This subplot shows the average MSE for participants without access to LLM explanation. The plot includes shaded regions representing the standard deviation and scatter points representing the data of individual participants.}
    \caption{MSE without LLM Explanation.}
    \label{fig:new-lineplot-mse-no-exp}
  \end{subfigure}
  \hfill
  \begin{subfigure}{0.48\textwidth}
    \includegraphics[width=\linewidth]{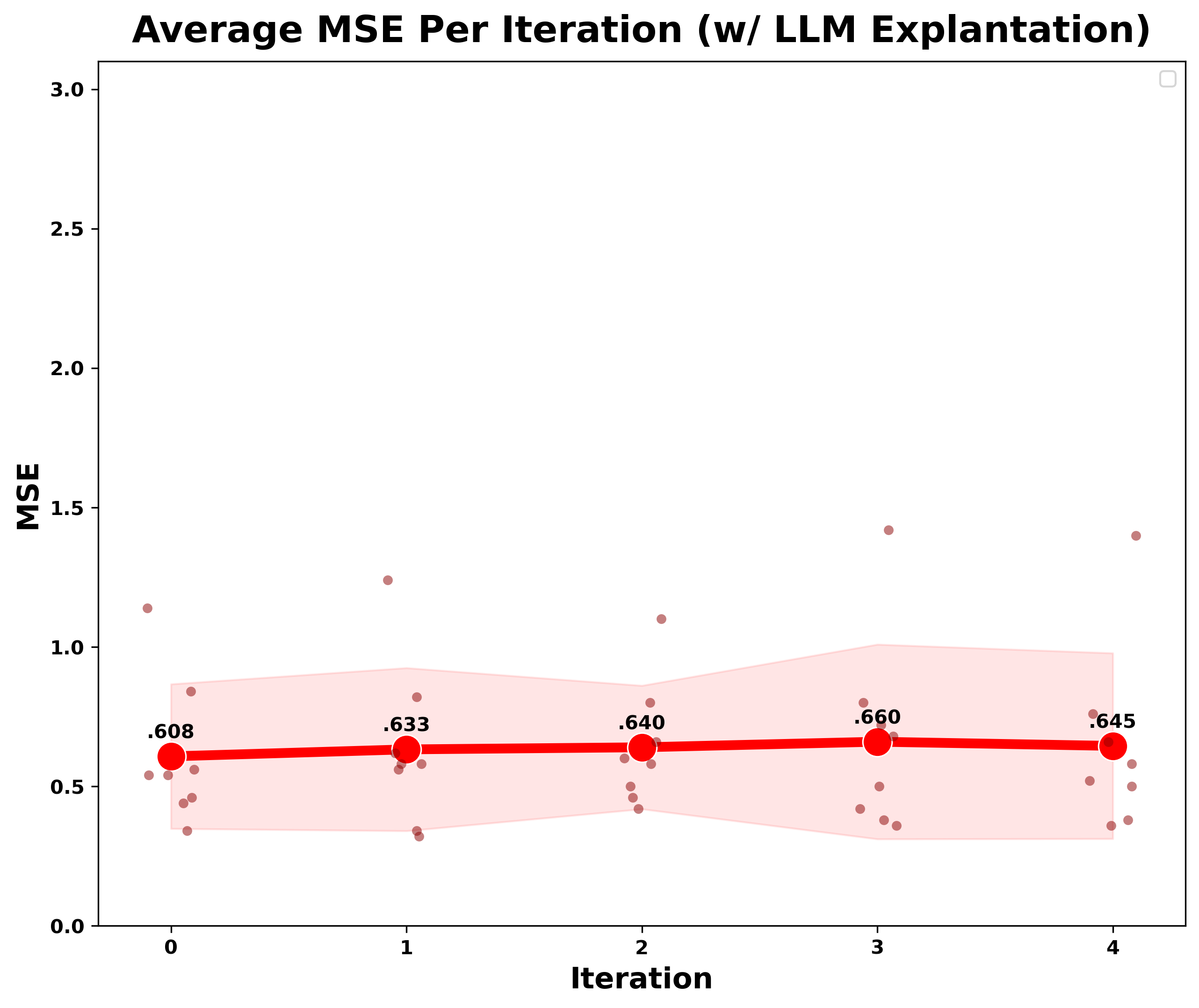}\Description{This subplot shows the average MSE for participants with access to LLM explanation. The plot includes shaded regions representing the standard deviation and scatter points representing the data of individual participants.}
    \caption{MSE with LLM Explanation.}
    \label{fig:new-lineplot-mse-exp}
  \end{subfigure}
  \caption{The average ACC and MSE for participants with or without LLM Explanation are presented in four subfigures, comparing ACC and MSE between the two conditions.}
  \label{fig:average-acc-mse-llm-explanation-performance-new}
\end{figure*}

Figure~\ref{fig:average-acc-mse-llm-instances-performance-new} and Figure~\ref{fig:average-acc-mse-llm-explanation-performance-new}  present the average ACC and MSE for participants under different conditions: comparing 10 vs. 50 instances per iteration and assessing the impact of LLM explanations.

%\steven{new: ACC comparison between 10 and 50 instances}

%\steven{new: ACC comparison between no explanation and explanation}

% \section{Survey Study for Understanding How Individuals Use LLMs in Data Annotation}
% \input{Tables/data-annotation-survey-questions}

\section{Questionnaires for Participants' LLM Background}\label{sec:appendix=participant-background}
\begin{table*}[h]
\footnotesize
\centering
\begin{tabular}{@{}lll@{}}
\toprule
\textbf{\#} & \textbf{Questionnaire Question} \\ \midrule
Q1 &  \begin{tabular}[c]{@{}l@{}}How would you rate your familiarity with using LLMs on a scale from 1 (not familiar) to 5 (very familiar)?\end{tabular} \\ \midrule
Q2 & \begin{tabular}[c]{@{}l@{}}Which of the following best describes your understanding of how LLMs work?\\ 
(1) I don’t understand at all.\\
(2) I have a basic understanding (e.g., they generate text based on input).\\
(3) I understand the general principles (e.g., machine learning, large datasets).\\
(4) I have a detailed understanding (e.g., specific architectures, training methods).
\end{tabular} \\ \midrule
Q3 & \begin{tabular}[c]{@{}l@{}}What are your years of experience in using LLMs? \\
(1) Less than a month\\
(2) 1-3 months\\
(3) 4 months to 1 year\\
(4) More than 1 year
\end{tabular} \\ \midrule
Q4 & \begin{tabular}[c]{@{}l@{}}How often do you use LLMs?\\
(1) Daily\\
(2) Weekly\\
(3) Bi-Weekly\\
(4) Monthly\\
(5) Never
\end{tabular} \\ \midrule
Q5 & \begin{tabular}[c]{@{}l@{}}How much time do you typically spend interacting with LLMs in a single session?\\
(1) Less than 5 minutes\\
(2) 5–15 minutes\\
(3) 15–30 minutes\\
(4) More than 30 minutes
\end{tabular} \\ \midrule
Q6 & \begin{tabular}[c]{@{}l@{}}In which contexts have you used LLMs? (Select all that apply)\\
(1) Academic research\\
(2) Professional work\\
(3) Personal projects\\
(4) Entertainment\\
(5) Other
\end{tabular} \\ \midrule
Q7 & \begin{tabular}[c]{@{}l@{}}For which tasks do you use LLMs? (Select all that apply)\\
(1) Data Labeling\\
(2) General Q\&A or research\\
(3) Writing assistance (e.g., drafting emails, reports)\\
(4) Programming or debugging\\
(5) Data analysis or visualization\\
(6) Creative tasks (e.g., storytelling, idea generation)\\
(7) Other
\end{tabular} \\ \midrule
Q8 & \begin{tabular}[c]{@{}l@{}}How confident are you in your ability to craft effective prompts for LLMs on a scale from 1 (beginner) to 5 (expert)?
\end{tabular} \\ \midrule
Q9 & \begin{tabular}[c]{@{}l@{}}How proficient do you feel in troubleshooting when an LLM generates incorrect or irrelevant responses on a scale from 1 (not proficient) to 5 \\(extremely proficient)?
\end{tabular} \\ \midrule
Q10 & \begin{tabular}[c]{@{}l@{}}Do you use any advanced techniques like prompt engineering or API integration with LLMs? If yes, please describe.
\end{tabular}
\\ \bottomrule
\end{tabular}
\caption{Questionnaire questions used for participants' LLM background.}
\label{tab:participants-llm-background-survey}
% \vspace{-5mm}
\end{table*}

\section{Post-Study Survey Questions with Likert-Scale Ratings\label{app:two-var-on-rating}}

The survey questions and the accompanying categories were rated on a seven-point Likert scale, 
as discussed in 
Section~\ref{sec:two-var-on-rating},
listed below:

\begin{itemize}
    \item 
    \textbf{(Q1) Understandable}: The annotation task was easy to understand.

    \item
    \textbf{(Q2) Ease of Use}: The annotation tool is easy to use.

    \item
    \textbf{(Q4) Intuitive System}: The interface of the annotation system is intuitive.

    \item
    \textbf{(Q5) Performance Satisfaction}: How satisfied are you with the performance of the system?

    \item
    \textbf{(Q6) Prompt Improvement}: This tool was helpful in improving my prompt. 
    
    \item
    \textbf{(Q7) Process Efficiency}: Using this tool made the process of prompt engineering more efficient.

    \item
    \textbf{(Q19) Task Completion}: I completed the annotation tasks efficiently.

\end{itemize}
%TC:endignore

\end{document}